\documentclass[iop,apj,twocolumn,numberedappendix]{emulateapj}

\usepackage{hyperref}
\usepackage[section,above]{placeins} 
\usepackage{float}                   
\usepackage{datetime}
\usepackage{xcolor}
\usepackage{graphicx,apjfonts,multirow,amsmath,bm}

\usepackage[normalem]{ulem} 
\newcommand{\del}[1]{}
\newcommand{\add}[1]{#1}
\newcommand{\note}[1]{}
\newcommand{\sacha}[1]{}

\newlength{\picwd}
\newcommand{\e}[1]{\ensuremath{\times 10^{#1}}} %
\newcommand{\un}[1]{\ensuremath{\ \mathrm{#1}}}

\newcommand{\rsun}{\ensuremath{\ R_\odot}}

\newcommand{\calS}{\mbox{\boldmath ${\cal S}$}}

\def\nab{\mbox{\boldmath $\nabla$}}

\def\rb{\bar{\rho}}

\def\tb{\bar{T}}

\def\Om{\mbox{\boldmath $\Omega_0$}}

\newcommand{\DD}{\mbox{\boldmath ${\cal D}$}}

\def\vec#1{\mbox{\boldmath$\displaystyle#1$}}

\newcommand{\degrees}{^\circ}

\newcommand{\curl}{\mbox{\boldmath $\nabla \times$}}
\newcommand{\cross}{\mbox{\boldmath $\times$}}

\slugcomment{To appear in Astrophysical Journal}

\shorttitle{Flux emergence in a magnetized convection zone}
\shortauthors{Pinto et al.}
 
\begin{document}

\title{Flux emergence in a magnetized convection zone}

\author{R. F. Pinto\textsuperscript{1,2}}

\author{A. S. Brun\textsuperscript{1}}
   
\affil{
  \textsuperscript{1}Laboratoire AIM Paris-Saclay, CEA/Irfu Universit\'e Paris-Diderot CNRS/ 
  INSU, 91191 Gif-sur-Yvette, France \\
  \textsuperscript{2}LESIA, Observatoire de Paris, 5, place Jules Janssen
92195 Meudon, France
}

\email{rui.pinto@cea.fr}

\begin{abstract}

  We study the influence of a dynamo magnetic field on the buoyant rise and emergence of twisted magnetic flux-ropes, and their influence on the global external magnetic field.
  We ran three-dimensional MHD numerical simulations using the ASH code and analysed the dynamical evolution of such buoyant flux-ropes from the bottom of the convection zone until the post-emergence phases.
  The global nature of this model can only very crudely \add{and inaccurately} represent the local dynamics of the buoyant rise of the implanted magnetic structure, but allows nonetheless to study the influence of global effects such as self-consistently generated differential rotation and meridional circulation, and the influence of the Coriolis forces. 
    Although motivated by the solar context, this model cannot be thought of as a realistic model of the rise of magnetic structures and their emergence in the Sun where the local dynamics are completely different.
  The properties of initial phases of the buoyant rise are determined essentially by the flux-rope's properties and the convective flows and are, in consequence, in good agreement with previous studies.
  However, the effects of the interaction of the background dynamo field become increasingly stronger as the flux-ropes evolve.
  During the buoyant rise across the convection zone, the flux-rope's magnetic field strength and scales as $B\propto\rho^{\alpha}$, with $\alpha\lesssim 1$.
  An increase of radial velocity, density and current density is observed to precede flux emergence at all longitudes.
  The geometry, latitude and relative orientation of the flux-ropes with respect to the background magnetic field influences the resulting rise speeds, zonal flows amplitudes (which develop within the flux-ropes) and the corresponding surface signatures.
  This influences the morphology, duration and amplitude of the surface shearing and the Poynting flux associated with magnetic flux-rope emergence.
  The emerged magnetic flux influences the system's global polarity, leading in some cases to a polarity reversal while inhibiting background dynamo from doing so in some others.
  The emerged magnetic flux is slowly advected poleward, while being diffused and assimilated by the background dynamo field.

\end{abstract}

\keywords{Sun: convection --- Sun: dynamo ---  Sun: magnetic fields}

\maketitle

%
\section{Introduction}

%
%
%
%
It is well known that the Sun undergoes recurrent phases of intense magnetic activity.
The most visible signature of magnetic activity is the presence of sunspots and active regions. These correspond to particularly strong concentrations of magnetic field which cross the surface of the Sun as a consequence of the underlying magneto-convective dynamics and the flux-emergence phenomena altogether.
Flux-emergence occurs, nevertheless, at a very broad range of spatial and temporal scales, with the sunspots contributing to only a fraction of the total photospheric magnetic flux \citep{schrijver_sustaining_1997}.
The active regions are often composed of a mixture of large (unipolar) spots and 
small-scale polarities having a broad distribution of life-times ranging from some days up to two solar rotation periods.
During this period of time, they are observed to rotate and have their main polarity pairs separate, giving rise to more complex and dynamic polarity distributions.
These magnetic structures are believed to be the surface tracers of twisted magnetic flux-ropes generated further below (at the tachocline) that rise buoyantly up to the surface.
They then emerge (at least partially) through the photosphere and provide the strong and coherent magnetic field structures composing the solar active regions (AR), which are prone to host eruptive events in the solar corona.
Such events are likely to involve the surface convection, the coronal dynamics, the flux-rope and coronal magnetic flux altogether.
Furthermore, it is now becoming clear that localised magnetic flux emergence is influenced by the large-scale magnetic field structure.
Emergence events feed magnetic flux into the corona and may lead to important reconfigurations of its magnetic field, either in a quasi-steady or in an impulsive way \citep[\emph{e.g},][]{liu_quasi-periodic_2012}.
Conversely, the triggering of flares (or even CME's) depends to a certain extent on the interaction of AR magnetic fields with its surroundings, namely on the orientation and gradients of the coronal field \citep{shibata_two-dimensional_1989,forbes_review_2000,kusano_magnetic_2012}.
Recent SDO observations indicate that flux-emergence on one location can indeed trigger eruptive events on very distant parts of the solar surface \citep{schrijver_long-range_2011}.
This emphasises the idea that some of the physical processes relating to flux-emergence are global in nature.
The causal link between such seemingly independent events remains illusive, and calls for the use of global-scale analysis.

The Sun has been going through its activity cycle for thousands of years, as revealed for instance by the modulation of $^{10}$Be concentration in Earth's polar ice cores \citep{beer_active_1998}.
This cycles show a quasi regular 11 year period (22 if one distinguishes between opposite polarities of the sun's global magnetic field) with a rising phase taking a shorter period of time than the decaying phase \citep{derosa_solar_2012}.
There are some cycle to cycle variations (duration and intensity).
For example, cycle 23 was longer than usual, with several months of unspotted solar surface between 2007 and 2009.
Large sunspots and complex active regions are now regularly appearing as cycle 24 becomes stronger.
Sunspots appearing more frequently during the rising phase of the cycle and at the maximum of activity contribute to the renewal of the coronal field and the global polarity reversal \citep{leighton_magneto-kinematic_1969, wang_magnetic_1991}.

%
%
%
%
%
%
Understanding the physics behind this wealth of inter-related phenomena certainly is a challenging affair.
Numerical MHD simulations have been employed recurrently as a tool to model different aspects of the general flux-rope rise and emergence problem.
The buoyant twisted flux-ropes are generally believed to arise from the tachocline (or nearby) as result of  hydro-magnetic instabilities taking place there.
These have been studied mostly using numerical models with local setups aiming specifically at resolving as well as possible the small-scale processes involved.
A remarkable exception is the work by \citet{nelson_buoyant_2011,nelson_magnetic_2013}, who where able to produce self-consistently several buoyant toroidal structures which develop into omega loops in a fast-rotating convection zone (three times the solar rotation rate).
Their breakthrough relied partly on the implementation of a numerical spatial scheme for the diffusive terms which allowed them to artificially reach substantially higher Reynolds and magnetic Reynolds numbers than otherwise.

Kinematic dynamo models \citep[e.g][]{cameron_solar_2007,jouve_solar_2008} have been shown to be able to reproduce the long time-scale properties of the global magnetic fields \citep[the ``butterfly diagram''][]{jouve_3-d_2007,charbonneau_dynamo_2010} and predict, to some extent, the latitudes and times of sunspot formation \citep{isik_magnetic_2011,nandy_unusual_2011,cameron_solar_2007,cameron_are_2012} but without accounting for the detailed magneto-convective dynamics.

The study of the late phases of the buoyant rise and emergence of magnetic flux-ropes has also resorted mostly to local high resolution settings 
\citep[e.g:][]{
archontis_flux_2010,
aulanier_equilibrium_2005,
cheung_solar_2009,
linton_helical_1996,
archontis_three-dimensional_2005,
cheung_simulation_2010,
hood_kink_1979,
komm_subsurface_2011,
galsgaard_heating_1997,
martinez-sykora_twisted_2008,
martinez-sykora_twisted_2009
}.
These studies use simulation domains which typically span a few$\un{Mm}$ above and below the photosphere.
Such simulations attempt to describe the very strongly stratified photospheric layers as finely as possible, which justifies in most cases the use of high-resolution local simulation domains.
One or multiple buoyant twisted magnetic flux-ropes are introduced near the lower boundary (typically a few\un{Mm} below the surface) and evolve in more or less turbulent media until they emerge.
We refer to the review by \citet{fan_magnetic_2009} for a more detailed discussion of local flux emergence.
Some of these works have been very successful at reproducing features observed at the surface of the sun, but neglect (or strongly simplify) the constraints imposed by the large scale dynamics.
The global background dynamo and magneto-convective processes at the origin of the large scale meridional and zonal flows cannot be fully catered by this type of model.

The properties of the buoyant rise of magnetic flux-ropes down from their assumed generation site up to the surface of the Sun require, to some extent, the use of global models of the convective zone.
Studies based on the thin flux-tube approximation \citep[][among many others]{weber_rise_2011,weber_comparing_2012} allow one to follow the evolution of slowly buoyant tubes for very long time-scales at the expense of neglecting all sources of magnetic or dynamical erosion caused by the background flows.
Global simulations resorting to finite-width magnetic flux-ropes conversely can take into account the interaction between the tube and the background convective motions and associated mean flows.
They are nevertheless restricted to moderate Reynolds numbers and to the study of sufficiently strong flux-ropes (as the buoyant rise time needs to be considerably shorter than the diffusive time-scales).
The limits of the global approach can in principle be explored in more detail by resorting to local setups reaching higher spatial resolutions and hence higher Reynolds numbers \citep[\emph{e.g}][]{hughes_rise_1998}.
But these cannot capture the effects of the large-scale flows on the trajectories of such flux-ropes, nor can self-consistently quantify the angular momentum transport phenomena related to the rise of such coherent and presumably self-connected structures.
The different approaches briefly listed in this paragraph are complementary and supply different pieces of the puzzle.
There are therefore two main options:  (a) study local models that embed a large-scale magnetic structure in a highly turbulent flow, or (b) study global models that embed a large-scale magnetic structure in a large-scale, laminar (or weakly turbulent) flow. 
In the former, the magnetic structures feel a more realistic (relative to the solar case) bombardment by small-scale turbulence, but the disadvantage is that self-consistent large-scale fluid motions or large-scale background magnetic fields are omitted \citep[although non-self-consistent versions could be added, e.g.][]{dorch_buoyant_2007}. 
The latter has the advantage that such large-scale effects can be included self-consistently.  However, the numerical restrictions that the global geometry imposes only allow the study of the rise of the magnetic structure in a laminar or weakly turbulent flow, where magnetic structure and velocity scales are comparable and diffusion, advection and transport times are all similar, thereby simulating a completely different problem from that encountered in the solar situation. 
We deliberately chose the latter option, as did \citet{jouve_three-dimensional_2009}, to which this paper is meant to be compared.

\citeauthor{jouve_three-dimensional_2009} (\citeyear{jouve_three-dimensional_2009}; hereafter JB09) specifically considered the influence of the global convective dynamics on the evolution of finite-width buoyant magnetic flux-ropes.
They performed a series of simulations of individual twisted magnetic flux-ropes inside a spherical convection zone possessing large scale mean flows (e.g. solar-like differential rotation and meridional circulation) and fully developed hydrodynamical convection.
They found that indeed latitudinal and longitudinal modulation of convective patterns and large scale flows have a direct influence on flux emergence of toroidal magnetic structures that are below $6-7\ B_{eq} \sim 4\e{5}\un{G}$, where $B_{eq}$ has been evaluated by computing the kinetic energy of the strongest downflows.
Above such a threshold the structures tends to rise and emerge almost as if they were embedded in a purely isentropic layer.
In such global-scale convective simulations, the differential rotation was also found to influence the rise trajectory of the flux-ropes.
As in \citet{wissink_numerical_2000,abbett_effects_2001}, the inertial (coriolis) forces due to the flux-rope displacement have an azimuthal component and a (secondary) component pointing towards the rotation axis which opposes the buoyant force (reducing its radial rise speed) and pushes the flux-rope polewards.
In JB09, the amplitude of this effect depends on the latitude at which the flux-ropes are initially placed, and more specifically on the differential rotation profile at that latitude.
They found that a solar-like background differential rotation makes flux-rope emergence harder at low latitudes.
They further confirmed the existence of thresholds for the amount of twist and initial field amplitude to obtain a radial rise of uniformly buoyant magnetic structures. 

The goal of this paper is to continue this effort by studying the evolution of twisted magnetic flux-ropes in a fully magnetised three-dimensional convection zone with a background magnetic field generated self-consistently by dynamo processes.
We will thus focus on the large-scale effects rather than the small-scale turbulent processes involved.
We introduce such twisted structures at the bottom of a magnetised turbulent convection zone and let it evolve for a long period of time, a few times longer than that of the buoyant rise time-scale.
This allows us to investigate whether and by what means the background magnetic field has an influence on the evolution of the coherent magnetic flux-ropes.
We note that while it might have been easier to impose a simpler steady global magnetic background field \citep[as in][]{dorch_buoyant_2007}, we believe that only a time varying multi-scale field as developed by dynamo action properly captures the dynamics involved in the issues we address here.
We also address the importance of the buoyant rise and emergence of such flux-ropes on the global energy and magnetic flux budgets.
Furthermore, we suggest surface diagnostics (evaluated at the top of the numerical domain) which precede and accompany the emergence of magnetic flux-ropes, and whether these depend on the properties of the background field.

\del{
  During such flux-emergence episodes in the Sun, a fraction of the magnetic flux initially connected to the coherent flux-rope is presumably transmitted to the solar corona, while the remaining flux is necessarily assimilated into the global dynamo field and reprocessed in a longer time-scale.
As a consequence, the global external magnetic field topology must be perturbed (if not strongly modified) during such a flux-emergence event, as would the corona in the real sun.
The exact form of the perturbations to the external environment most likely depends on the relative orientation of the background field in respect to the flux-rope's field, their relative amplitude, and the associated surface flows.
We investigate here -- within the limits inherent to our methodology -- the contribution of flux-rope emergence phenomena to the global coronal (\emph{i.e, external}) magnetic flux and the spatial and temporal correlations between surface magnetic fields and flows.
}

\del{Our simulations can also contribute to understanding how the spot's polarities move polewards and contribute to the overall reversal of the solar global poloidal field} \citep{wang_magnetic_1991,schrijver_photospheric_2003,isik_magnetic_2011,benevolenskaya_polar_2004}.

The remainder of this manuscript is organized as follows:
Sect. \ref{sec:setup} described the model equations and numerical setup, discusses the parameter-space explored and the scope and limitations of our model, Sect. \ref{sec:ropeevol} describes the early phases of the evolution of the twisted flux-ropes (namely the buoyant rise through the CZ), Sect. \ref{sec:laterevol} describes the later phases (flux-emergence, post-emergence and re-assimilation) and Sect. \ref{sec:coronal} discusses the consequences of the flux-emergence episodes in the corona.
A general discussion follows in Sect. \ref{sec:discussion}.

\section{Model Setup}
\label{sec:setup}

\subsection{Anelastic MHD equations}
\label{sec:equations}

\begin{figure*}
  \centering
  \includegraphics[angle=90,height=6.5cm]{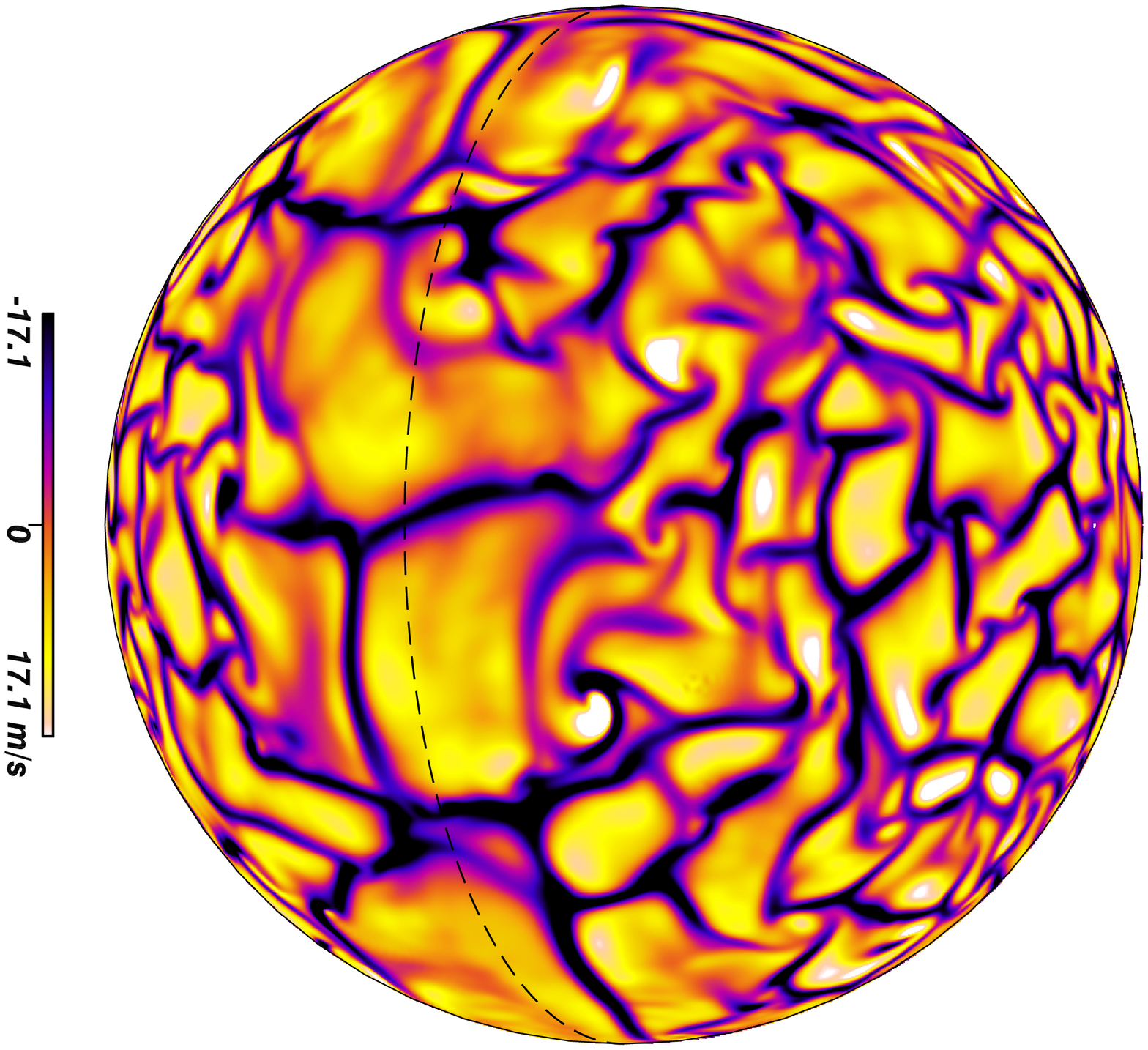}
  \includegraphics[angle=90,height=7cm]{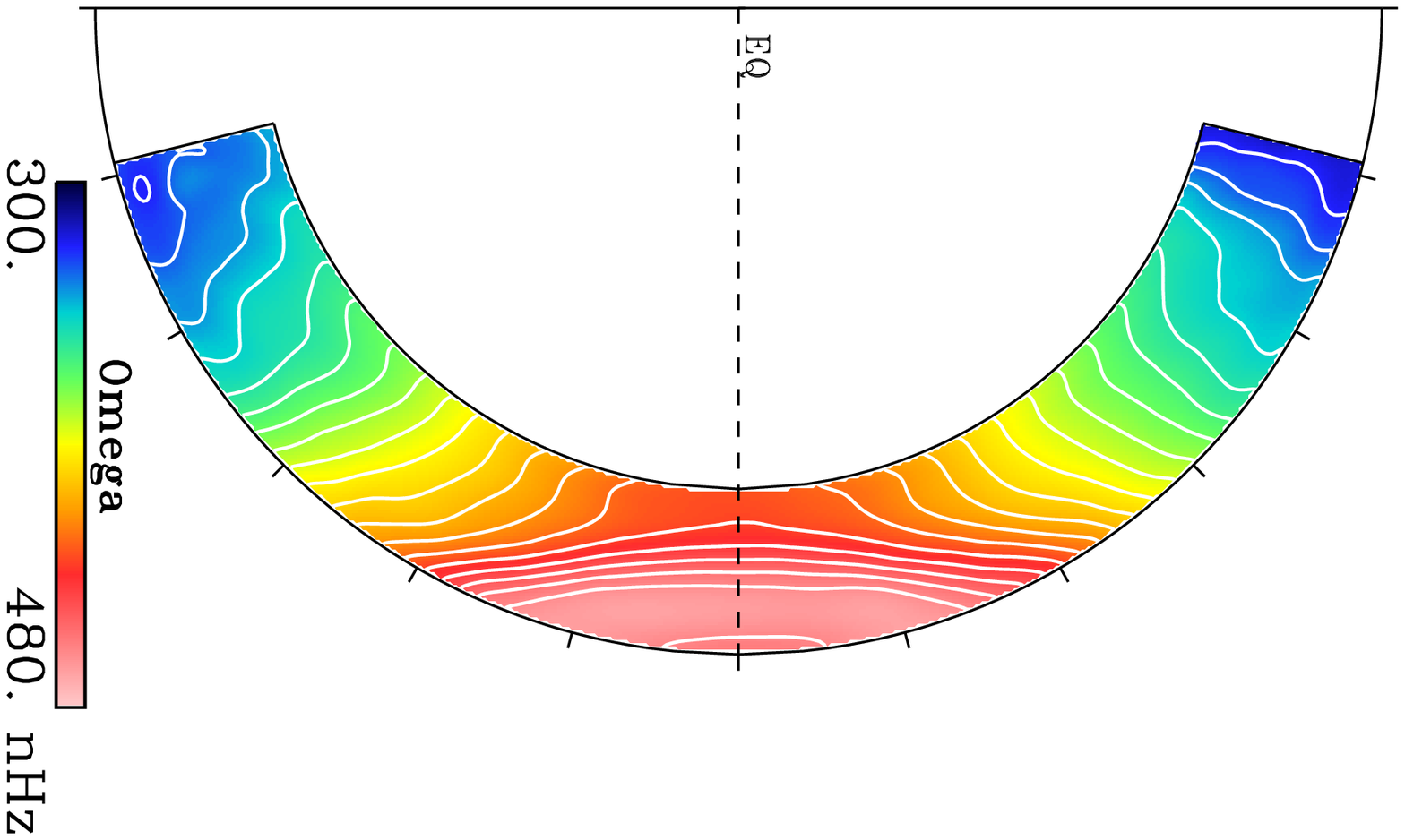}
  \includegraphics[height=7cm]{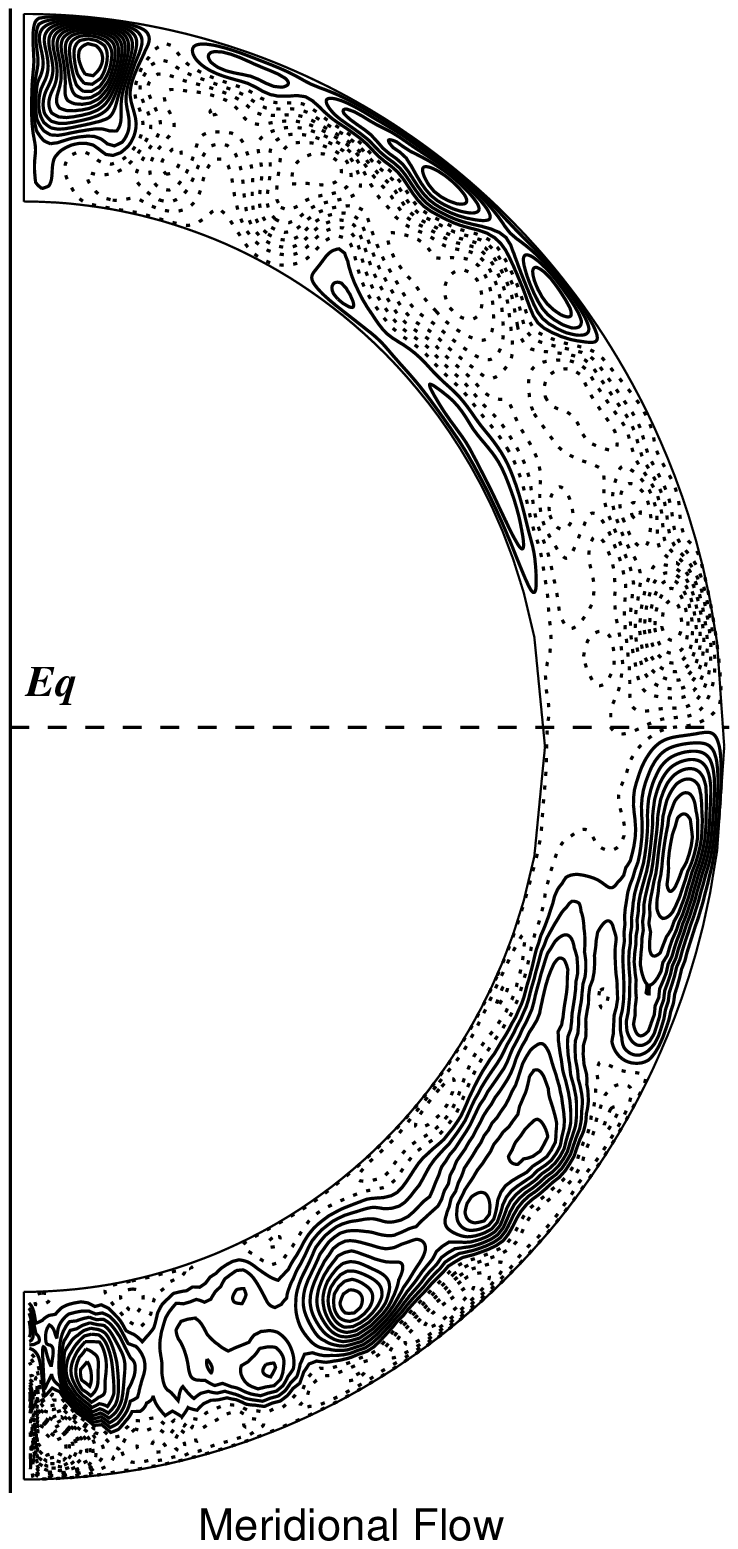} 
  \caption{Convective
    motions and mean flows in our model.
    The first panel shows the radial velocity profile near
    the top of the shell, with yellow and dark blue tones representing respectively upflows and downflows.
    The second panel shows the differential
    rotation profile and the right panel shows the meridional
    circulation.
    The last two panels show longitude and time ($272$ days) averaged data.
    For the meridional flow, dashed (plain) lines represent
    counterclockwise (clockwise) circulation and the intensity
    varies approximately between about $-20$ and $20 \un{m.s^{-1}}$.}
  \label{fig:bgfieldflows}
\end{figure*}

\begin{figure}
  \centering
   \includegraphics[width=.8\picwd]{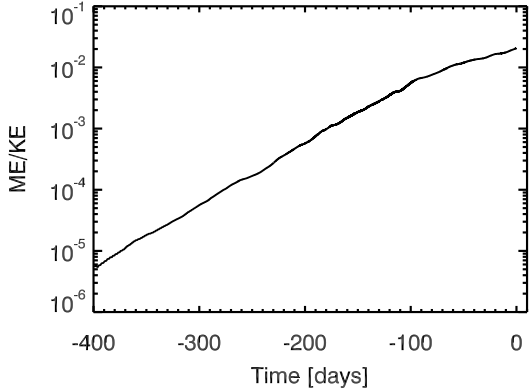}
  \caption{Evolution of the ratio of magnetic energy (ME) to kinetic energy (KE) for the background dynamo field.
    The instant $t=0$ corresponds to the time of introduction of the magnetic flux-rope.}
  \label{fig:bgfieldme}
\end{figure}

\begin{figure}
  \centering
  \includegraphics[width=\picwd]{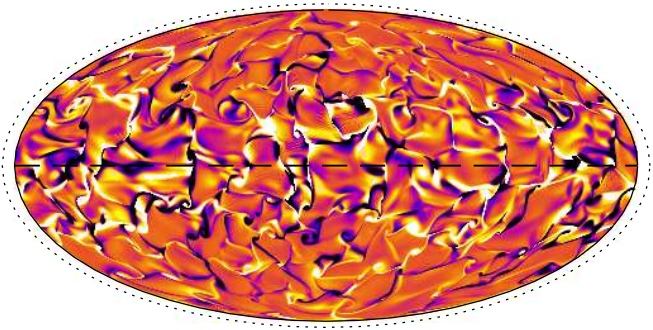}
  \caption{Mollweide projection of the radial component of the background dynamo magnetic field near $r=0.96 R$ at $t=0$.
    Dark tones denote negative polarity.
    The colour table scales from $-170$ to $+170\un{G}$.
    We note the presence of mixed field polarity in the downflow lanes.
    The dotted black line indicates the position of the spherical surface placed at $r=1\rsun$.}
  \label{fig:bgfield}
\end{figure}

The simulations described here were performed with the anelastic
spherical harmonic (ASH) code. ASH solves the three-dimensional
MHD anelastic equations of motion in a rotating spherical shell using a
pseudo-spectral semi-implicit approach
\citep{clune_computational_1999,brun_global-scale_2004}. 
The effects of compressibility on the convection are taken into 
account by means of the anelastic approximation, which describes well the advective dynamics 
while filtering out sound waves that would otherwise severely limit the time steps allowed by the simulation.
ASH further uses a Large-eddy Simulation (LES) approach, with
parametrisation to account for subgrid-scale (SGS) motions. These equations
are fully nonlinear in velocity and magnetic fields.
The thermodynamic variables are linearised with respect to a spherically symmetric mean
state having density $\bar{\rho}$, pressure $\bar{P}$, temperature
$\bar{T}$, specific entropy $\bar{S}$.
Perturbations are denoted as $\rho$, $P$, $T$ and $S$.
The equations being solved are
\begin{equation}
  \nab\cdot\left(\rb {\bf v}\right) = 0\ ,
\end{equation}
\begin{equation}
  \nab\cdot\mathbf{B} = 0\ ,
\end{equation}
\begin{eqnarray}
  \label{eq:momentum}
  \bar{\rho} \left[ \frac{\partial{\bf v}}{\partial t}+\left({\bf v}\cdot{\bf \nab}\right){\bf v}\right. &+& \left.\phantom{\frac{}{}} 2\Omega_0\times {\bf v} \right] = -{\bf \nab} P +\rho{\bf g} \\ \nonumber
  &+& \frac{1}{4\pi}{\bf \left(\nab \times B\right) \times B}-{\bf \nab \cdot \cal D}-\left[{\bf \nab}\bar{P}-\bar{\rho}{\bf g}\right]\ ,
  \label{eqNS}
\end{eqnarray}
\begin{eqnarray}
  \label{eq:energy}
  \bar{\rho}\bar{T}\frac{\partial S}{\partial t}&+&\bar{\rho}\bar{T}{\bf v}\cdot{\bf \nab}(\bar{S}+S)={\bf \nab}\cdot [\kappa_r\bar{\rho}c_p{\bf \nab} (\bar{T}+T)\\ \nonumber
  +\kappa_{0}\bar{\rho}\bar{T}{\bf \nab}\bar{S}&+&\kappa \bar{\rho} \bar{T}{\bf \nab} S]+\frac{4\pi\eta}{c^2}{\bf j}^2+2\bar{\rho}\nu\left[e_{ij}e_{ij}-\frac{1}{3}({\bf \nab \cdot v)}^2\right]\ ,
\end{eqnarray}
\begin{eqnarray}
  \label{eq:induction}
  \frac{\partial \mathbf{B}}{\partial t} = \curl{\left(\mathbf{v} \cross \mathbf{B} \right) - \nab\cross\left(\eta \curl{\mathbf{B}} \right)}\ ,
\end{eqnarray}
where $c_p$ is the specific heat at constant pressure, ${\bf v}=(v_r,v_{\theta},v_{\phi})$ is the local velocity in spherical geometry in the rotating frame of constant angular velocity $\Om = \Omega_0 {\bf \hat{e}_z}$, 
$\mathbf{B} = \left(B_r, B_\theta, B_\phi \right)$ is the magnetic field, $\mathbf{j}=\frac{c}{4\pi}\curl\mathbf{B}$ is the current density, 
${\bf g}$ is the gravitational acceleration, 
$\kappa_r$ is the radiative diffusivity and 
$\DD$ is the viscous stress tensor, with components
\begin{eqnarray}
  \label{eq:viscoustensor}
  {\cal D}_{ij}=-2\rb\nu\left[e_{ij}-\frac{1}{3}\left(\nab\cdot{\bf v}\right)\delta_{ij}\right],
\end{eqnarray}
where $e_{ij}$ is the strain rate tensor.
As mentioned above, the ASH code uses a LES formulation where $\nu$, $\kappa$ and $\eta$ are assumed to be, respectively, an effective eddy viscosity, an eddy diffusivity and a magnetic diffusivity (chosen to accommodate the resolution) that represent unresolved SGS processes.
The thermal diffusion $\kappa_0$ acting on the mean entropy gradient occupies a narrow region in the upper convection zone.
Its purpose is to transport heat through the outer surface where radial convective motions vanish \citep{gilman_compressible_1981, wong_comparison_1994}.
To close the set of equations, linearized relations for the thermodynamic fluctuations are taken as
\begin{equation}\label{eos}
  \frac{\rho}{\rb}=\frac{P}{\bar{P}}-\frac{T}{\tb}=\frac{P}{\gamma\bar{P}}
  -\frac{S}{c_p},
\end{equation}
assuming the ideal gas law
\begin{equation}\label{eqn: gp}
  \bar{P}={\cal R} \rb \tb ,
\end{equation}
where $\cal R$ is the ideal gas constant, taking into account the mean molecular weight $\mu$ corresponding to a mixture composed roughly of 3/4 of Hydrogen and 1/4 of Helium per mass.
The reference or mean state (indicated by overbars) is derived from a one-dimensional solar structure model and is regularly updated with the spherically symmetric components of the thermodynamic fluctuations as the simulation proceeds \citep{brun_seismic_2002}.
It begins in hydrostatic balance so the bracketed term on the right-hand side of Eq.\ref{eq:momentum} initially vanishes.
However, as the simulation evolves, both the turbulent and magnetic pressures drive the reference state slightly away from strict hydrostatic balance.

Finally, the boundary conditions for the velocity are impenetrable and stress-free at the top and bottom of the shell.
We impose a constant radial entropy gradient at the top and bottom.
A latitudinal entropy profile is imposed at the bottom, as in \citet{miesch_solar_2006}.
We match the magnetic field to an external potential magnetic field both at the top and the bottom of the shell \citep{brun_global-scale_2004}.

\subsection{Background dynamo model}
\label{sec:dynamomodel}

\begin{table}[]
  \centering
  \begin{tabular}{r c}
    \hline\hline
    domain          & $0.72 - 0.97 \rsun$ \\
    grid resolution ($N_r \times N_\theta \times N_\phi$) & $256 \times 1024 \times 2048$ \\
    diffusive coefficients (mid-CZ) & $\nu=1.13\e{12} \un{cm^2s^{-1}}$ \\
                                    & $\kappa=4.53\e{12} \un{cm^2s^{-1}}$ \\
                                    & $\eta=2.83\e{11}\un{cm^2s^{-1}}$ \\
    dimensionless numbers           & $P_r=0.25,\ P_m=4$  \\
                                    & $Ra=1.85 \times 10^5>Ra_c$ \\
                                    & $R_e=120,\ R_m\sim 480$ \\
                                    & $T_a=1.8\e{6},\ R_{oc} = 0.63$ \\
    \hline
  \end{tabular}
  
  \caption{Summary of the background dynamo parameters. 
    $\nu$, $\kappa$ and $\eta$ are the effective viscosity, thermal and magnetic diffusivities.
    $P_r$ and $P_m$ are the Prandtl and magnetic Prandtl numbers.
    $Ra$ and $Ra_c$ are the Rayleigh and critical Rayleigh numbers.
    $R_e$ and $R_m$ are the Reynolds and magnetic Reynolds numbers.
    $T_a$ and $R_{oc}$ are the Taylor and convective Rossby numbers.
    See the Sect. \ref{sec:dynamomodel} for more details.}
  \label{tab:dynamopars}
\end{table}

\begin{figure}[]
  \centering
  \includegraphics[width=\picwd]{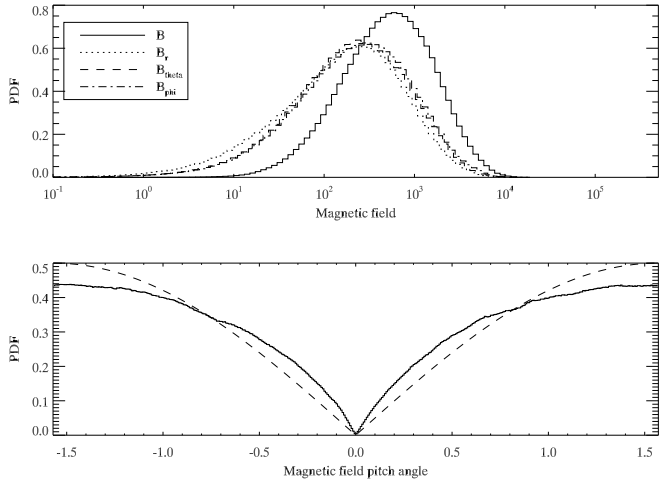}
  \caption{PDF of the magnetic field amplitude $B = \sqrt{\left(B_r^2 + B_\theta^2 + B_\phi^2\right)}$ at $t=0$ on the latitudinal interval $\left[25^{\circ} N, 45^{\circ} N\right]$ covered by the flux-rope's evolution (top panel; also, all three $\mathbf{B}$ components are superimposed).
  Note that the histogram bins are distributed logarithmically.
  The global dynamo field peaks between $10^2\un{G}$ and $10^3\un{G}$ and the FWHM of the distribution is of about one order of magnitude around that value.
  The bottom panel shows the PDF for the magnetic field's pitch angle in the same latitudinal interval (thick continuous line) and the expected distribution for a uniform random distribution of poloidal and toroidal fields (thin dashed line).}
  \label{fig:bfield_pdf}
\end{figure}

\begin{figure*}[]
  \centering
  \includegraphics[width=\picwd]{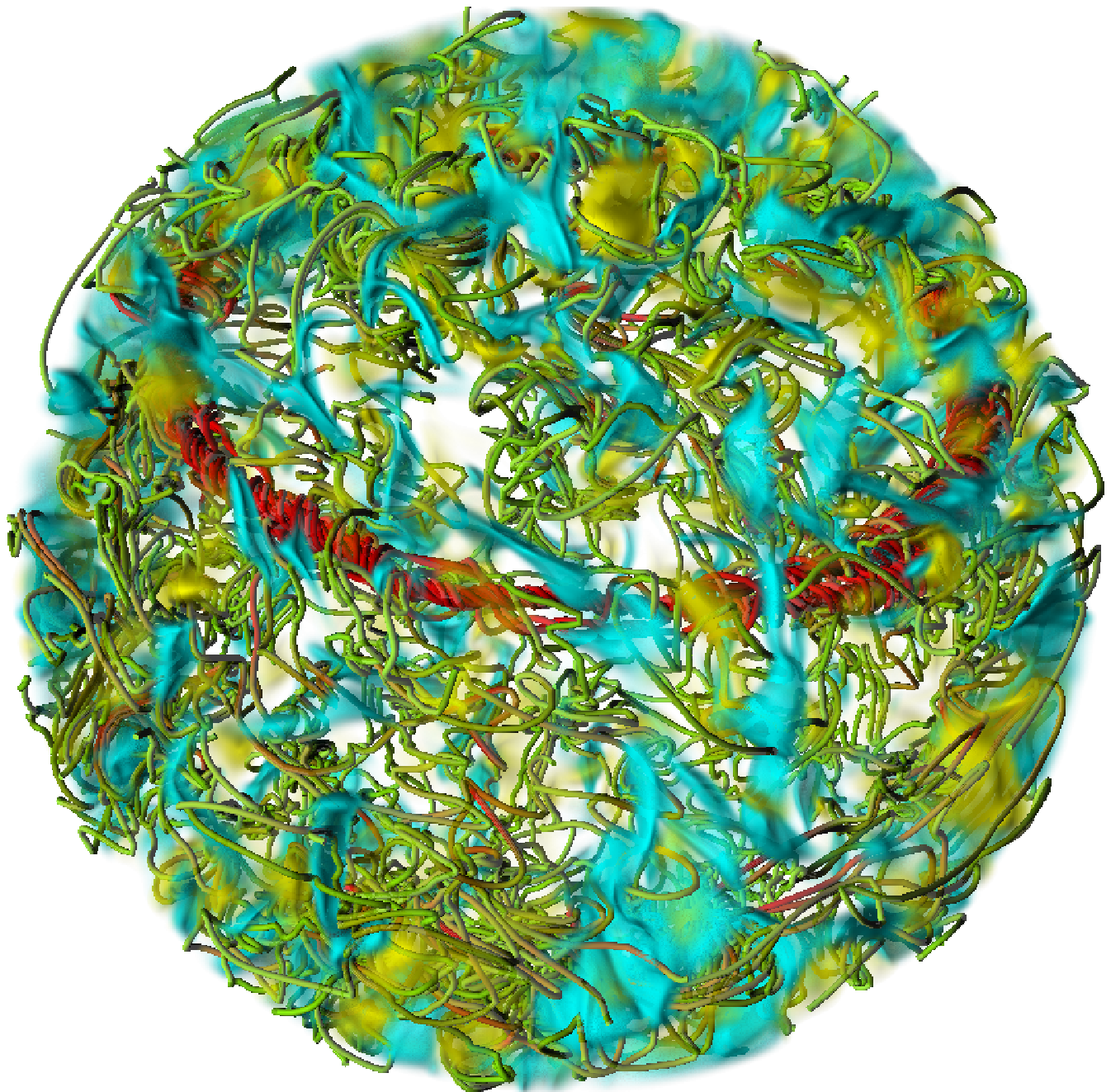}
  \includegraphics[width=\picwd]{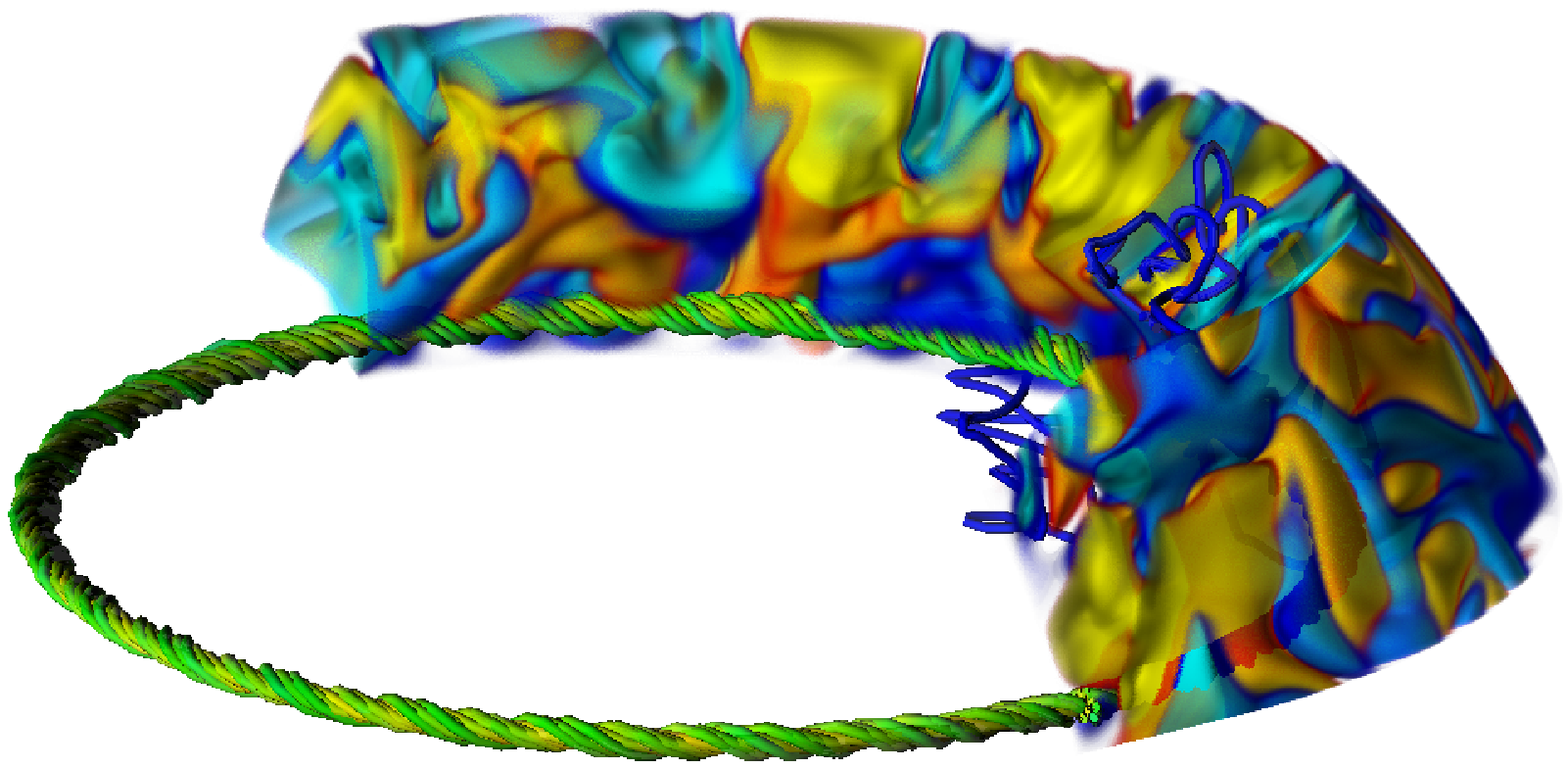}
  \caption{Initial Conditions.
    Snapshot of a twisted magnetic flux-rope introduced at the bottom of the CZ in the dynamo run described in Sect. \ref{sec:dynamomodel}.
    The left panel shows a sample of magnetic field lines in all the numerical domain (for which red/green tones represent strong/weak $\mathbf{B}$-field).
    The right panel shows the flux-rope in more detail, over a small sub-domain of the numerical setup.
    In both panels blue and yellow volumes represent, respectively, convective upflows and downflows (only the strongest down and upflows are represented on the left panel).}
  \label{fig:fluxrope_init}
\end{figure*}

Our experiments consist in introducing a toroidal magnetic flux-rope at the base of the convection zone in a thermally equilibrated convection model in which a dynamo is operating. 
The study of the purely hydrodynamic case was the subject of JB09.

Our numerical model presents a simplified description of the physical processes acting on the magnetized solar convection zone.
Solar values are taken for the heat flux, rotation rate, mass and radius and a perfect gas is assumed since the upper boundary of the shell lies below the H and He ionisation zones.
Contact is made with a seismically calibrated 1D solar structure model for the radial stratification.
Table \ref{tab:dynamopars} summarises the model's main parameters, which we describe in detail hereafter.
The computational domain extends from about $0.72 \rsun$ to $0.97 \rsun$.
The numerical domain uses $256$ grid points in the radial direction, $1024$ in the latitudinal 
direction and $2048$ in azimuth.
The reference state was obtained through the 1D CESAM stellar evolution code \citep{morel_cesam:_1997,brun_seismic_2002} which uses a classical mixing-length treatment calibrated on solar models to compute convection.
We are dealing with the central portion of the convection zone but neglect for this work the
penetrative convection below that zone and the atmosphere above (which is stable with respect to convection).
See \citet{browning_dynamo_2006,brun_modeling_2011,pinto_3d_2011}; Warnecke et al. (2012) for simulations of solar convection coupled to lower or upper stable layers.

The effective viscosity, thermal and magnetic diffusivity $\nu$, $\kappa$ and $\eta$ are here taken to be functions of radius alone and are chosen to scale as the inverse of $\bar{\rho}^{1/3}$. We use the values: $\nu=1.13\e{12} \un{cm^2s^{-1}}$, $\kappa=4.53\e{12} \un{cm^2s^{-1}}$ and $\eta=2.83\e{11}\un{cm^2s^{-1}}$ at mid-CZ, corresponding to a Prandtl number $P_r=0.25$ and a magnetic Prandtl number $P_m=4$.
The diffusive coefficients are, inevitably, much higher than those believed to represent the real solar conditions.
The $P_m$ value stated above is, in consequence, considerably higher than the Solar one, but eases the development of a sustained dynamo.
The magnetic Prandtl number is held fixed in all the runs described in this manuscript.
In all cases, the spherical shell is initially rotating at the rate $\Omega_0=2.6\e{-6}\un{rad.s^{-1}}$ (corresponding to a rotation period of $28$ days).

We start from a spherically symmetric convection zone with a realistic density stratification profile and solid-body rotation.
The density contrast in this convective case is about $24$ between the top and the bottom of
the domain.
The convection zone is initially in hydrostatic equilibrium but convection is readily triggered, as the background plasma is convectively unstable. 
The entropy gradient is $dS/dr=-10^{-7}$ and the Rayleigh number was chosen to be supercritical $Ra=1.85 \times 10^5>Ra_c$ \citep[the critical Rayleigh number being $Ra_c \sim 10^4$;][]{jones_compressible_2009}.
The system then relaxes (after about a viscous time scale or hundreds of convective overturning times) to a statistically stationary state with a well balanced radial energy flux throughout the whole domain (see the bottom panel of Figure 2 of JB09).
The convection is moderately turbulent, with a rms Reynolds number $R_e=v_{conv}(r_{top}-r_{bot})/\nu_{midCZ}=120$,  where the characteristic length scale is chosen to be the depth of the CZ and $v_{conv} = 80 \un{m.s^{-1}}$. 
In the simulations, the Taylor number is $T_a=1.8\e{6}$ and the convective Rossby number is then $R_{oc}=Ra/(T_a P_r)=0.63 < 1$, thus ensuring a prograde differential rotation \citep{brun_turbulent_2002}.  
Figure \ref{fig:bgfieldflows} displays the radial convective velocity along with the differential and meridional circulation achieved self-consistently in the simulation.
As described in more detail in JB09, convection is dominated at low latitudes by elongated patterns (the so-called banana convective cells), whereas high latitude convective patterns are more isotropic.
We note at mid latitude a zone of strong horizontal shear associated with the large axisymmetric differential rotation realized in the simulation.
The radial and latitudinal profile of the angular velocity are solar-like when compared to helioseismic inversions \citep[][]{schou_helioseismic_1998,thompson_internal_2003} with a fast equator, slow pole and conical profile at mid latitude. 
We however acknowledge that we do not model the tachocline nor the near surface shear layer.
For the former we have adopted here the latitudinal thermal wind forcing used in \citet{miesch_solar_2006} \citep[see however][for a more self-consistent approach]{brun_modeling_2011}.
Finally we also display the axisymmetric meridional flow present in the simulation.
It is mostly poleward near the surface, we one large dominant cell of order $20\un{m.s^{-1}}$.
Small counter cells are also seen near the boundaries and in the polar cap where azimuthal averages are harder to perform due to the small lever arm there.

Starting from such an hydrodynamical convective state, we then add a seed magnetic field which evolves by the action of the dynamo processes driven by the turbulent motions described above.
A weak seed magnetic field was chosen to be a $\{l=5, m=4\}$ multipole introduced in the convection zone.
The system was then evolved for $800$ days, that is for about $28$ solar rotations.
With our choice of parameters the magnetic Reynolds number is $R_m=Re \times P_m \sim 480$.
This is well above the threshold of about $300$ for dynamo action in stratified and rotating solar convective shells, as evaluated by \citet{brun_global-scale_2004}.
Figure \ref{fig:bgfieldme} displays the ratio between magnetic (ME) and kinetic (KE) energies over last $400$ days of the simulation.
We indeed see that dynamo action takes place, as $ME/KE$ increases steadily.
We clearly see the exponential growth and the change of slope when nonlinear feedback from the Lorentz force starts to be felt by the flow.
We have run the simulation until ME reaches an amplitude of order $2\%$ of KE.
We chose to stop the simulation at that instant ($t=0\un{d}$ in the figure) as we did not want the averaged background field to be too intense and the large scale axisymmetric flows to be influenced by the growing magnetic field.
This somewhat arbitrary choice allows us to have a multi-scale background magnetic field while keeping the background large-scale flows unchanged, hence letting us establish a direct link with previous works \citep[e.g][]{jouve_three-dimensional_2009}.
In particular the differential rotation remains solar-like as the one shown on Fig. \ref{fig:bgfieldflows}, since Maxwell stresses are not yet strong enough to modify significantly the redistribution of angular momentum \citep[see][for a more detailed discussion]{brun_global-scale_2004}.

We show on Figure \ref{fig:bgfield} the surface radial magnetic field at that time ($t=0\un{d}$). This corresponds to the instant of introduction (addition to the background field) of our twisted magnetic flux-ropes (see Sect. \ref{sec:fluxrope}).
As already discussed in \citet{brun_global-scale_2004}, the radial component of the field is mostly found in the downflow lanes and ME can locally be more intense than KE \citep[see also][]{cattaneo_origin_1999}.

Figure \ref{fig:bfield_pdf} shows more precisely the distribution of the magnetic field's strength on the section of the CZ traversed by the flux-ropes (see Sect. \ref{sec:fluxrope} for more details) at $t=0$.
The average background field amplitude is about $0.7\e{3}\un{G}$, and the upper tail of the distribution extends up to about $5\e{4}\un{G}$.
In the region traversed by the magnetic flux-ropes, the radial component of the magnetic field is on average slightly smaller than the horizontal components.
To show this better, the bottom panel in Fig. \ref{fig:bfield_pdf} displays the background magnetic field's pitch angle distribution in the same sub-domain.
The pitch angle $\psi$ is defined here as the angle between the magnetic field vector and the azimuthal direction, such that
\begin{equation}
  \tan\psi = \frac{\sqrt{B_r^2 + B_\theta^2}}{B_\phi}\ .
\end{equation}
If the magnetic field vectors had random spatial orientations everywhere in the domain, then 
the probability distribution function for the pitch angle $f\left(\psi\right)$ would be such that
\begin{equation}
  f\left(\psi\right) d\psi = \frac{2\pi B^2 \left| \sin\psi\right| d\psi}{4\pi B^2} = \frac{1}{2}\left|\sin\psi\right| d\psi\ .
\end{equation}
The function $f\left(\psi\right)$ is represented by the dashed line in the bottom panel of Fig. \ref{fig:bfield_pdf}.
This shows that the background field is slightly more toroidal than poloidal in this region.

For comparison purposes, we also prepared a purely hydrodynamical background having the same properties (stratification, meridional flows and rotation pattern) as the dynamo runs but without a background magnetic field.

\subsection{Introduction of a flux-rope}
\label{sec:fluxrope}

\begin{figure}[]
  \centering
  \includegraphics[width=\picwd]{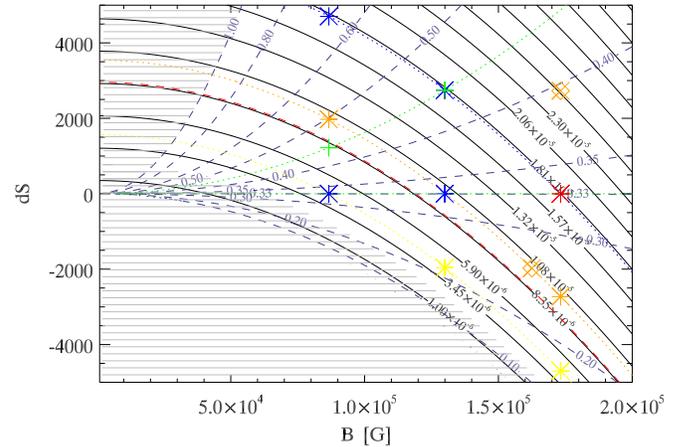}
  \caption{Flux-rope parameter space $\left|B\right|$ vs. $\Delta S$.
    The continuous black lines are contours of the perturbation $\Delta\rho/\rho$ at the centre of the twisted flux-ropes, which corresponds to a given pair $\{\left|\mathbf{B}_0\right|, \Delta S \}$.
    The blue dashed lines represent $\sin\psi_t$, where $\psi_t$ is the twist threshold for a given flux-rope.
    The filled regions (grey horizontal lines) are excluded \emph{a priori}, corresponding to $\sin\psi \geq 1$ or for magnetic fields below the minimum field strength threshold ($B = 6.1\e{4}\un{G}$ for $\Delta S = 0$) found in \citet{jouve_three-dimensional_2009}.
    The red dashed line represents an empirical re-evaluation of this threshold, as an outcome of our simulations.
    The red asterisk represents our standard case.
    The dotted blue line follows the contour of constant $\Delta\rho/\rho$ which contains the standard case.
    The blue and red asterisks represent the cases which are more thoroughly discussed in this manuscript; the other symbols represent runs used to verify the consistency our approach.
  }
  \label{fig:parameter_space}
\end{figure}

\begin{table}[ht]
\centering
\begin{tabular}{c c c c c c c c c}
  \hline\hline
  run  & $B_{\phi}^{max}\ \left(10^5\un{G}\right)$  & $A_0\ \left(10^3\right)$ & $q$ & $A_s$ \\
  \hline
  standard flux-rope           &  1.73 &  3.30 &  20 &  0.00 \\
  negative flux-rope           & -1.73 & -3.30 &  20 &  0.00 \\
  left-handed flux-rope        &  1.73 &  3.30 & -20 &  0.00 \\
  left-handed neg. flux-rope   & -1.73 & -3.30 & -20 &  0.00 \\
  weak flux-rope               &  8.66 &  1.65 &  20 &  0.00 \\
  medium flux-rope             &  1.30 &  2.48 &  20 &  0.00 \\
  no dynamo, std. flux-rope    &  1.73 &  3.30 &  20 &  0.00 \\
  hi. buouy., medium field     &  1.30 &  2.48 &  20 &  52.3 \\
  hi. buouy., weak field       &  8.66 &  1.65 &  20 &  23.2 \\
  \hline
\end{tabular}
\caption{Flux-rope parameters for the runs discussed in the text.}
\label{tab:fluxropepars}
\end{table}

We introduced axisymmetric toroidal twisted magnetic flux-ropes near the lower boundary of the computational domain.
The (background) initial conditions are those resulting from the evolution of a dynamo run, as presented above (Sect. \ref{sec:dynamomodel}).
We performed several runs, whose parameters are listed in Table \ref{tab:fluxropepars}.
Fig. \ref{fig:fluxrope_init} shows the initial state of a typical run (our standard case).

The magnetic geometry of the flux-rope is expressed by the two potential functions $A$ and $C$
\begin{eqnarray}
  A\left(r, \theta\right) & = & -A_0 r\times\delta\left(r,\theta\right)\ ,\label{eq:potA} \\
  C\left(r, \theta\right) & = & -\frac{A_0 a^2q}{2}\times\delta\left(r,\theta\right)\ , \label{eq:potC}
\end{eqnarray}
where
\[
  \delta\left(r, \theta\right)  =  \exp\left[-\left(\frac{r - R_t}{a} \right)^2 \right] \left[ 1 + \tanh\left(2\frac{\theta - \theta_t}{a/R_t} \right)  \right] \ .
\]
These potential functions define a divergenceless and axi-symmetric magnetic field $\mathbf{B}$ in terms of the poloidal-toroidal decomposition
\begin{equation}
  \label{eq:btorpol}
  \mathbf{B} = \curl \curl \left(C \mathbf{e}_r \right) 
  + \curl \left(A \mathbf{e}_r \right) \ ,
\end{equation}
whose amplitude at the axis of the structure ($r=R_t,\ \theta=\theta_t$) is given by
\begin{equation}
  \label{eq:maxbphi}
  B_{0} \equiv B_\phi^{max} = 2 A_0 \frac{R_t}{a}\ .
\end{equation}
The pitch angle $\psi$ of the twisted magnetic field lines with respect to the azimuthal direction at the flux-rope's boundary is such that
\begin{equation}
  \label{eq:tanpsi}
  \tan\psi \approx \frac{qa}{R_t + a}\ .
\end{equation}

We also apply the following entropy perturbation to the background field
\begin{equation}
  \label{eq:delta_entropy}
  \delta S = -A_S \frac{2 R_t}{a} \exp\left[ -\left(\frac{r - R_t}{a} \right)^2 \right] \left[\cosh^2\left( 2\frac{\theta - \theta_t}{a/R_t} \right) \right]^{-1}\ ,
\end{equation}
its maximum amplitude being $\Delta S = -2 A_s R_t / a$ at the flux-rope's axis.
The flux-ropes are initially set-up in pressure equilibrium with the surrounding plasma.
This condition constraints the density deficit $\Delta\rho/\rho$ (relative to the background density) to be
\begin{equation}
  \label{eq:deltarho}
  \frac{\Delta\rho}{\rho} = 1 - \exp\left[-\frac{\Delta S}{c_p}\right]\left[1 - \frac{B_0^2}{8\pi P}\right]^{1/\gamma} \ .
\end{equation}
We only consider in this paper $\delta\left(r, \theta\right)$ and $\delta S$ perturbations which are axisymmetric, such that the flux-ropes buoyancy is independent of the azimuthal coordinate \citep[unlike in][Jouve et al., 2013,]{fan_three-dimensional_2008}.
%

As shown by \citet{emonet_physics_1998}, there is a threshold in magnetic twist the flux-rope must have to be able to maintain its coherence during its rise through the convection zone.
This threshold is expressed in terms of a minimum pitch angle $\psi$ at the tube's periphery
\begin{equation}
  \label{eq:sinpsi}
  \sin\psi_t = \frac{ \left| B_{poloidal} \right| }{ \| \mathbf{B} \|} \geq 
  \left[ \frac{a}{H_p} \left|\frac{\Delta\rho}{\rb}\right| \frac{\beta}{2} \right]^{1/2}\ ,
\end{equation}
where $a$ is the flux-rope's radius, $H_p$ the background pressure scale-height, $\Delta\rho$ the density deficit relative to the background density $\rb$ and $\beta$ is the ratio of gas to magnetic pressures.
In simple terms, a flux-rope is able to rise cohesively if its magnetic tension is able to counteract the torque applied by the flows around the magnetic structure as it rises buoyantly.
Equation (\ref{eq:sinpsi}) describes this balance in the plane orthogonal to the flux-rope's axis, that is, the poloidal plane in an axisymmetric setup.
The overall picture of the conditions for the cohesive rise of such flux-ropes in three-dimension convective shells was subject of previous studies \citep[][among others]{archontis_three-dimensional_2005,jouve_three-dimensional_2009}.
Detailed discussions of the evolution of twisted flux-ropes near, above and below this threshold can be found in \citet{jouve_three-dimensional_2009}.
In this paper, we will only consider magnetic structures which are expected to rise cohesively, following those studies' conclusions.
The focus here is on how the globally magnetised medium, rather than a purely hydrodynamic one, interacts with and affects the evolution of such magnetic structures (as explained in Sect. \ref{sec:dynamomodel}).

Figure \ref{fig:parameter_space} shows the parameter space $\left|B\right|$ vs. $\Delta S$ used to define the runs listed in Table \ref{tab:fluxropepars} and summarises the relations between all the flux-ropes parameters. 
Contours of constant $\Delta\rho/\rho$ (black continuous lines) and $\sin\psi_t$ (blue dashed lines) were over-plotted to show how the different runs relate to each other in terms of these two fundamental parameters.
The density deficit $\Delta\rho/\rho$ (and consequently the buoyant rise speed) increases from left to right and from bottom to top.
For a given $B_0$, a positive $\Delta S$ translates into a density deficit higher than that in the $\Delta S=0$ case.
The buoyant rise speed increases accordingly, and the so does the corresponding threshold pitch angle $\psi_t$.
A negative $\Delta S$ produces the opposite effect.
It is also possible to explore different flux-rope magnetic field strengths while keeping the same density deficit.
Decreasing the magnetic field amplitude $B_0$ then implies increasing $\Delta S$.
This corresponds to a upwards and leftward trajectory over one of the $\Delta\rho/\rho$ contour lines in the plot, and therefore to an increase in the pitch angle threshold.
Indeed, torque balance in the meridional plane (or in the plane orthogonal to the flux-rope's axis) requires that the flux-ropes keep the same poloidal magnetic field amplitude if the buoyant driver remains the same (having all other parameters --- $a$ and $H_p$ --- fixed).
The only way to satisfy this condition while decreasing $B_0$ is to increase the flux-rope's pitch angle.
Increasing the magnetic field amplitude $B_0$ while keeping $\Delta\rho/\rho$ constant has the opposite consequences, namely implying a decrease in $\Delta S$ and pitch angle threshold $\psi_t$.
The same type of reasoning can be applied to paths of constant $\psi_t$ in the $\{B_0, \Delta S\}$ parameter space.
The areas of the parameter space for which it is impossible to form cohesive buoyant flux-ropes are greyed out.
These correspond namely to $\{B_0, \Delta S\}$ pairs implying $\sin{\psi_t}>1$ and to density deficits not strong enough to counteract the strongest convective downflows in the CZ.
We note that in practise the latter limit actually underestimates the minimum value for $\Delta\rho/\rho$ which can be used.
The red dashed line represents an empirical re-evaluation of the $\Delta\rho/\rho$ threshold, as an outcome of our simulations.
Very slowly rising buoyant flux-rope's may require extremely low diffusivities, such that the corresponding diffusive time-scales are larger than their long buoyant rise time-scales.
It is of course easier to follow numerically the evolution of strongly-buoyant flux-ropes than weakly-buoyant ones (as the diffusive coefficients cannot be lowered arbitrarily), and our model presents no exception to this general rule.

For completeness, we show below the scaling relations between $\Delta S$, $B$, $\Delta\rho/\rho$ and $\sin\psi_t$, which follow from the equations (\ref{eq:delta_entropy}), (\ref{eq:deltarho}) and (\ref{eq:sinpsi}):
\begin{eqnarray}
  \label{eq:scalingtubes}
  \Delta S \left(\frac{\Delta\rho}{\rho},B\right)&=& -c_p \ln\left[1-\frac{\Delta\rho}{\rho}\right] + \frac{c_p}{\gamma}\ln\left[1-\frac{B^2}{8\pi P}\right] \\
  \frac{\Delta\rho}{\rho}\left(B, \psi_t\right) &=& \sin^2\psi_t\left[\frac{a}{H_P}\frac{4\pi P}{B^2}\right]^{-1}  \\
  \Delta S\left(B, \psi_t\right)&=& -c_p \ln\left[1-\sin^2\psi_t\left(\frac{a}{H_P}\frac{4\pi P}{B^2}\right)\right] + \nonumber \\
  & & \frac{c_p}{\gamma}\ln\left[1-\frac{B^2}{8\pi P}\right]\ .
\end{eqnarray}
Figure \ref{fig:parameter_space} was built assuming invariant $a=2.87\e{-2}\rsun$ and $R_t=0.75\rsun$, which are the ones we chose to use in our simulations.
The values of $P$, $H_P$ and $c_p$ are then implicitly defined by the former parameters (the flux-rope's initial position) and the convective background, and are therefore also invariant.
Changing these would affect the exact slopes and positions of the $\Delta\rho/\rho$ and $\sin{\psi_t}$ contour lines, slightly deforming the diagram but still maintaining its main properties.

\subsection{Scope and limits of our model}
\label{sec:scope_limits}

Let us better clarify the scope and limits of our methods before moving on to the more detailed discussion of our work. We use a global three-dimensional numerical model of a convective spherical shell which solves a set of MHD equations, in the frame of the anelastic approximation (discussed in detail in Bannon 1996; Lantz \& Fan 1999).   The maximum numerical resolution we use (giving a spatial resolution of 1.5 to 2 Mm horizontally, 0.01 to 1 Mm vertically) restricts our studies to laminar or weakly turbulent flows and large cross-sectional area flux-ropes. 
The effective viscosity and resistivity are necessarily much higher than those in the real Sun, and our magnetic structures are resolved with about 64 points in radius and 25 points in latitude. Note that previous local studies have required at least 200 points in each direction  (Fan 2008;  Dorch 2007) and that some authors (e.g. Hughes and Falle 1998) argue that resolution and the local Reynolds numbers are critical to an accurate portrayal of the rise of a structure.   The necessity of making the magnetic structure large in conjunction with high diffusivities leads to lengthscales and timescales of the convective flows and background magnetic fields that are comparable to the those of the magnetic structure.  The magnetic structures occupy approximately $1/4$ of the convection zone by radius, and evolve on a timescale of about 25 days, which may be compared with the convective turnover time of 35 days and the magnetic diffusion time of 80 days.  Therefore, due to our choice 
of a
global geometry and the constraints it places on numerical resolution, we are necessarily restricted to studying the brief rise of a large magnetic structure in a large-scale weakly turbulent magneto-convective background where diffusive processes clearly play a significant role.  This means that what we study then is clearly {\bf not} the solar problem, wherein a magnetic structure that is much larger than the typical lengthscales of both the highly-turbulent velocity and magnetic fields rises quickly on a timescale that is far shorter than any diffusive timescale.  Furthermore, the similarity of all our lengthscales really demands that numerous realizations of the simulations should be performed to create accurate statistics, but such an ensemble is unfortunately too expensive to perform in our global geometry.  
   Our magnetic background is also somewhat arbitrary.  In our simulations, idealised twisted magnetic flux-ropes are added to a magneto-convective setup - a dynamo - which we let evolve 400 days beforehand. The magnetic energy spectrum reaches a fully developed profile after about 150 days, with the later evolution (between 150 and 400 days) translating simply into an increase of amplitude of the whole spectrum. We choose to stop the dynamo run when the magnetic energy reached about 2\% of the kinetic energy. This deliberate but completely arbitrary choice ensures that the background flows retain the same global properties (meridional circulation and prograde differential rotation) and hence lets us establish a more clear comparison with the progenitor hydrodynamic case (Jouve \& Brun 2009).  Different background dynamo setups could perhaps lead to different results, and such effects should be addressed in the future.
   Furthermore, it should be noted that the top of the numerical domain (the ``surface'') is placed at $0.97\rsun$ thereby omitting a very dynamic layer of the actual solar surface.  Thus our conclusions concerning ``flux-emergence'' are strictly related to emergence from the weakly stratified interior of the calculation into the highly stratified surface convective layers above. 
Significant modification of any characteristics detected in our simulations should be expected and therefore any predictions must be considered very carefully before their relation to the solar context, if any, is implied (as in Abbett et al. 2001, for example). 
\add{However, due to the first point above, these simulations should not be considered as solar. in any way, shape or form in the first place anyway.}


\section{Buoyant rise in a dynamo field}
\label{sec:ropeevol}

\begin{figure*}[]
  \centering
  \includegraphics[width=2\picwd]{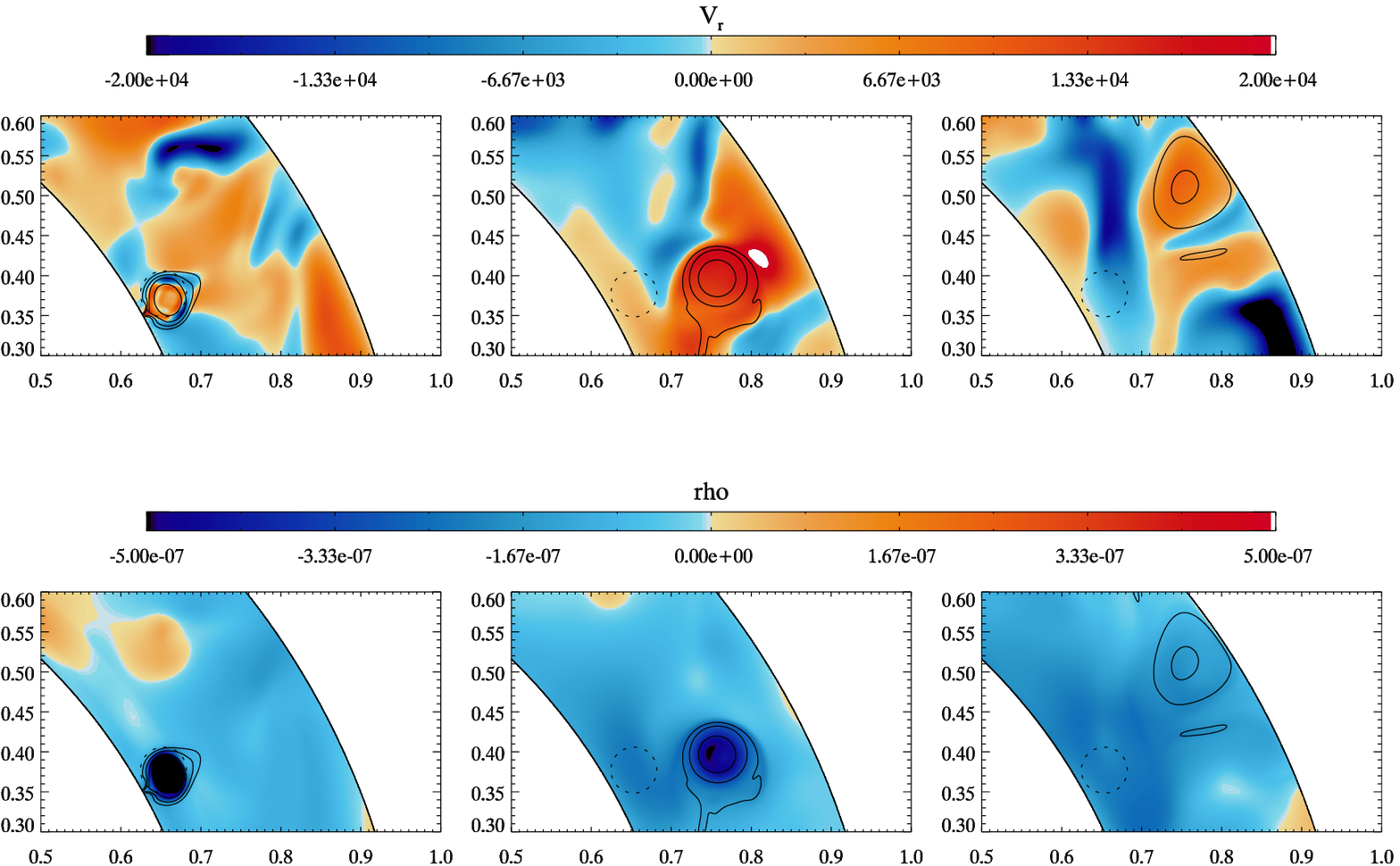}
  \caption{Sequence of snapshots of the flux-rope in standard case (polar slices at $\phi=90^\circ$) at three different instants of its temporal evolution ($t\approx 1, 10, 20$ days).
    The colour-scale shows $v_r$ in the top row and $\Delta\rho$ in the bottom row.
    The axis coordinates are in units of $\rsun$, with origin at the centre of the Sun.
  The black continuous lines are contours of $B_\phi$ and the dotted line represents the initial position of the tube.}
  \label{fig:fluxrope_snapshots}

  \vspace{0.1\picwd}

  \includegraphics[width=2\picwd]{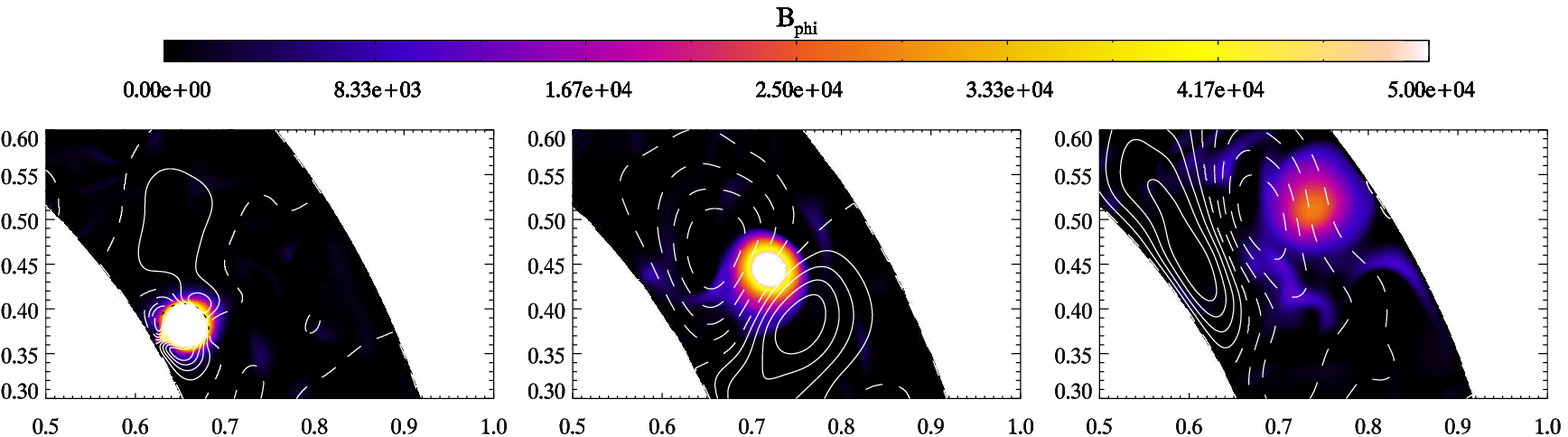}
  \caption{Sequence of snapshots (polar slices at $\phi=90^\circ$) at the same instants of its temporal evolution as in Fig. \ref{fig:fluxrope_snapshots} ($t\approx 1, 10, 20$ days) and in the same sub-domain.
    The colour-scale represents the amplitude of $B_\phi$, thus tracing the flux-rope's position.
    The white lines are streamlines of the poloidal massflux $\rho\mathbf{v}$ (\emph{i.e.}, contours of its streamfunction), the continuous lines representing CW flows and the dashed lines CCW flows.
}
  \label{fig:fluxrope_massflux}
\end{figure*}

\begin{figure}[]
  \centering
  \includegraphics[width=\picwd]{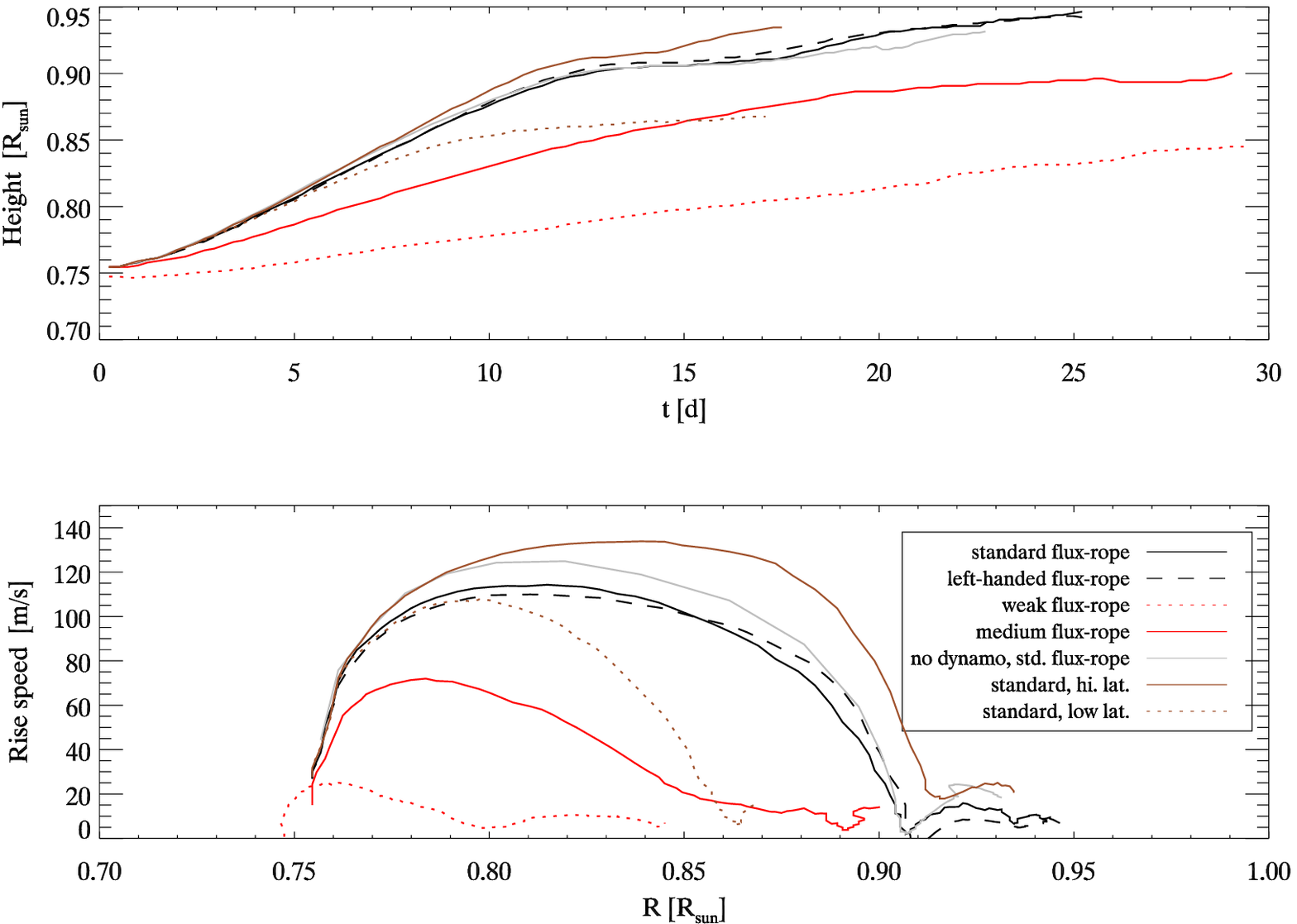}
  \caption{Height (radius) and rising velocity of the centre of the magnetic flux-rope.
  The axis of the flux-rope is defined here as the position of the maximum of $B_\phi$ in the meridional plane.
    Each curve corresponds to one of the runs listed in Table \ref{tab:fluxropepars} (\emph{cf.} the inset key for reference).
}
  \label{fig:fluxrope_height}

  \vspace{0.1\picwd}

  \centering
  \includegraphics[width=\picwd]{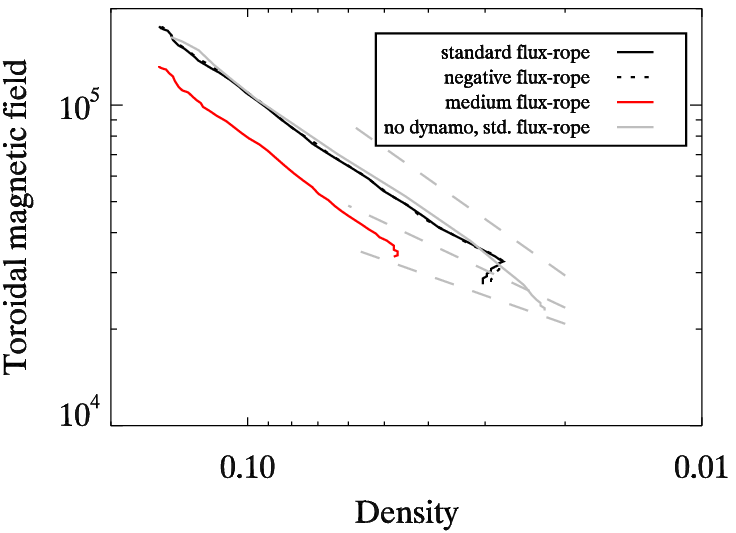}
  \caption{Magnetic field strength at the centre of the twisted flux-ropes $B_c$ as a function of the total central density $\rho$ (log-log plot) during the whole runs (buoyant rise, emergence and later phases).
    During the buoyant rise, most curves are well fitted by a power-law $B_c \propto\rho_c^\alpha$ with an index $\alpha\lesssim 1$; the exceptions are the weak-field cases, which are severely distorted by the convective flows and lose their spatial coherence.
    The dashed grey guidelines indicate the slopes for $\alpha=1, 2/3$ and $1/2$.}
  \label{fig:b_rho}
\end{figure}

\begin{figure}[]
  \centering
  \includegraphics[width=\picwd]{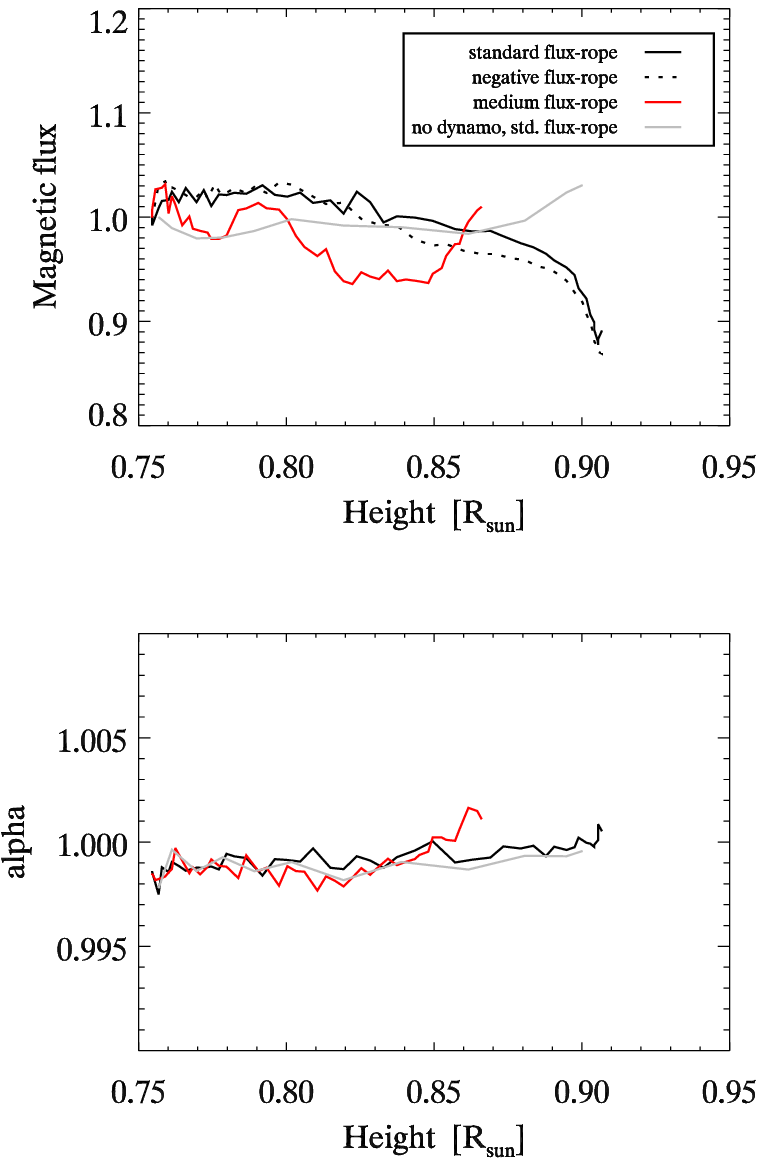}
  \caption{Top panel: Relative variation of the magnetic flux contained in the twisted flux-ropes as a function of the position of the flux-rope's axis.
    Bottom panel: index $\alpha$ as obtained from Eq. \ref{eq:phi_m_ratio}, using the ratio of magnetic flux to mass in the buoyant flux-ropes.
  }
  \label{fig:tubeflux}
\end{figure}

\begin{figure*}[]
  \centering
  \includegraphics[width=2\picwd]{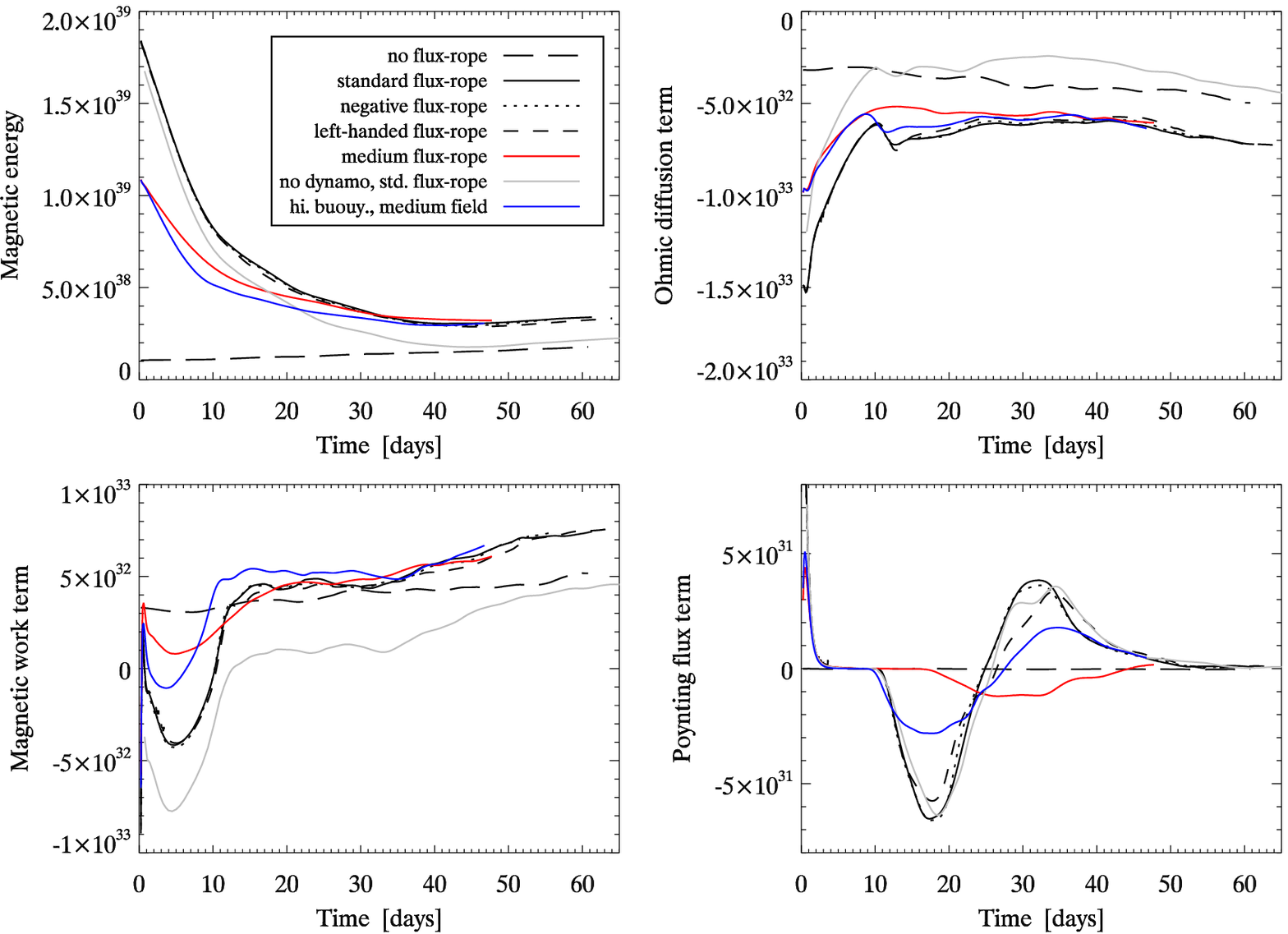}
  \caption{Evolution of the total magnetic energy in the whole numerical domain for different cases presented here (standard case, medium $B$-field cases and standard case in an hydrodynamical convection zone).
    The first panel (top, left) shows the total magnetic energy as a function of time, while the other panels shows the source/sink terms in the equation for $\partial_t \int B/8\pi dV$, namely the rates of ohmic diffusion, the rate of work done by the Lorentz force and the Poynting flux.
    The abscissas represent time in days with $t=0$ at the moment when the flux-rope was introduced.
    The long term evolution of the total magnetic energy in the background dynamo model is also shown (long-dashed line).}
  \label{fig:fluxrope_me}
\end{figure*}

Let us now describe the evolution of the twisted magnetic flux-ropes introduced near the bottom boundary of the background model described above (Sect. \ref{sec:setup}).


\subsection{Buoyant rise properties}
\label{sec:buoyantrise}

Figure \ref{fig:fluxrope_snapshots} shows three snapshots of the evolution of a flux-rope as it rises through the convection zone, namely at $t \approx 1, 10$ and $20$ days.
The top row shows the radial velocity $v_r$ while the second row shows the density deficit $\Delta\rho$ in the flux-rope.
The black lines are contours of the toroidal magnetic field $B_\phi$.
For simplicity, only the standard case is represented in the figure (strong-field highly buoyant case).
All other cases whose flux-ropes manage to emerge share the same qualitative properties; the differences will be discussed later on in the text.
The flux-rope expands as it rises mostly due to the background pressure stratification and partially due to magnetic diffusion.
We emphasise that the flux-rope's diffusive time-scale $a^2/\eta$ is $80$ days, which is larger than the typical buoyant rise time ($\sim 10\un{d}$, except for the very weakly buoyant cases).
The flux-rope section is perfectly circular at $t=0$, but its periphery suffers small and intermittent deformations.
At about $t=1\un{d}$ the buoyant force has provided enough upwards momentum for the flux-rope to start rising coherently.
The top left panel in Fig. \ref{fig:fluxrope_snapshots} shows that the whole core of the flux-rope has a positive $v_r$ at this moment, while its periphery still shows a different (yet transient) behaviour.
The first panel in the second row in the figure shows how the density deficit $\Delta\rho$ (which the buoyant force is proportional to) is stronger at the core than at the periphery, partly explaining the aforementioned different behaviours (see also the definitions in equations \ref{eq:potA} to \ref{eq:deltarho}).
Also, the magnetic tension responsible for keeping the flux-rope together against the work done by the external flows is weaker at its periphery, as the magnetic field strength is weaker there and field-line curvature radii are larger.
The magnetic tension force in the poloidal plane scales as $F_T \propto B_p^2/R_c$ for a flux-rope layer with circular cross-section, where $R_c$ is the curvature radius and $B_p$ is the poloidal magnetic field amplitude.
The second panels in both rows of Fig. \ref{fig:fluxrope_snapshots} show the same flux-rope close to mid-height in the convection zone.
At this time ($t\approx 10\un{d}$), the flux-rope's rise speed is close to its maximum value ($\sim 115\un{m/s}$; see Fig. \ref{fig:fluxrope_height}).
The core of the flux-rope maintains its circular cross-section, which has expanded to a larger radius, and the density deficit $\Delta\rho$ as decreased accordingly.
It is worth noting that while the flux-rope's periphery is continuously dragged with the surrounding flows, the well-known double-tailed profile is not clearly identified in the dynamo runs as it is in the case with a hydrodynamic background 
\citep[cf. e.g][particularly the first panel in Fig. 19 of the latter, for the case with $P_m = 4$]{emonet_physics_1998,hughes_rise_1998,jouve_three-dimensional_2009}.
In fact, a wake continuously forms behind the rising flux-ropes, but it is promptly assimilated by the magnetic background field (the magnetic field in the former and the latter are of the same magnitude).
The peripheral layers of the flux-ropes show a rather erratic pattern in all runs with a dynamo background, suggesting magnetic flux is continuously exchanged with the magnetised surroundings.
The third panel of Fig. \ref{fig:fluxrope_snapshots} shows the flux-rope after it has reached the top boundary of the domain and stopped rising buoyantly.
At the instant represented ($t\approx 20\un{d}$), the system has gone past a flux-emergence episode (see Sect. \ref{sec:emergence}).
The flux-rope maintains its spatial coherence for a period of time much longer than the buoyant rise time, though.
The flux-rope remains in the highest layers of the CZ while being slowly advected poleward by the mean meridional flows and sheared by the horizontal surface flows (see Sect. \ref{sec:surfacefields} for details on the poleward drift and surface shearing).
At the same time, a part of its magnetic flux progressively reconnects and mixes with the ambient magnetic field.
Section \ref{sec:laterevol} describes in greater detail the later evolution phases.
The flux-rope generates a dipolar flow around itself as it rises buoyantly through the CZ, as shown in Figure \ref{fig:fluxrope_massflux}.
The colour-scale represents the toroidal field $B_\phi$ and the white lines are streamlines of the azimuthally-averaged poloidal mass flux $\bar{\rho}\langle\mathbf{v}_{p}\rangle_{\phi}$.
Dashed and continuous lines represent, respectively, counter-clockwise and clockwise circulations.
The instants represented are the same as in Fig. \ref{fig:fluxrope_snapshots}.
This circulation flow has the following characteristics: a strong radially oriented upflow above the flux-rope, a shallower upflow in the wake of the flux-rope and a return flow which encircles the flux-rope.
This flow pattern appears as soon as the flux-rope starts rising buoyantly as a whole, as can be seen in the first panel ($t\approx 1\un{d}$).
At this stage, the dipolar flow encompasses a small area, of radius roughly twice as big as the radius of the flux-rope.
The spatial extent of this pattern increases with time, as the flux-rope rise speed increases (and the flux-rope magnetic field is strong enough for it to keep its coherence).
This is clear in the second panel ($t\approx 10\un{d}$), when the flux-rope is close to its maximal rise speed.
The dipolar flow is destroyed as the rope slows down when it reaches the surface (third panel, $t\approx 20\un{d}$).

\subsection{Buoyant rise speed}
\label{sec:buoyantrisespeed}

Fig. \ref{fig:fluxrope_height} shows the height of the flux-rope's axis as a function of time (top panel) and its rising velocity as a function of radius (bottom panel) for the standard case (strong magnetic field), the case with a hydrodynamic background and a standard flux-rope, the case with a left-handed flux-rope and the cases with intermediate and weak field and low buoyancy.
We also considered a low-latitude and a high-latitude cases, corresponding to flux-ropes similar to those in the standard case, but his time placed initially at $15\un{N}$ and $45\un{N}$, respectively.
Among the standard latitude cases,
the flux-rope evolving in a purely hydrodynamical background has the highest buoyant rise speed (continuous grey line line in Fig. \ref{fig:fluxrope_height}), slightly higher than the standard case's speed (continuous black line in Fig. \ref{fig:fluxrope_height}).
Both curves are qualitatively similar, but the former has a maximum value of about $\sim 125\un{m/s}$ while the later peaks at $\sim 115\un{m/s}$, that is, a $\sim 9\%$ deviation.
The left-handed flux-rope (having the same $B_0$,$\delta S$ and $\Delta\rho/\rho$ as the standard case, dashed black line in the figure) attains a maximum rise velocity of $\sim 110\un{m/s}$, $\sim 5\%$ less than the standard case.
More importantly though, the rise speed profile as a function of height is skewed towards higher radii relatively to the standard case.
The left-handed flux-rope's speed is initially lower (in the first half of the domain) but it becomes higher afterwards.
The case with negative polarity (negative $B_0$) follows the standard case very closely and was left out of the plot for simplicity.
The cases with weaker $B_0$ and $\Delta\rho/\rho$ (red lines in the figure) attain lower buoyant rise speeds, as expected.
The weakest of them (dotted red line) does not manage to attain the top of the domain, being disrupted before (roughly at mid-height in the CZ).
In the first days of the simulation, this flux-rope actually acquires a slightly negative speed, hardly managing to counter-balance the convective downflows.
This case led us to re-evaluate the threshold for the flux-ropes buoyancy.
We ran a few extra cases, at different density deficits, to better constraint this threshold.
The re-evaluated threshold if represented in Fig. \ref{fig:parameter_space} with a red dashed line.

Overall, the background magnetic field seems to exert an enhanced drag over the flux-ropes.
Furthermore, this effective drag depends on the relative orientation of the background and the flux-rope magnetic fields (\emph{i.e}, on the way the flux-rope connects with the external magnetic field).

The bottom panel of Fig. \ref{fig:fluxrope_height} further shows that the flux-rope traversing an hydrodynamic convection zone evolves smoothly, while the dynamo cases show a more irregular behaviour.
The amplitude of these variations is small in comparison to the average rise speeds, and it is hardly discernible in the figure.
We interpret this feature as the signature of a flux-rope opening its way upwards by consecutive episodes of reconnection with the background field \citep[\emph{cf.}][albeit in a simpler scenario]{dorch_buoyant_2007}.
It is likely that these small deviations to the flux-tube trajectories become larger if higher turbulence levels are attained, as for example in \citet{hughes_rise_1998}.

Variations in the initial flux-rope latitude produce a very wide range of buoyant rise speeds, independently of the other flux-rope geometrical parameters.
This behaviour was already found by \cite{jouve_three-dimensional_2009} both in their convective and isentropic cases with hydrodynamic backgrounds.
Similar flux-ropes (with the same density deficit, magnetic strength and twist) will experience different effective buoyancy forces as a function of the background rotation rate, which is non-uniform across the meridional plane (see Fig. \ref{fig:bgfieldflows}, middle panel).
The buoyant force applied to the flux-ropes is, per unit mass, proportional to $g - r\sin^2{\theta}\Omega^2\left(r, \theta\right)$.
Low-latitude flux-ropes cross zones with rotation rates necessarily higher than those placed at higher-latitudes.
This effect is particularly expressive In the cases shown in Fig. \ref{fig:fluxrope_height}, initially placed at $15\un{N}$ and $45\un{N}$.
The high-latitude case crosses a region of the CZ with roughly uniform $\Omega$, while the low-latitude case crosses layers with increasing $\Omega$ as it rises through.
This makes the difference in flux-rope vertical acceleration actually increase during the buoyant rise, further accentuating the differences.
The instantaneous flows each flux-rope encounters during the buoyant rise further contributes to increasing this spread in rise speed.
In fact, the trajectory of the flux-rope placed at $45\un{N}$ crosses zones for which the background convective flows with $\langle v_r\rangle_{\phi, t} > 0$, while its low-latitude counterpart crosses through a more balanced distribution of upflows and downflows both in longitude and in time.
This is somewhat specific to the instant chosen to be the initial time $t_0$ in our model.
But, since our simulation had at that instant already achieved a statistically mature state in regards to its convective flows, the flow properties encountered by the flux ropes in our runs are on average representative of any instant.

\subsection{Flux-rope expansion, magnetic flux and density}

The magnetic flux-ropes expand as they rise buoyantly through the CZ.
The flux-rope's magnetic field strength and density are then both expected to vary as a result of this expansion.
It has been suggested that the evolution of these two quantities is well described by the simple power-law
\begin{equation}
  B_c \propto\rho_c^\alpha\ ,
  \label{eq:powerlaw}
\end{equation}
where $B_c$ is the flux-ropes toroidal magnetic field strength and $\rho_c$ is the total density inside the flux-ropes.
\citet{cheung_simulation_2010}, in particular, have verified that an $\alpha$-index between $1/2$ and $2/3$ allows for a correct assessment of the combined variations of magnetic field strength and density in their simulations of the later phases of the emergence of a horizontal flux-rope.
Their study was based on a numerical model which covers a small cartesian domain spanning roughly $90\un{Mm} \times 50\un{Mm}$ horizontally and $10\un{Mm}$ vertically across the photosphere, in which an horizontal magnetic flux-rope was introduced.
We assess now if this result still holds for self-connected toroidal flux-ropes as those in our setup and  whether it is extensible to the whole of the buoyant rise (from the bottom to the top of the CZ).

Figure \ref{fig:b_rho} shows a log-log plot of the magnetic field strength $B_c$ as a function of the flux-rope's total density $\rho_c$ (both measured at the axis of the flux-rope) for some of the runs we performed (the legend identifies each of the cases represented).
The axis of the flux-rope was defined as the position in the poloidal plane where $\left|B_\phi\right|$ is maximum.
The grey dashed lines indicate the slopes corresponding to $\alpha=1/2$, $2/3$ and $1$.
It is readily visible that $B_c$ scales as $\rho_c^\alpha$, with $\alpha$ very close to but smaller than $1$.
Each one of the curves in Fig. \ref{fig:b_rho} was fitted to the aforementioned power-law.
We restricted the fitting procedure to the moments when the flux-rope height is in the interval between $0.8\rsun$ and $0.92\rsun$ in order to avoid spurious effects due to the proximity of the numerical boundaries.
This excludes (for all runs) the initial acceleration at the bottom of the CZ and the braking at the top.
%
The strong-field and more buoyant cases are best fitted with an index $\alpha=0.998\pm 0.001$ (black continuous and dotted lines in Fig. \ref{fig:b_rho}).
The case with a hydrodynamical background (no dynamo) shows a similar behaviour, but with a smoother profile and smaller fitting error.
These cases are those for which the toroidal symmetry and coherence are better maintained during their whole evolution.
Cases with a lower magnetic field strength and lower buoyancy show a slightly steeper power-law index (slightly higher $\alpha$; red curve in Fig. \ref{fig:b_rho}) but, perhaps more importantly, show a considerably higher fitting error.
Cases with weak fields but strong buoyancy (not represented in Fig. \ref{fig:b_rho} for simplicity) remain on average very close to their strong-field counterparts, but showing stronger variations (their profile is less smooth).
As a side note, we also observed that the evolution of $\rho_c\left(r\right)$ shows very little differences for the different cases, while the dispersion in the $B_c\left(r\right)$ curves is noticeably higher; the evolution of the density at the axis of the flux-ropes seems to be defined by the background stratification almost on its own.

Let us now discuss the physical significance of the $B_c\propto\rho_c^\alpha$ scaling law found.
We will, for simplicity, consider the magnetic flux tube to be perfectly axisymmetric toroidal structures of circular cross-section.
Let us define $a$ as the minor radius of the toroidal tube and $R$ the major radius (the distance between the centre of the torus to the centre of the tube).
The position of the centre of the tube is given by $r$ and $\theta$, with the usual meaning in spherical coordinates.
The cross-section surface of the tube is $A=\pi a^2$ and its length is $L = 2\pi R = 2\pi r\sin\theta$.
The volume of the toroidal flux-tube is $V = A\times L = 2\pi^2 a^2 r\sin\theta$.
The mass and magnetic flux contained in the flux-tube are, at all moments,
\begin{eqnarray}
  \label{eq:mass_flux}
  M    &=& \rho A\times L = \rho_c 2\pi^2 a^2 r\sin\theta \nonumber \\
  \Phi &=& B_c A = B_c \pi a^2 \ .
\end{eqnarray}
These two quantities are conserved during the buoyant rise as long as the mass and magnetic flux exchanges with the environment are negligible, and so it the ratio $\Phi/M$
\begin{equation}
  \label{eq:phi_m_ratio}
  \frac{\Phi}{M} L = \frac{B_c}{\rho_c} = \rho_c^{\alpha - 1}\ ,
\end{equation}
with the last equality implying that the power-law in Eq. \eqref{eq:powerlaw} holds.
Equivalently,
\begin{equation}
  \label{eq:bphi_over_rho}
  \frac{B_c}{\rho_c} \propto L = 2\pi r\sin\theta\ .
\end{equation}
That is, both $B_c$ and $\rho_c$ depend on the variations in flux-tube cross-section $a$ (poloidal expansion), but only $\rho_c$ is sensitive to variations in the tube's length $L = 2\pi r\sin\theta$ (toroidal expansion).
The contribution of the poloidal flux-tube expansion vanishes in Eq. \eqref{eq:bphi_over_rho}, as both $M$ and $\Phi$ are proportional to $A=\pi a^2$ (see Eq. \ref{eq:mass_flux}).
If the flux-rope varied its cross-section $a$ while keeping its length $L$ constant, then the index $\alpha$ would be exactly equal to $1$.
Conversely, if the flux-rope underwent a purely toroidal expansion (constant $a$, growing $L$), then $0< \alpha < 1$.
In our simulations, the total flux-rope expansion is a combination of both the toroidal and poloidal components.
The torus length $L\propto r$ increases during the buoyant rise, and so does the flux-rope's cross-section (due to the background pressure radial profile).
As stated above, $\alpha\lesssim 1$ in all our runs, meaning the poloidal component of the flux-rope expansion dominates over the toroidal component.

Of course, these arguments rely on the assumption of conservation of mass and magnetic flux inside the buoyant flux-rope (and more specifically, on the conservation of the ratio $\Phi/M$).
We need to verify if this assumption fails, or at least how much the leakage of mass and magnetic flux affects the results.
As stated before (Sect. \ref{sec:buoyantrise}), a fraction of the peripheral magnetic flux is continuously dragged within the wake which forms behind the flux-rope as it rises buoyantly.
It is also likely that reconnection occurring around the flux-ropes and any other diffusive process altogether lead to a flux exchange with the surroundings.
The top panel in Figure \ref{fig:tubeflux} shows the evolution of the magnetic flux $\Phi$ contained within the twisted flux-ropes as a function of height, as the ropes rise through the CZ.
The curves do indicate that the losses (to the surroundings) are substantial, specially in the later phases of the buoyant rise.
The stronger inflection of the black continuous and dotted lines in the upper part of the domain is due to the flux-emergence episode; the curves represent the interval $t=0$ to $t\sim 15\un{d}$, with the emergence episode starting roughly at $t=12\un{d}$.
Weaker-field flux-ropes suffer stronger losses and the flux-rope evolving in a hydrodynamic background is the one better verifying flux conservation.
The evolution of the mass contained in the flux-ropes correlates very well with that of the magnetic flux (i.e, magnetic flux exchange is dominated by processes also implying mass exchange).
The bottom panel in Fig. \ref{fig:tubeflux} summarises the combined effect of the variations of $\Phi$ and $M$ on $\alpha$.
The index $\alpha$ is obtained by re-arranging Equation \eqref{eq:phi_m_ratio}, that is
\begin{equation}
  \label{eq:const_ratio}
  \alpha = \frac{\log{\left(\Phi L \middle/ M\right)}}{\log{\rho_c}} + 1.
\end{equation}
This confirms the $\alpha\lesssim 1$ value found above.
This result holds for all cases studied, for different flux-rope buoyancy forces and initial latitudes.

\subsection{Magnetic energy balance}
\label{sec:energy_balance}

The evolution of the total magnetic energy is described by the equation
\begin{equation}
  \label{eq:magenergy}
  \frac{d}{dt}\int_V \frac{B^2}{8\pi} dV = -\int_V \frac{J^2}{\sigma} dV - \frac{1}{c}\int_V \left(\mathbf{J}\times\mathbf{B} \right)\cdot\mathbf{v} dV - \oint_S \calS dS\ ,
\end{equation}
where the terms on the r.h.s represent, respectively, the ohmic diffusion rate, the rate of work done by the Lorentz force and the energy loss rate transported by the Poynting flux across the domain's top and bottom boundaries.
The ohmic diffusion term is always an energy sink, while the other two terms on the r.h.s of Eq. \eqref{eq:magenergy} can either act as energy sources or sinks.
The magnetic work rate term $- \frac{1}{c}\int_V \left(\mathbf{J}\times\mathbf{B} \right)\cdot\mathbf{v} dV$ is, for example, positive when magnetic structures induce flows in the system and negative when the flows shear the magnetic structures with little or no feedback on the former.
The Poynting vector (in the Poynting flux term) is defined as
\begin{equation}
  \calS = \frac{c}{4\pi} \mathbf{E}\times\mathbf{B}
\end{equation}
with $\mathbf{E}$ being the electric field vector, and quantifies the electromagnetic energy transfer rate across a surface, that is, the total electromagnetic energy which leaks into or out of the system.
Considering a finite conductivity $\sigma = c^2/\left(4\pi\eta\right)$ and the Ohm's law $\mathbf{J}/\sigma = \mathbf{E} + \mathbf{v}\times\mathbf{B}$, this vector becomes
\begin{equation}
  \label{eq:poynting_vector}
  \calS = \frac{\eta}{c}\mathbf{J}\times\mathbf{B} +
  \frac{c}{4\pi}\left[ B^2\mathbf{v} -
  \left(\mathbf{v}\cdot\mathbf{B}\right)\mathbf{B}\ \right] .
\end{equation}
The second and third terms in the r.h.s. of Eq. \eqref{eq:poynting_vector} describe, respectively, the advective transport of the magnetic energy density and to the propagative energy transport supported by the magnetic tension (e.g.: Alfv\'en waves excited by transverse motions).
The first term in the r.h.s. is most often neglected in ideal MHD studies, and we observe it to be always much smaller than the other terms in our simulations in spite of the finite diffusivity.
The radial component of this vector (\emph{i.e} the component orthogonal to our domain's boundaries) is, in CGS units,
\begin{equation}
  \label{eq:poynting_r}
  \mathcal{S}_r = \frac{\eta}{c}\left[J_\theta B_\phi - J_\phi B_\theta \right] + 
  \frac{c}{4\pi}\left[ v_r\left(B_\phi^2+B_\theta^2\right) - 
    B_r\left(v_\phi B_\phi + v_\theta B_\theta\right) \right]\ .
\end{equation}
Analysing the Poynting fluxes obtained from numerical simulations requires some special precautions.
These are most often computed at the numerical boundaries of the computational domains, and therefore the specific choice of boundary conditions certainly influences the Poynting flux amplitudes.
Eq. \eqref{eq:poynting_r} separates the effects of the conditions imposed on components of the velocity parallel and perpendicular to the boundary (in our setup $v_\theta$ and $v_\phi$ are parallel while $v_r$ is perpendicular).
It also shows how the Poynting flux strongly depends on the parallel components of the magnetic field; matching the boundary magnetic field to an external radial field imposes a null poynting flux there, for example \citep[\emph{cf.}][]{brun_global-scale_2004}.
In our setup, the magnetic field is matched to an external potential field, thus allowing for magnetic energy leakage through the boundaries.
The parallel components of the velocity are also allowed to vary in time at the boundaries.
The advective component of the Poynting flux is neglected (due to the boundary condition imposed for $v_r$), but this term is probably smaller in comparison to the third term in the r.h.s of Eq. \eqref{eq:poynting_r} at the top of the CZ in the real Sun.
We tested the consistency of the magnetic energy balance in our models in two ways.
First, we computed the radial Poynting fluxes at all points (in radius) of our model's domain and verified that there were no noticeable boundary layer effects.
Second, we compared directly the l.h.s and the r.h.s of Eq. \eqref{eq:magenergy} and verified that the equality holds down to the numerical precision at all times, meaning that leakage and diffusion are consistently accounted for in the simulations.

Figure \ref{fig:fluxrope_me} shows the temporal evolution of the total magnetic energy (top left panel) for some of the runs in Table \ref{tab:fluxropepars} during a time interval of up to $60$ days after the introduction of the magnetic flux-ropes, together with its source/sink terms.
The background dynamo run (with no flux-rope introduced) is superimposed for comparison.
The flux-ropes introduced in the system correspond to very strong concentrations of magnetic field, and so the total magnetic energy is much higher than that of the background dynamo at the moment of introduction ($t=0$ in Fig. \ref{fig:fluxrope_me}, top left panel).
This is true even for the weaker magnetic cases we ran (red and blue lines).
This initial perturbation in the magnetic energy is, of course, a consequence of our method and it would not take place if such flux-ropes were generated self-consistently by the magneto-convective flows in the CZ as in \citet{nelson_buoyant_2011,nelson_magnetic_2013}, for fast rotating solar-like stars.
This issue is beyond the scope of this paper, though, and requires future investigation.

The magnetic energy decays quickly during the buoyant rise of the flux-ropes (the first $12$ days for the standard flux-rope).
During this phase, the energy losses are initially dominated by the ohmic diffusion (top right panel),
but the work done by the Lorentz force grows as the flux-rope gains momentum due to the buoyant force.
Let us explain this effect.
The left bottom panel shows that the magnetic work term is indeed negative during the whole buoyant rise phase for all cases represented, peaking at about the moment when the buoyant rise speed is maximal.
That happens because the rising flux-rope induces a flow around itself in the CZ, as shown in Sect. \ref{sec:buoyantrise}.
This translates into work done by the Lorentz force, transferring magnetic energy into kinetic energy.
Note that the sign of this term changes afterwards, as the buoyant rise ends.
In the later phases, much after the flux emergence episode (happening at about $t=12\un{d}$ for the standard case), the magnetic energy curve inflects and starts varying almost in parallel with the unperturbed dynamo case.
At this point, the dynamo processes have taken over the perturbations forced by the buoyant flux-ropes in the background flows, and the magnetic work term is consistently positive.

The bottom right panel of Fig. \ref{fig:fluxrope_me} shows the total Poynting flux $\mathcal{S} = \mathcal{S}_{top} - \mathcal{S}_{bottom}$.
The initial spike visible in all the Poynting flux curves corresponds to the contribution of the flux crossing the lower boundary of the domain $\mathcal{S}_{bottom}$, which is only important in the beginning of the simulations (during the first $4$ days) and has a positive contribution to the system's magnetic energy.
The Poynting flux curves peak again during the flux emergence episode ($t=10 - 22\un{d}$ for the standard case), this time being dominated by the flux crossing the upper boundary $\mathcal{S}_{top}$.
This peak has a negative sign, meaning magnetic energy is being transferred outward to the corona.
The amplitude of the Poynting flux depends both on the flux-ropes magnetic field strength and buoyant rise speed.
This is understandable in view of Equations \eqref{eq:poynting_vector} and \eqref{eq:poynting_r} and the fact that the amplitude of the surface flows induced by the flux-rope rise and emergence depends directly on the buoyant rise speed.

The black and blue lines in the plot correspond to flux-ropes with the same buoyancy but different $B_0$, the former having a stronger magnetic field (standard case) than the latter.
The weaker flux-rope produces a Poynting flux which is smaller than the standard case by a factor $2$, even if the initial $B_0$ is only smaller by a factor $1.25$, which is consistent with 
$\mathcal{S} \propto B^2$ (the surface flow velocities being the same).
Note that the match is not perfect; we are comparing initial state parameters with later diagnostics and each of the flux-ropes suffer different effects from the environments in their evolution from the bottom to the top of the CZ.
The blue and red lines correspond to cases with the same initial magnetic field strength, but with the former being more buoyant than the latter.
Both the time delay of about $10\un{d}$ and the difference in amplitude relate directly with the different buoyant rise speeds across the CZ.


The total magnetic energy evolution depends weakly on the flux-rope's polarity and handedness, even though these properties lead to important differences in the surface signatures of flux emergence (as we will see later on).
The flux-rope's magnetic field does interact with the surrounding dynamo field; differences in amplitude, polarity and handedness induce changes in the evolution of the flux-ropes themselves but translate into a small effect in the evolution of the magnetic energy of the whole system.

Figure \ref{fig:fluxrope_me} also shows a few other interesting points.
The magnetic energy in the case with a purely hydrodynamical convective background decays faster than all the others, as in the latter the energy decay due to the flux-rope rise and expansion accounts only for a fraction of the total magnetic energy.
But, later in its evolution (for $t\geq 40 \un{d}$), the magnetic energy starts growing at a temporal rate close to that of the dynamo run.
At this stage, the flux-rope is already well past the emergence episode and the associated magnetic fields have already been strongly disrupted by the shearing flows at the upper layers of the CZ (see Sect. \ref{sec:laterevol}).
The flux-rope's magnetic flux which did not emerge is now seeding a (local) dynamo.

The cases where weaker magnetic flux-ropes were added to the dynamo background show a somewhat surprising behaviour.
Their associated magnetic energy decays more slowly than that in the strong field cases, even if they are less robust to disruption by the background convective flows (smaller magnetic stresses to counterbalance deformations imposed by the flows) and dynamo field (reconnection events should take away a higher fraction of the flux-rope's own field).
The most important parameter here seems to be the buoyant rise speed.
Lower rising speeds mean that the cross-section $A$ of the flux-ropes will expand more slowly in time (as the flux-ropes takes longer to attain the lesser dense layers of the CZ) and therefore the nominal magnetic field $B_0$ will also decrease more slowly (as $B\propto A^{-1}$, the total magnetic flux in the flux-ropes being approximately conserved).
The flux-rope's contribution to the total magnetic field scales as $B_0^2$, so this effect becomes important.
Note how flux-ropes with the same initial magnetic field strengths but different buoyancy forces assume different decay rates (the red/blue pairs of lines, with the blue representing cases with the same magnetic field strength but higher buoyancy).
The buoyant rise speeds for each case can be inferred from the $\Delta\rho/\rho$ values in the diagram in Fig. \ref{fig:parameter_space} and verified in Fig. \ref{fig:fluxrope_height}.
The blue and red lines bifurcate in Fig. \ref{fig:fluxrope_me}, such that the blue lines (more strongly buoyant) start decaying faster, at a rate close to that of the ``strong-field'' cases,  which have the same initial $\Delta\rho/\rho$ (black lines in the plot).

\section{Post buoyant rise phases}
\label{sec:emergence}

\begin{figure*}[]
  \centering
  \includegraphics[width=2\picwd]{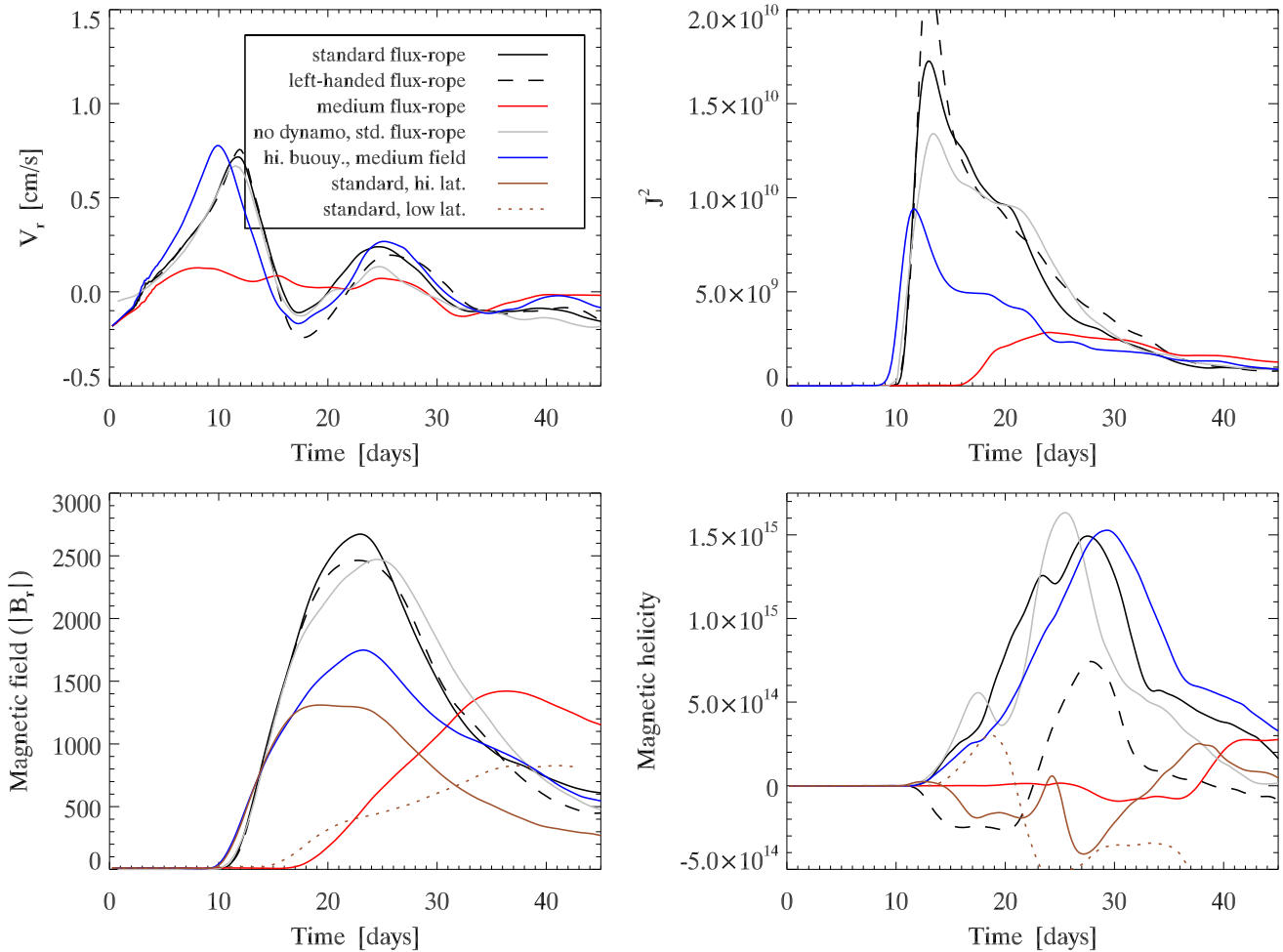}
  \caption{Surface variations (relative to the background state) of 
    the radial velocity, unsigned magnetic field strength, current density squared and magnetic helicity before and during the flux-emergence episode averaged over the latitude interval where magnetic flux emerges.
    Note how the maximum of the quantities $v_r$ and $J^2$ precede that of the other ones.
    The amplitude of the $v_r$ signal is mostly determined by the flux-rope's rise speed.
    There is also a signal in surface density (not shown here) which correlates very well with $v_r$, but has a small amplitude.
    Surface magnetic field strength and current density are determined by the flux-rope's strength; the magnetic helicity further depends on the handedness of the emerging rope.
  }
  \label{fig:precursors}
\end{figure*}

In this section we focus on the manifestations of the arrival of the buoyant magnetic flux-ropes at the top of the domain, after they crossed the convection zone.
For simplicity, we will hereafter use the term ``surface'' to refer to the top of the numerical domain and define flux-emergence as strong and localised enhancements of magnetic field at the top of the domain caused by the arrival of rising flux-ropes.
We also define magnetic flux-emergence episode as the time interval from the first signs of flux-rope related variation of the magnetic flux crossing the surface up to the peak in surface magnetic flux and energy.
These moments correspond, respectively, to $t=10\un{days}$ and $t=22\un{days}$ for the standard case in our simulations (note that these timings could be different in real solar events).

\subsection{Emergence precursors}
\label{sec:precursors}

We will focus here on identifying features which consistently precede the actual flux-emergence episode in our simulations.
Figure \ref{fig:precursors} shows the temporal evolution of the radial velocity $v_r$, the unsigned magnetic field $\left|B_r\right|$, the current density squared $J^2$ and the magnetic helicity $\mathbf{A}\cdot\mathbf{B}$ evaluated at the surface and within the latitudinal interval where flux emergence occurs as a function of time.
The earliest tracer of flux-rope rise and emergence is the radial velocity $v_r$, which also correlates well in time with the surface density fluctuations $\rho$.
The latter has a very weak signal, though, and is not represented in Fig. \ref{fig:precursors} for simplicity.
These gradually increase during the whole flux-rope buoyant rise and peak at the moment when the flux emergence episode starts (that is, when the surface magnetic flux and energy start increasing).
That is about $12\un{days}$ before magnetic flux and energy due to the flux-rope emergence reach their maximum value in the standard case.
The same time delay is observed for the cases with weaker flux-ropes magnetic fields but the same buoyancy.
The $v_r$ signal has the same amplitude as the standard case but the surface magnetic fields naturally peaks at a lower value.
Lower buoyancy flux-ropes generate much weaker $v_r$ signals; the red line in the top left panel in the figure represents a flux-rope with an initial $B_0$ which is $0.75$ times the standard case's one (black continuous line).
The magnetic field signal keeps an amplitude close to that of all other flux-ropes with the same initial $B_0$ but different buoyancies.
The delay between $v_r$ and $\left|B_r\right|$ is higher in this case, of order $15\un{days}$.
The surface $v_r$ (and also $\rho$, albeit with a very weak signal) are, therefore, observable precursors to the formation of solar active zones.
The physical cause for this behaviour lies on the dipolar flow self-consistently generated by the flux-rope itself as it rises through the CZ (see Sect. \ref{sec:ropeevol} and Figs. \ref{fig:fluxrope_snapshots} and \ref{fig:fluxrope_massflux}).
Namely, the radial upflow which forms above the flux-rope extends radially up to the surface.
A transient over-dense patch forms there as the upflow is stronger closer to the flux-rope than faraway from it, meaning plasma accumulates just below the surface increasingly faster than it manages to be evacuated horizontally.
This phenomenon ends as soon as the flux-rope starts slowing down at the upper part of the CZ (the buoyancy is reduced there).
The amplitude of the surface $v_r$ and $\rho$ signals is directly related to the flux-rope's buoyant rise speed (that is, to $\Delta\rho/\rho$), with more buoyant flux-ropes producing stronger signals.

An additional interesting feature is the temporal evolution of $J^2$, as it starts growing roughly at the same time as the other magnetic quantities (that is, when the flux emergence episode starts) but then grows much faster than the latter.
Its peak value occurs at about $2\un{days}$ after $v_r$ and $\rho$ peak and $10\un{days}$ before the magnetic flux and energy do in the standard case.
These time-delays are similar in case with a hydrodynamical background (grey line in the plots).
Also, the amplitudes of the $J^2$ signals is naturally stronger for the cases with stronger field flux-ropes.
In fact, $J^2$ is a good tracer of the boundary of the flux-rope, as that is where the magnetic gradients are the strongest.
The time-lags (of $\sim 10 \un{days}$ in the standard case) between the peaks of $J^2$ and $\left|B_r\right|$ are consistent with the flux-ropes rising speeds and the physical distance between its axis and its periphery.
The magnetic helicity $\mathbf{A}\cdot\mathbf{B}$ signal, contrary to the previous diagnostics, depends also on the handedness of the emerging flux-rope.
Right-handed and left-handed flux-ropes produce quasi-symmetric signatures during the emergence episode.
In the post-emergence phases this symmetry is broken, though, as both left and right-handed magnetic flux-ropes are sheared in the same direction by the convective surface flows.
The last panel in Fig. \ref{fig:precursors} clearly shows the initially symmetric evolution for the right and left handed flux-ropes ($t=10-12\un{days}$) and the inflexion happening thereafter.
The two lower panels in Fig. \ref{fig:precursors} also show the magnetic field amplitude and helicity for the high and low latitude cases, as they exhibit remarkably different behaviours.
The high latitude case produces a perturbation in $\left|B_r\right|$ which is similar to that of the standard case, while the low latitude case shows a weaker and more delayed perturbation.
The low latitude flux-rope effectively rises more slowly (see Fig. \ref{fig:fluxrope_height}) and is distorted to a larger extent by the convective flows.
The magnetic helicity evolution perturbations have initially opposite signs, even though both flux-ropes ( high and low latitude) have the same polarity and handedness.
The difference lies in the specific surface shearing motions (the differential rotation) encountered at the latitude of emergence.

The results presented in this section offer a viewpoint complementary to that proposed by 
local helioseismology studies \citep{ilonidis_detection_2011,kosovichev_local_2012,hanasoge_seismic_2012}, which have been producing increasingly more detailed constraints about the subsurface flows in the first few$\un{Mm}$ below the solar surface and possible early detection of rising flux-ropes.
As such, these new methods may provide in the near future the means to verify our findings.
The upper boundary of our numerical domain lies just below, at $r=0.97\rsun$.
Furthermore, the quantitative details of the temporal profiles in Fig. \ref{fig:precursors} are probably influenced by the boundary conditions imposed there.
Despite these reasons, we believe that the qualitative properties of the results discussed in this section may be generalised, to some extent, to the solar surface.
Note that we do not aim here at producing reliable quantitative predictions of such surface flux-emergence precursors.
This problem is currently being addressed using models whose numerical boundary lies above the photosphere and which account for the radiative losses taking place there \citep[see][]{pinto_3d_2011}.
Preliminary results suggest that the qualitative properties of the flux-emergence precursors described here are correct (in the sense that they are not a direct consequence of the domain truncation and upper boundary conditions).
Please note that while the exact time delays and amplitudes we found may not transpose directly to the Sun, the temporal ordering between the different diagnostics should remain the same.


\subsection{Surface fields}
\label{sec:surfacefields}

\begin{figure*}[!ht]
  \centering
  \includegraphics[width=0.5\picwd]{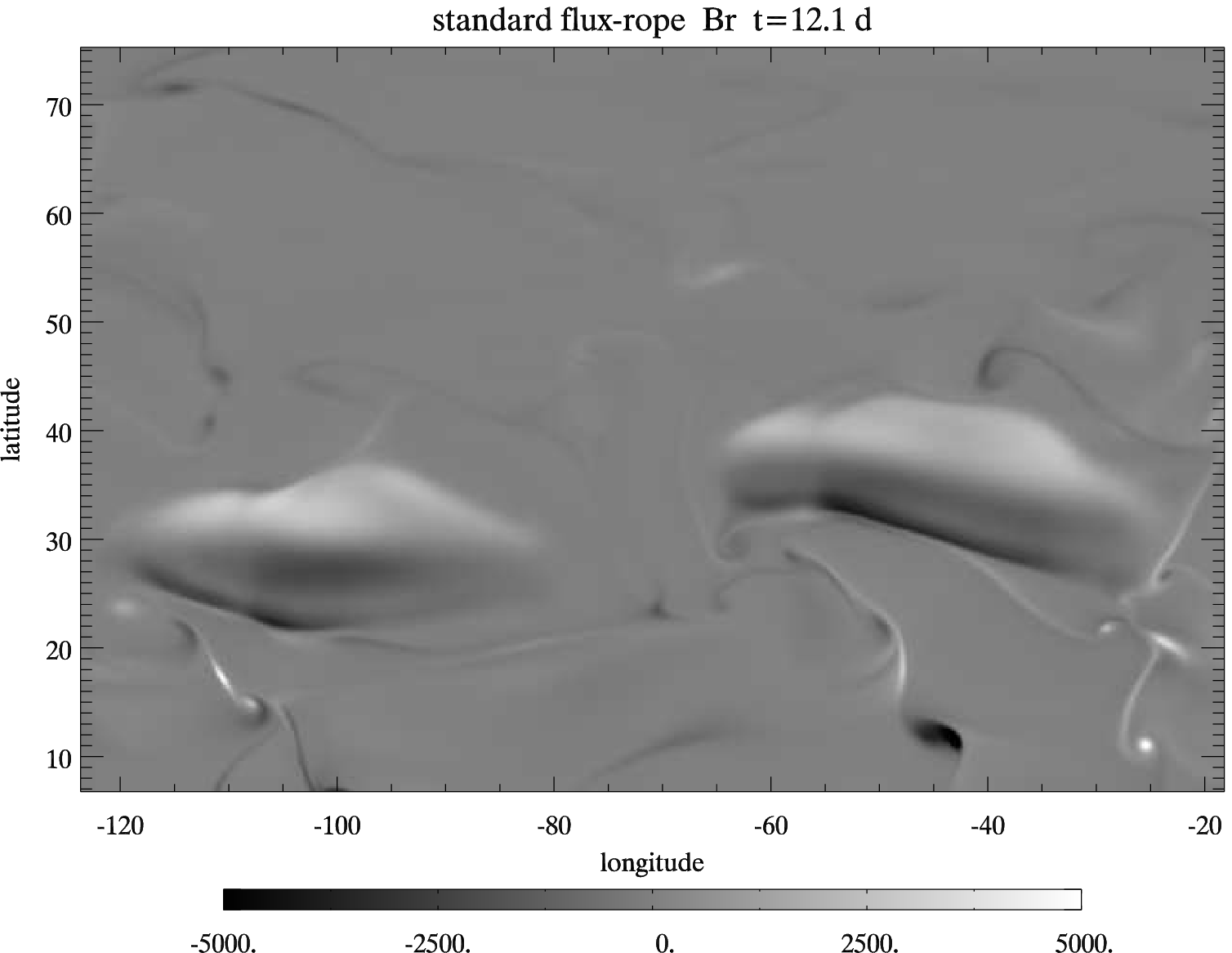}
  \includegraphics[width=0.5\picwd]{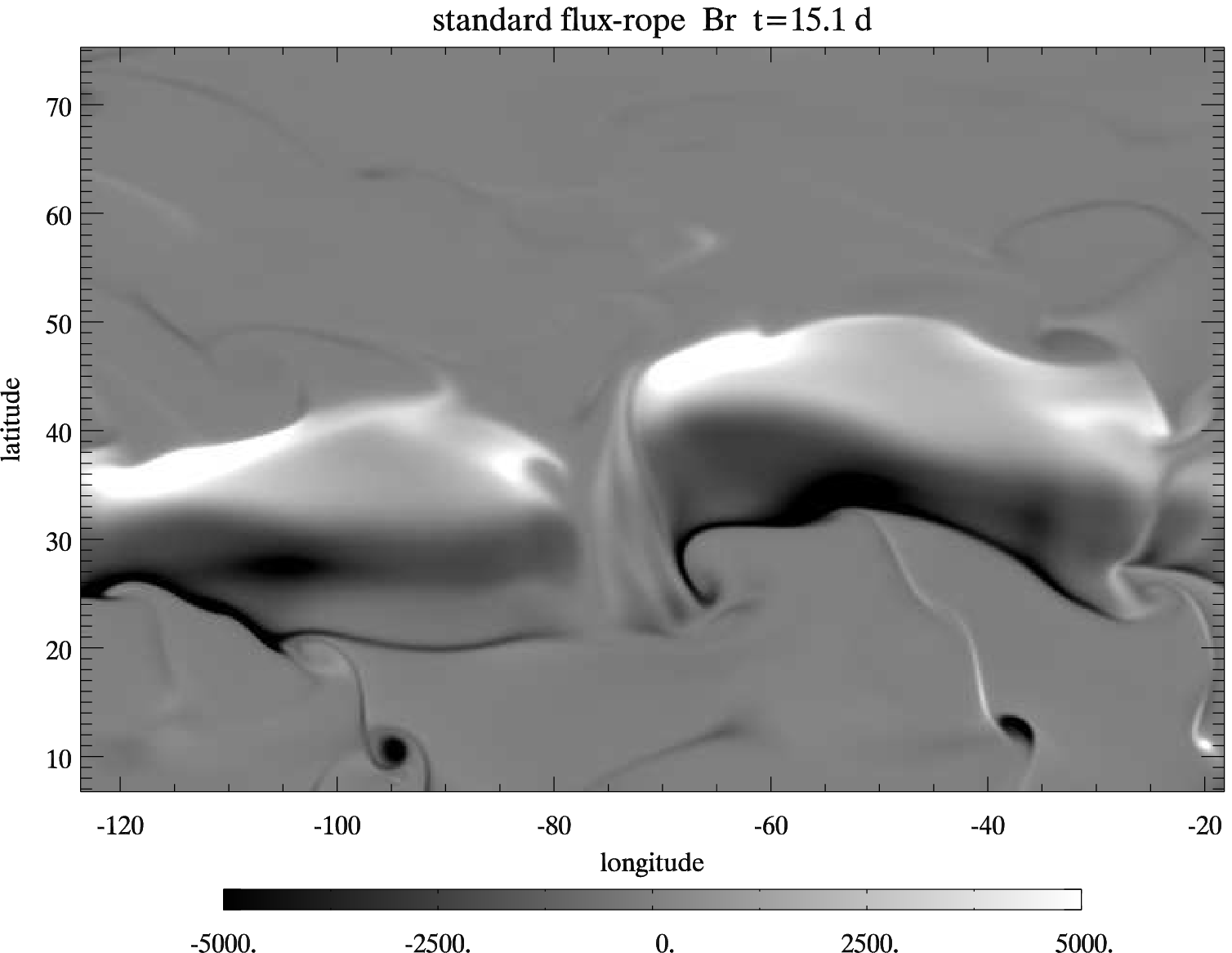}
  \includegraphics[width=0.5\picwd]{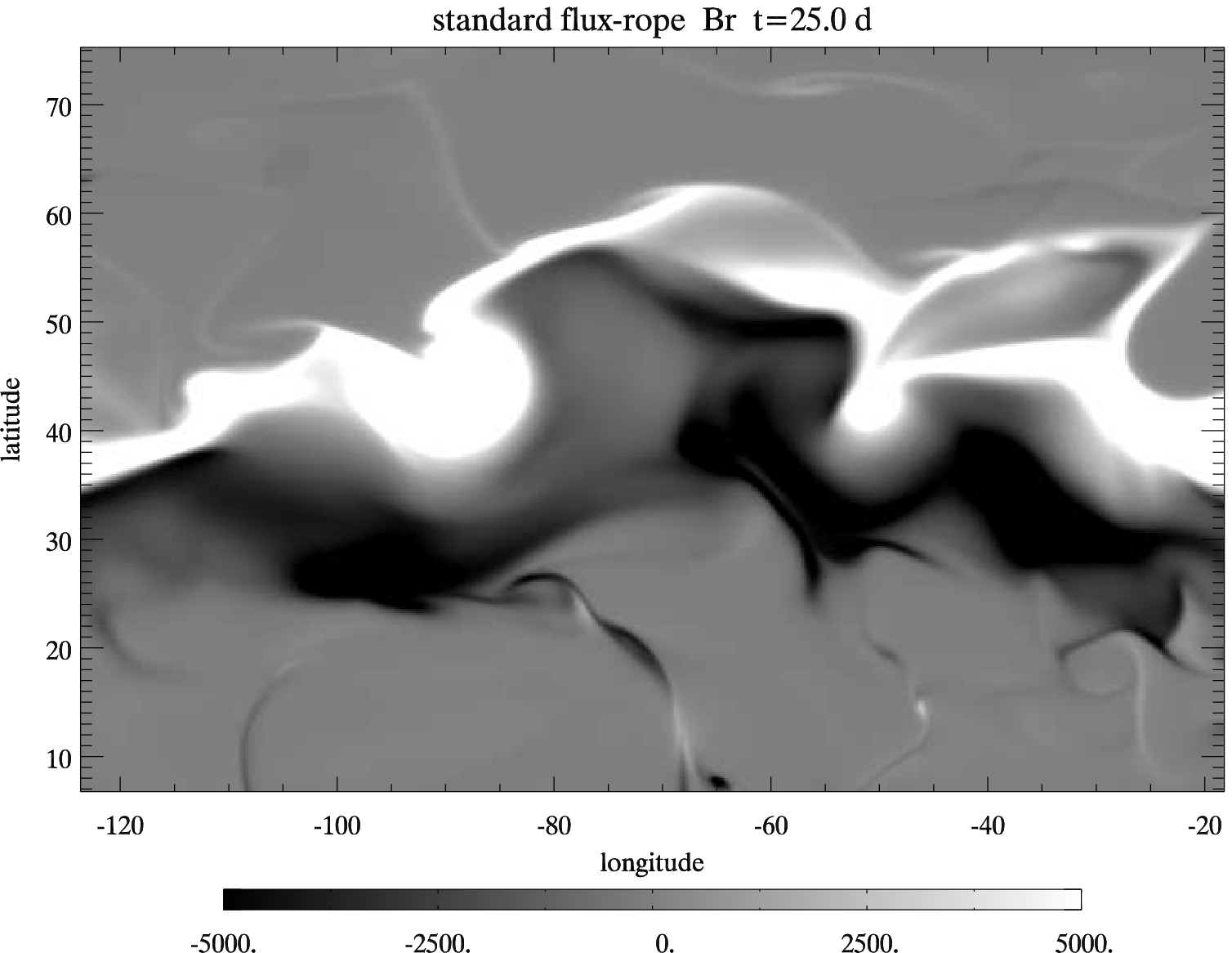}
  \includegraphics[width=0.5\picwd]{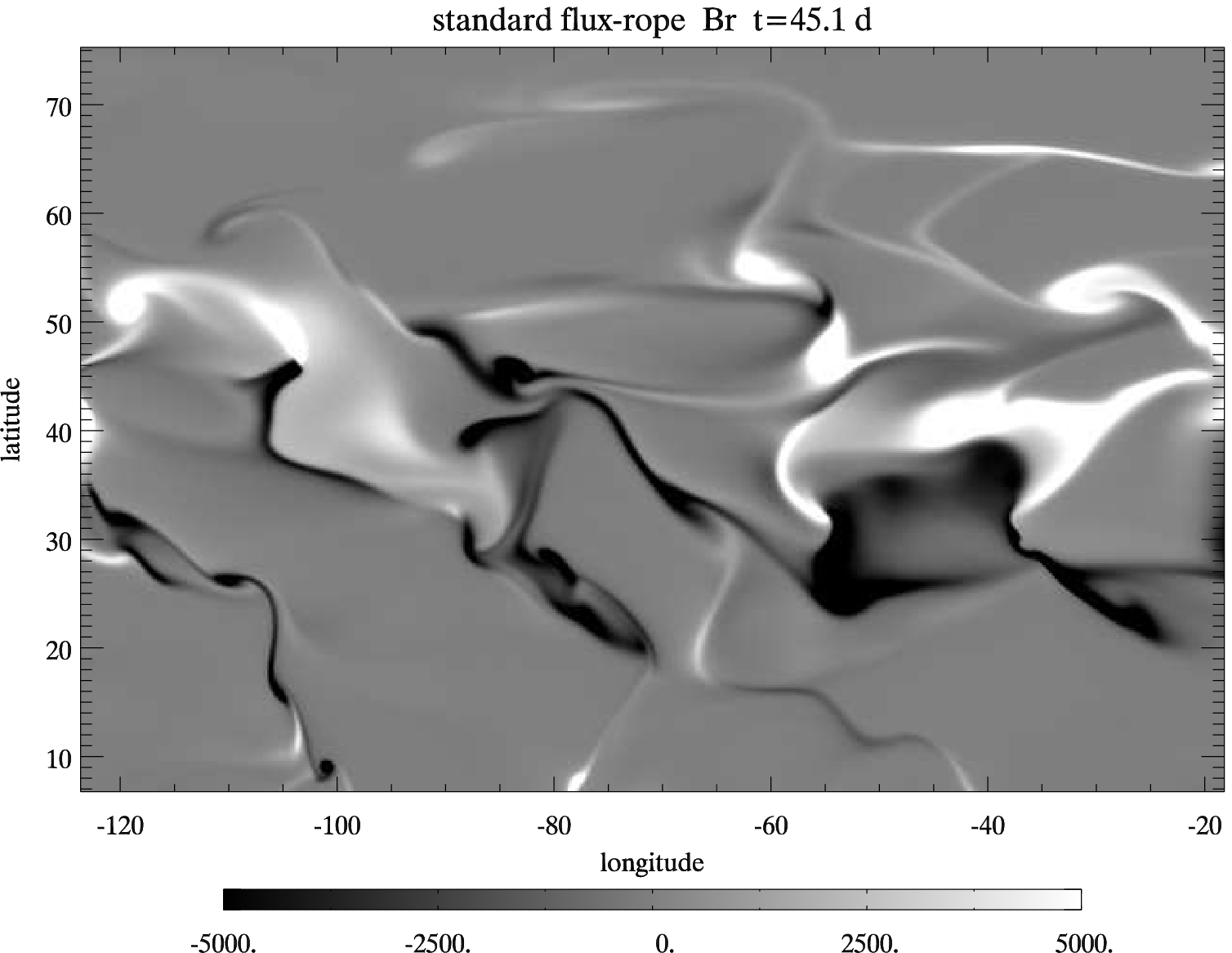} \\

  \vspace{0.02\picwd}

  \includegraphics[width=0.5\picwd]{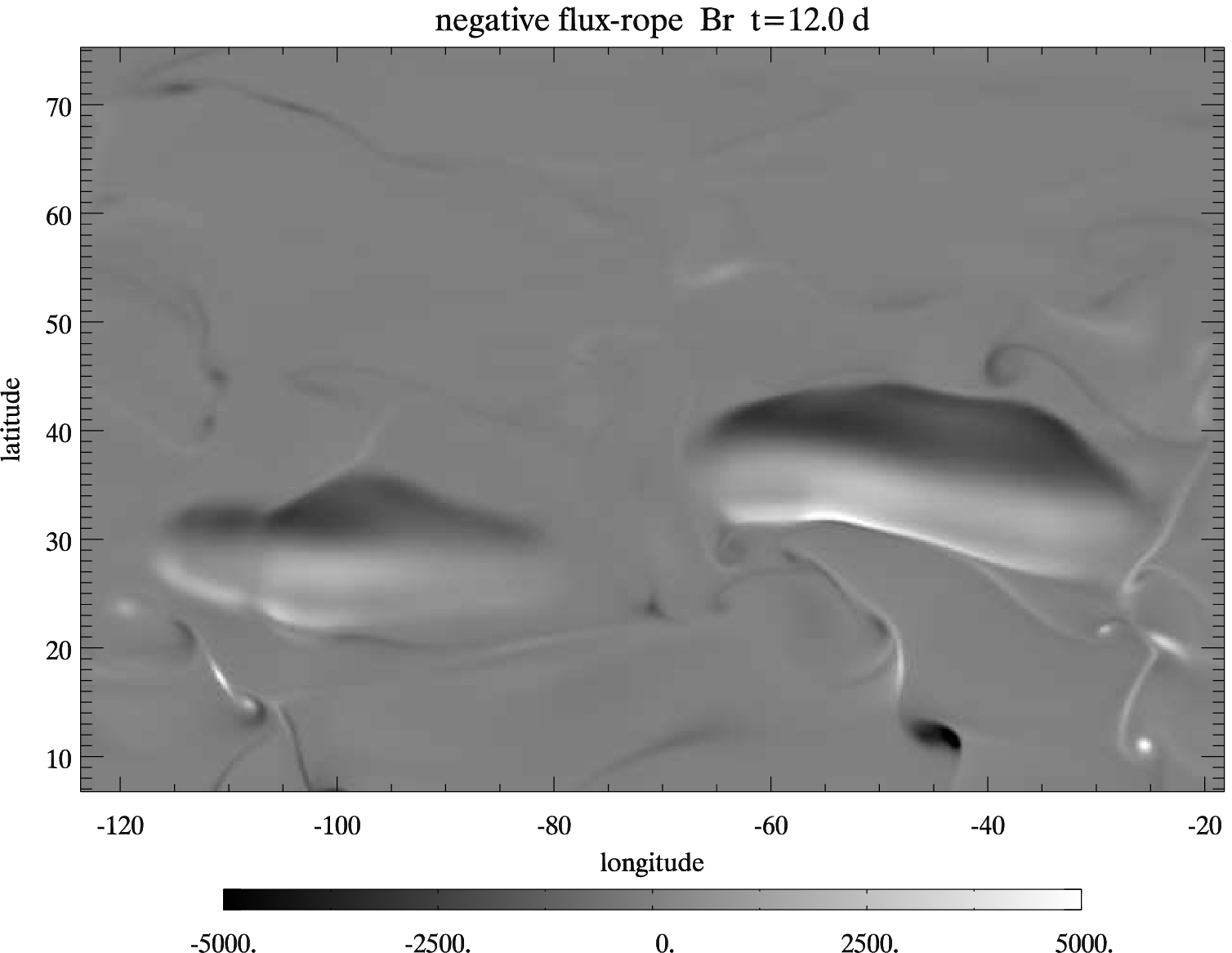}
  \includegraphics[width=0.5\picwd]{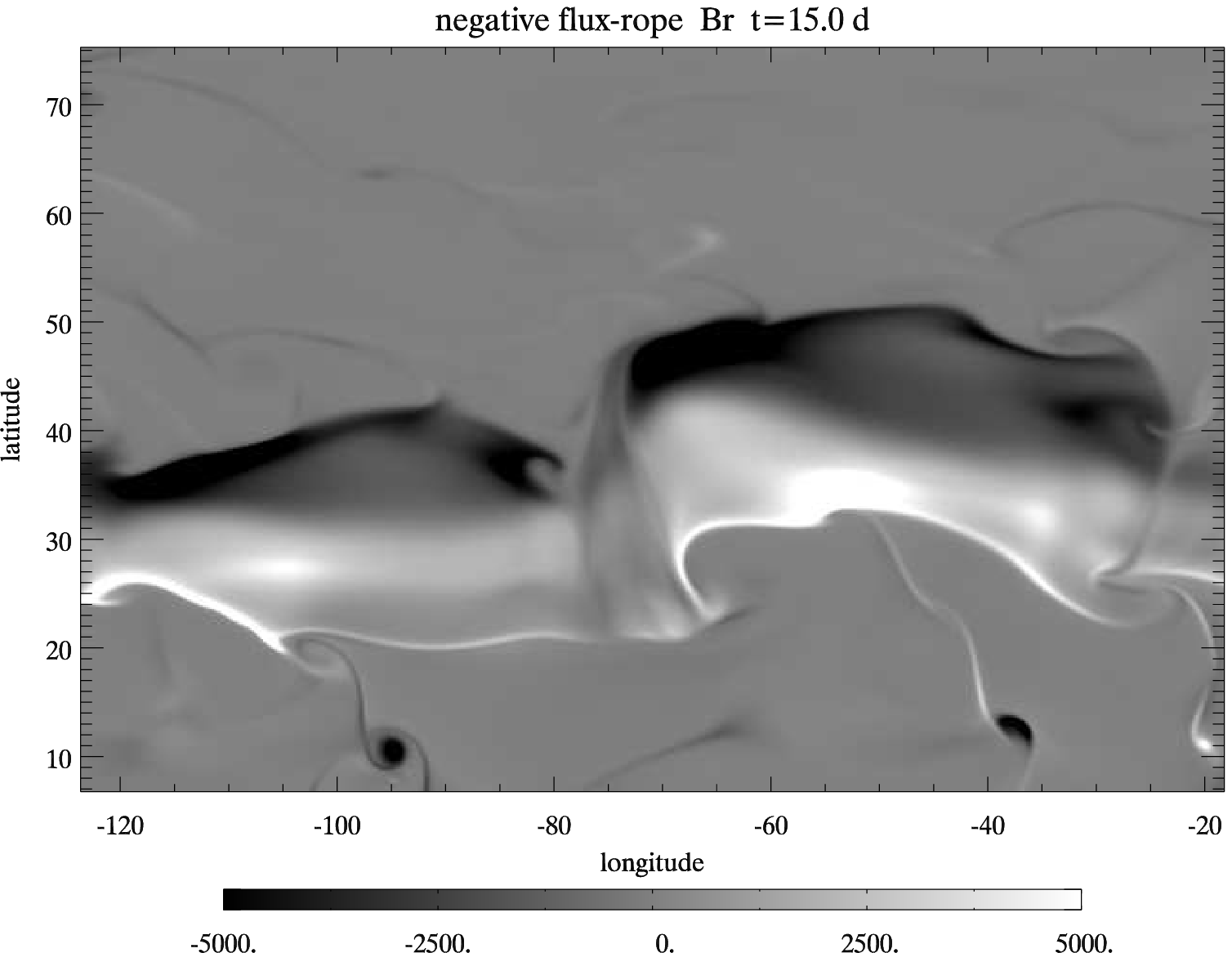}
  \includegraphics[width=0.5\picwd]{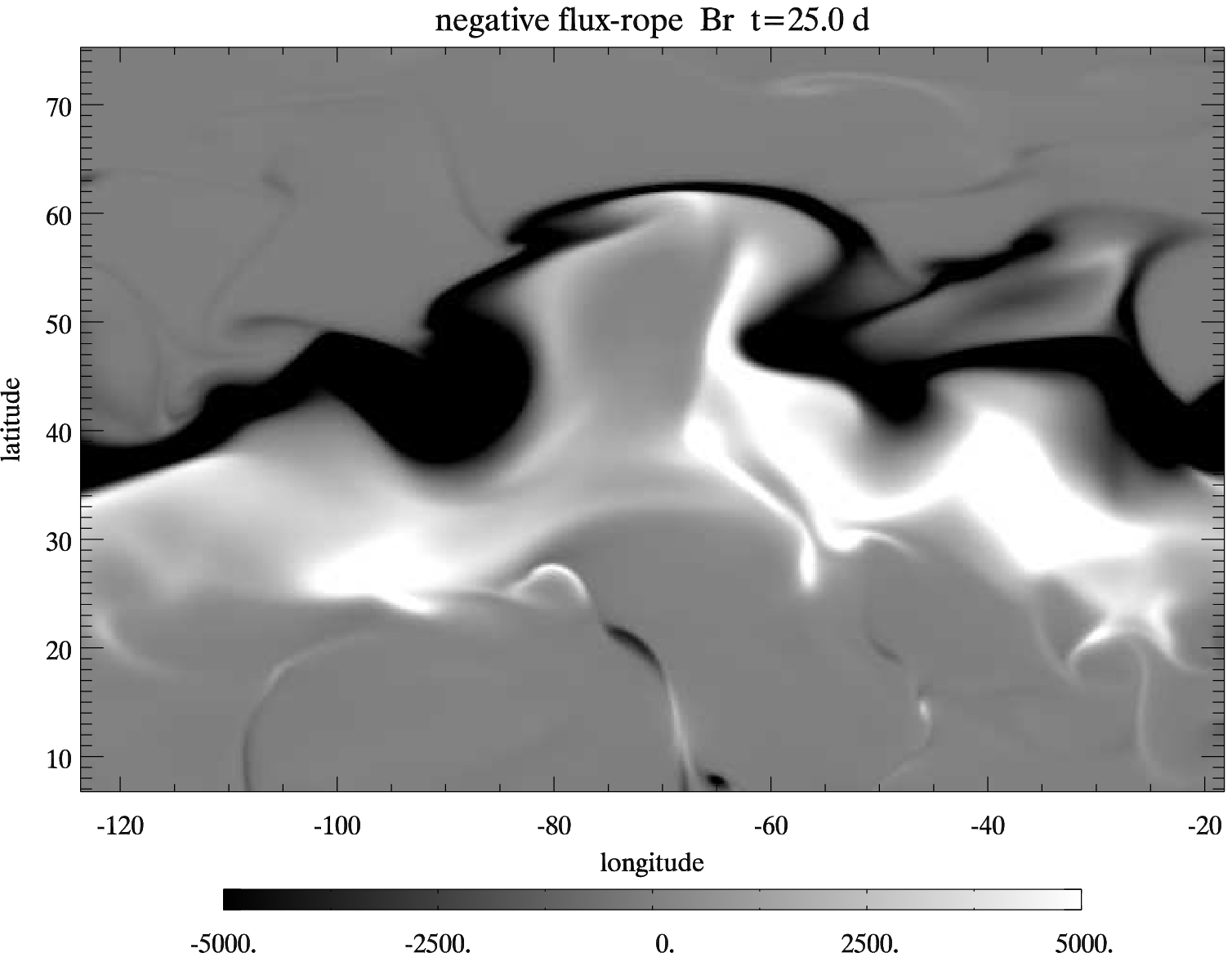}
  \includegraphics[width=0.5\picwd]{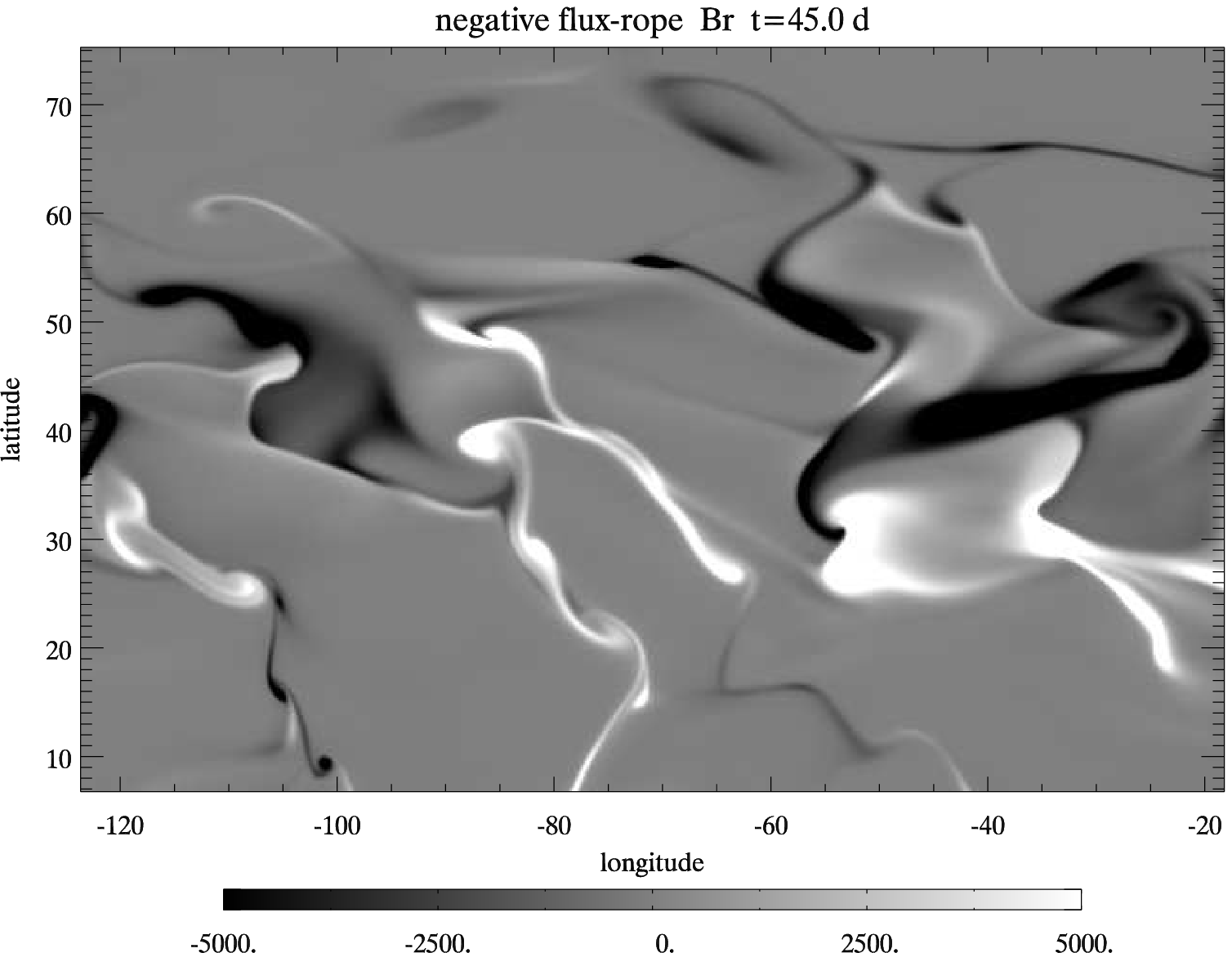} \\

  \vspace{0.02\picwd}

  \includegraphics[width=0.5\picwd]{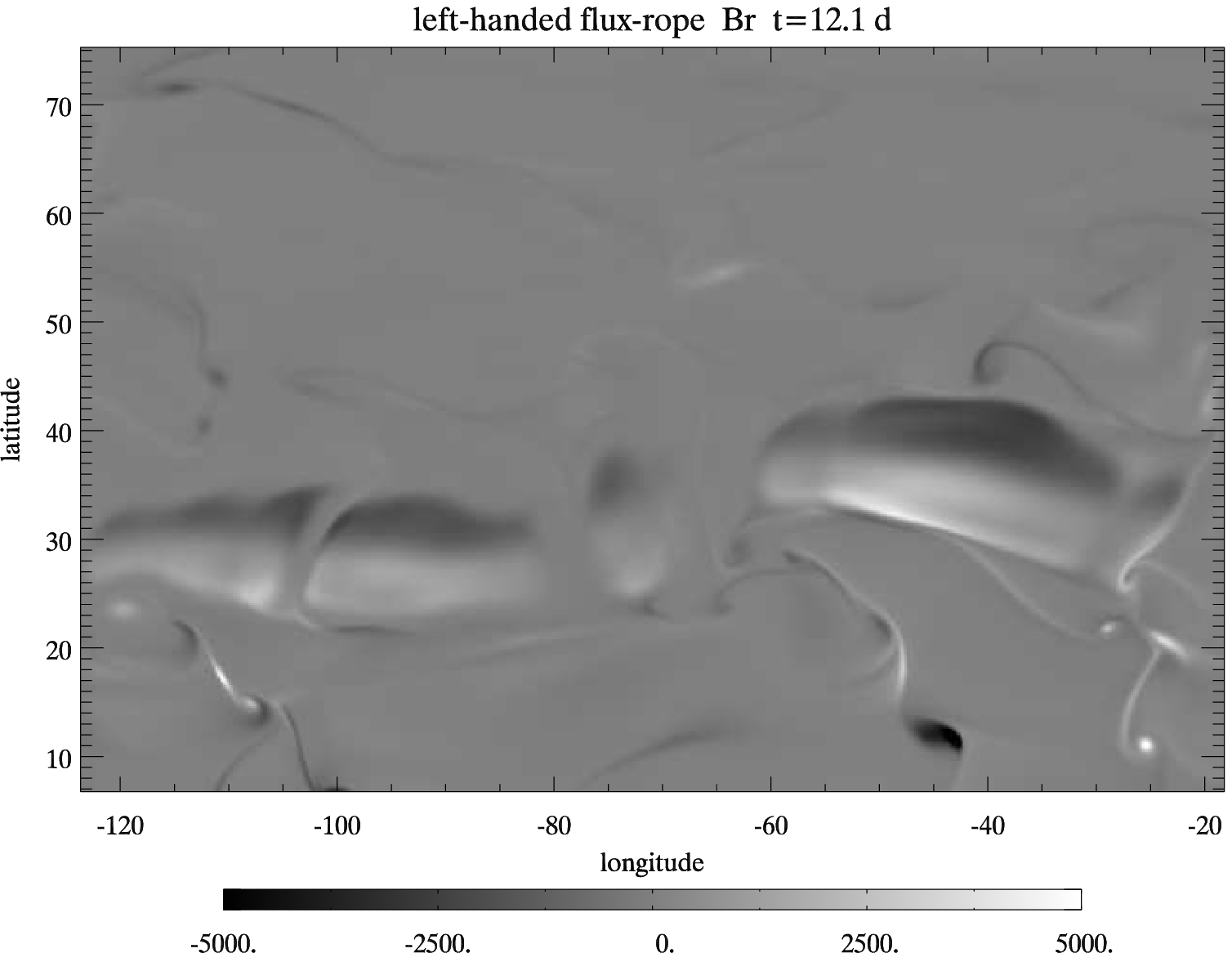}
  \includegraphics[width=0.5\picwd]{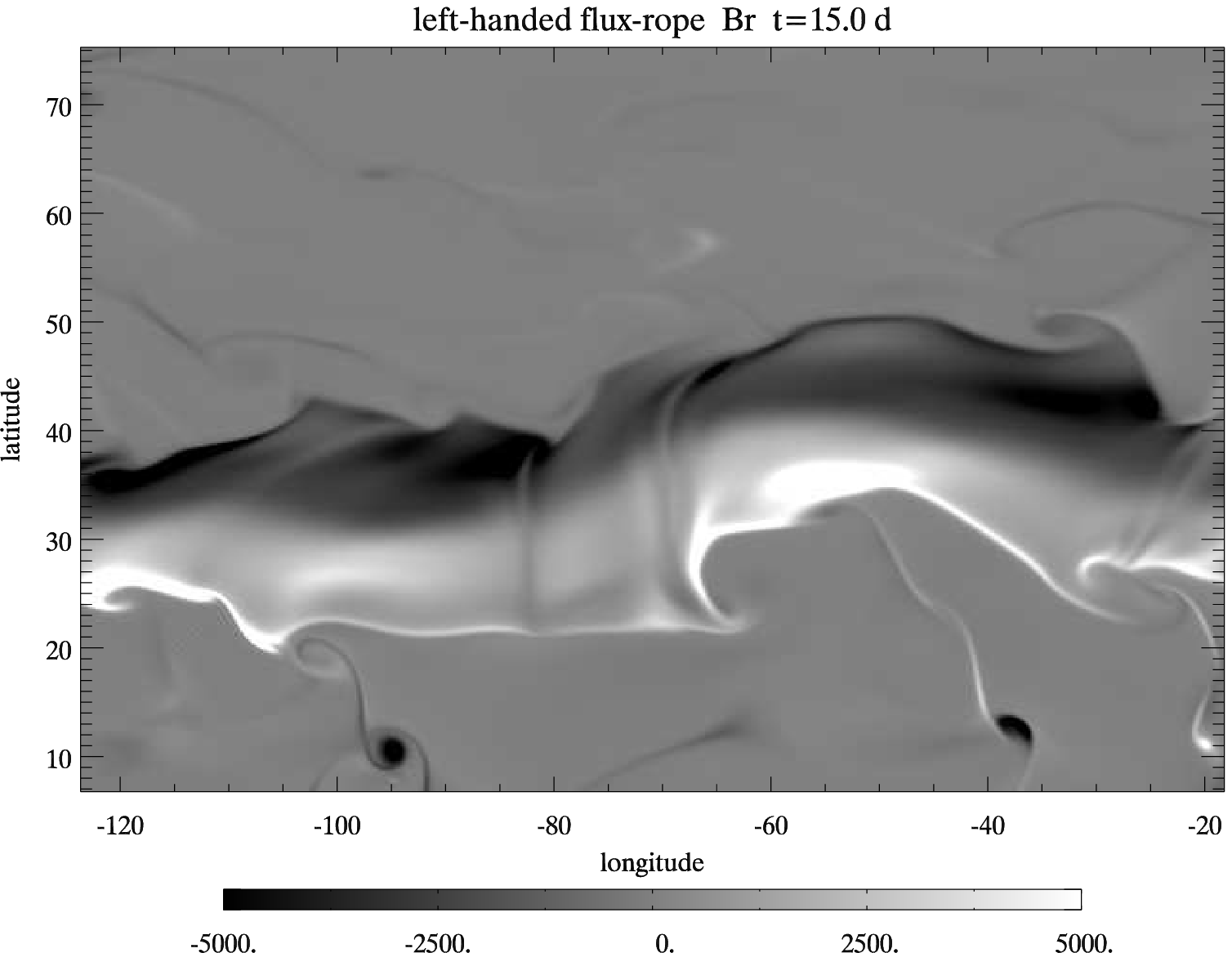}
  \includegraphics[width=0.5\picwd]{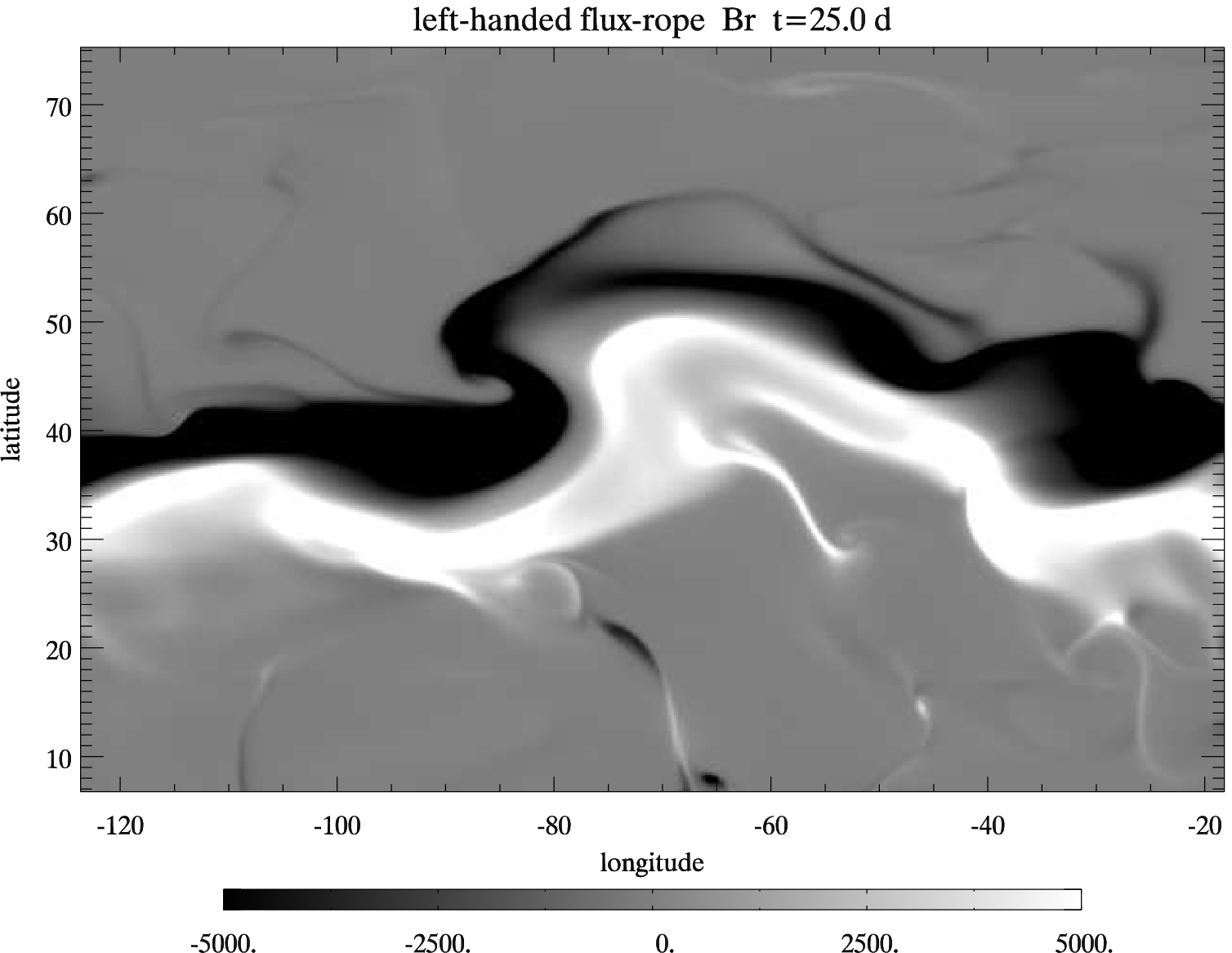}
  \includegraphics[width=0.5\picwd]{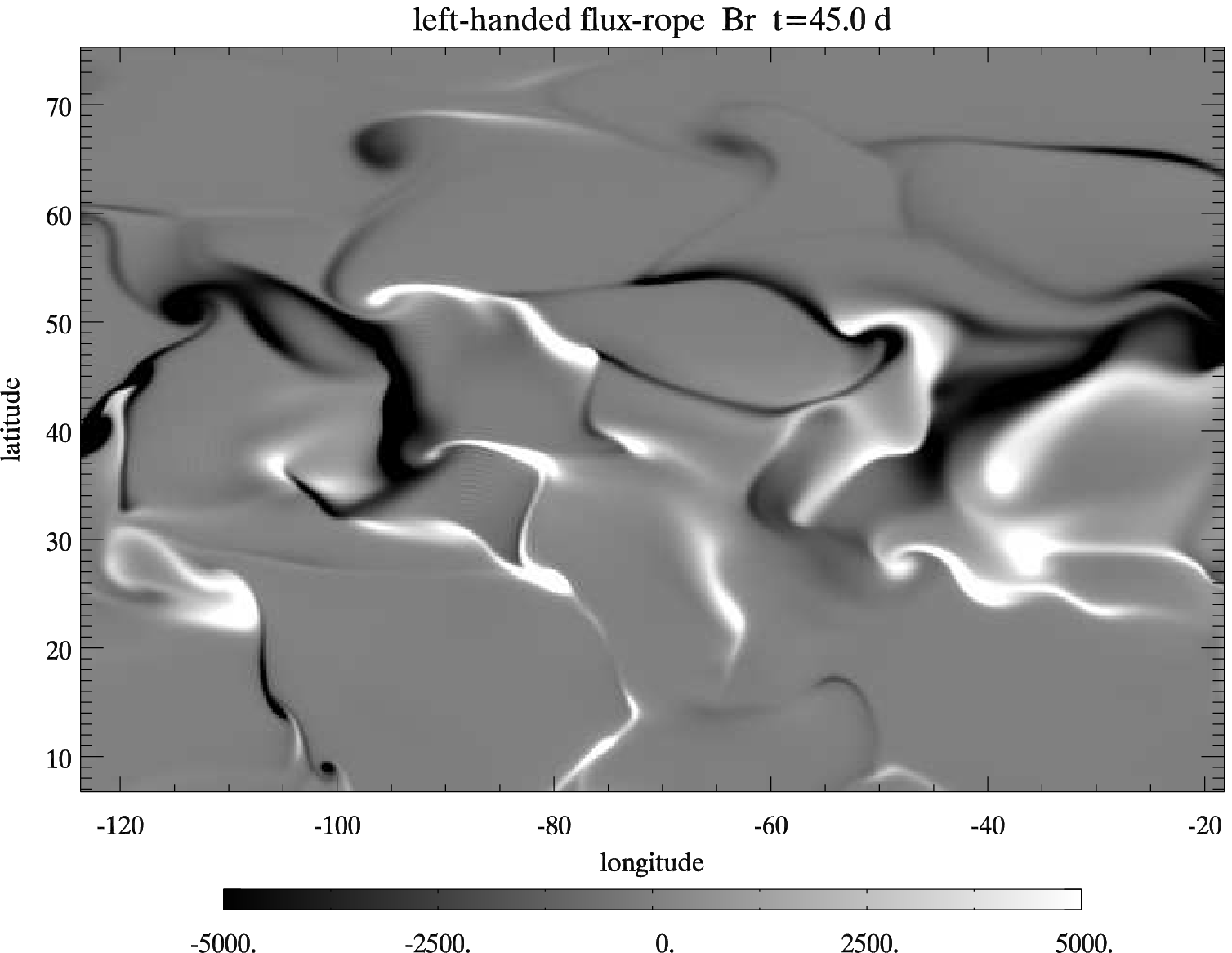} \\
  \caption{
    Close-up view of the surface radial magnetic field $B_r$ during the flux-emergence episode at different instants (from left to right, $t=12,15,25,45\un{d}$). 
    The first row represents the standard case, the 2\textsuperscript{nd} row the negative polarity case, and the 3\textsuperscript{nd} row the left-handed case.
    The colour-scale is saturated at $\pm 5\un{kG}$
  }
  \label{fig:magnetogram}

  \vspace{0.15\picwd}

  \includegraphics[width=0.5\picwd]{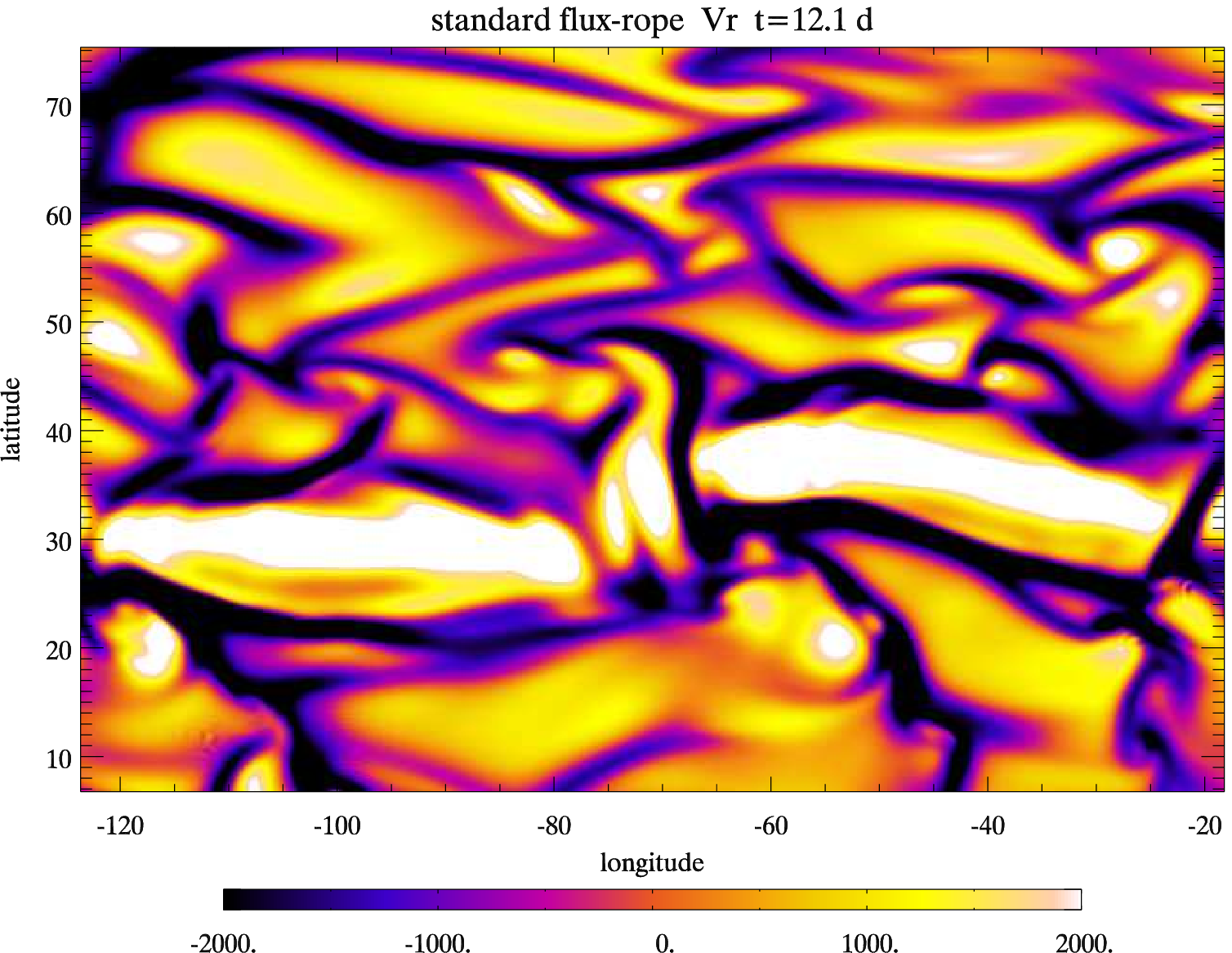}
  \includegraphics[width=0.5\picwd]{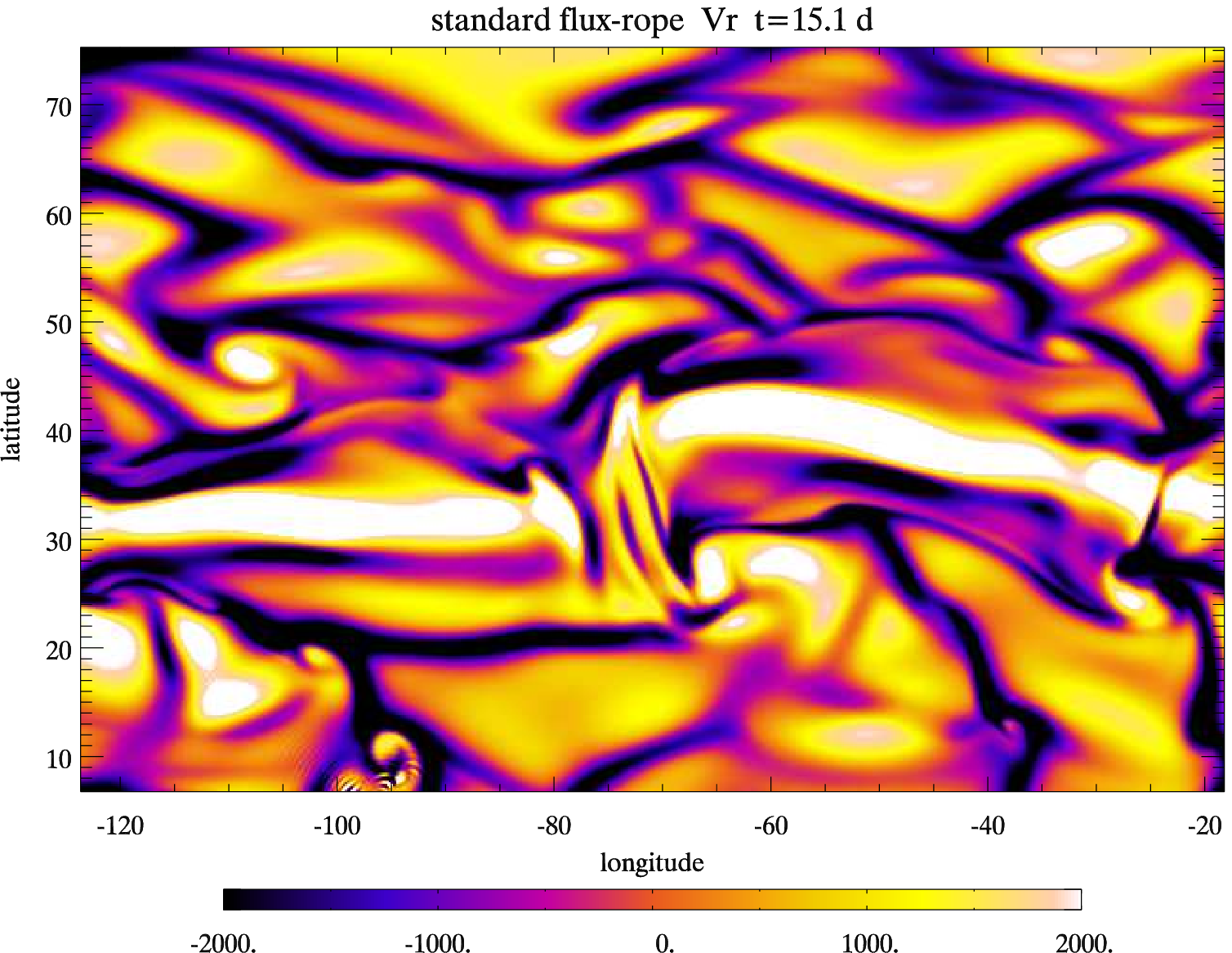}
  \includegraphics[width=0.5\picwd]{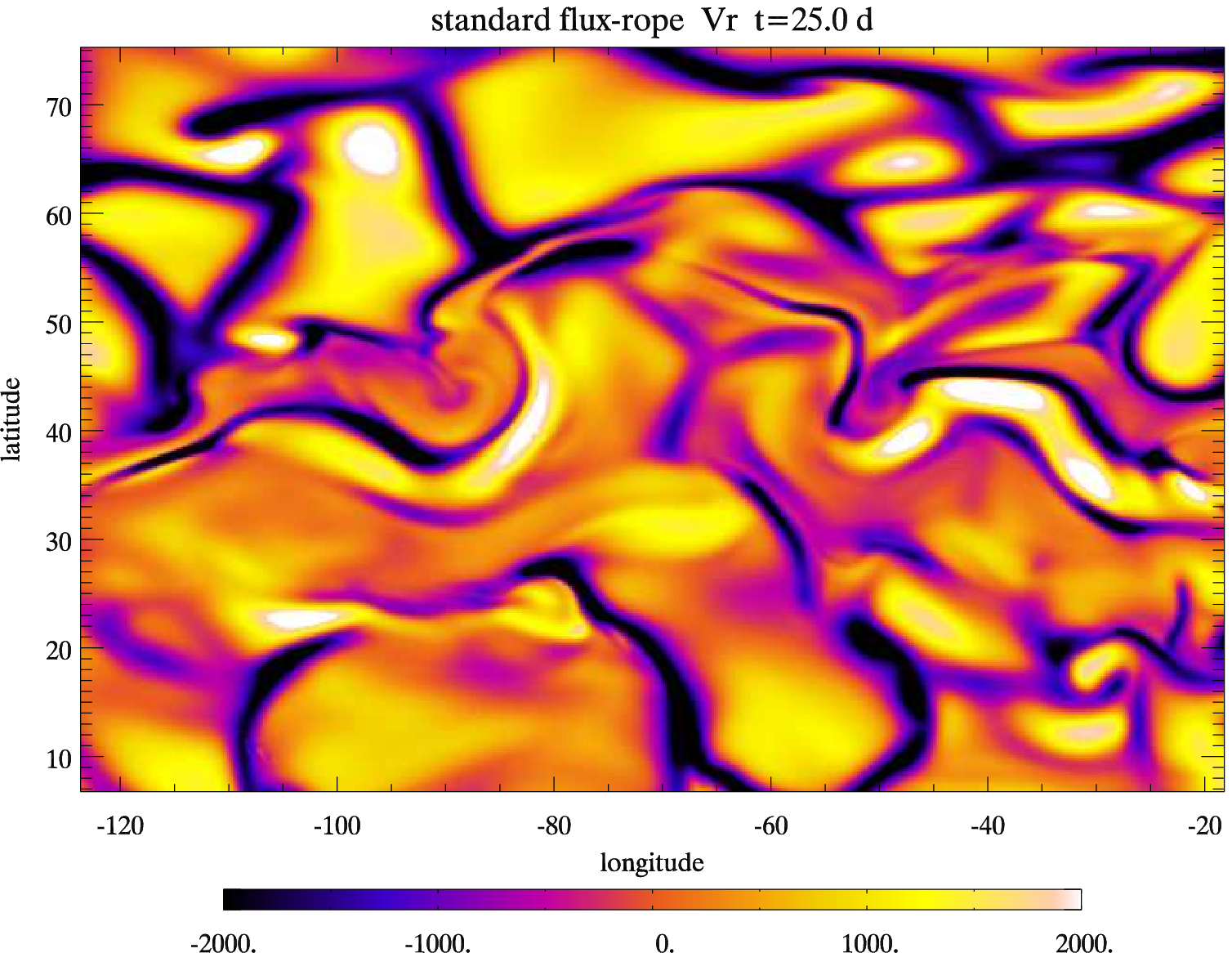}
  \includegraphics[width=0.5\picwd]{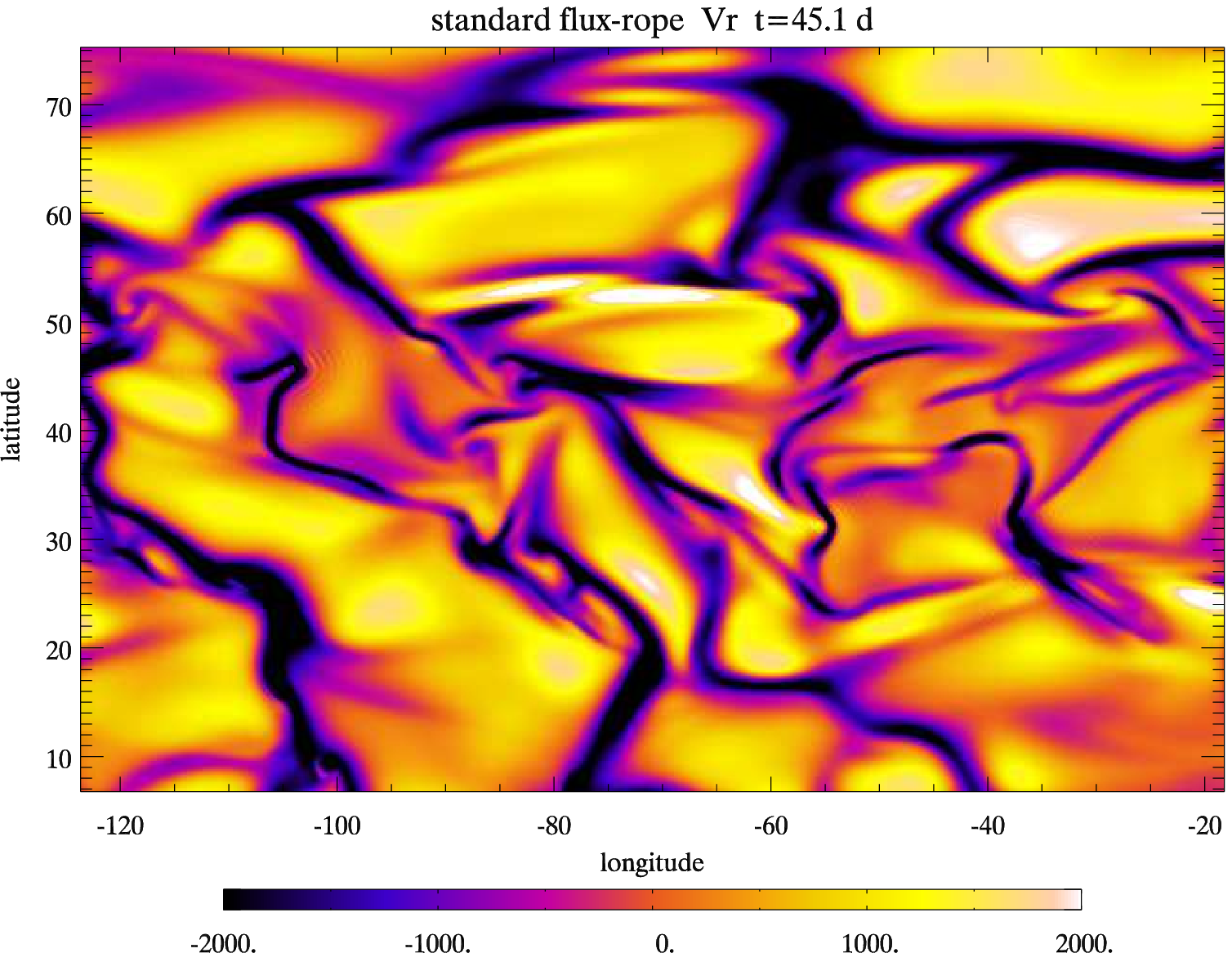} \\

  \vspace{0.02\picwd}

  \includegraphics[width=0.5\picwd]{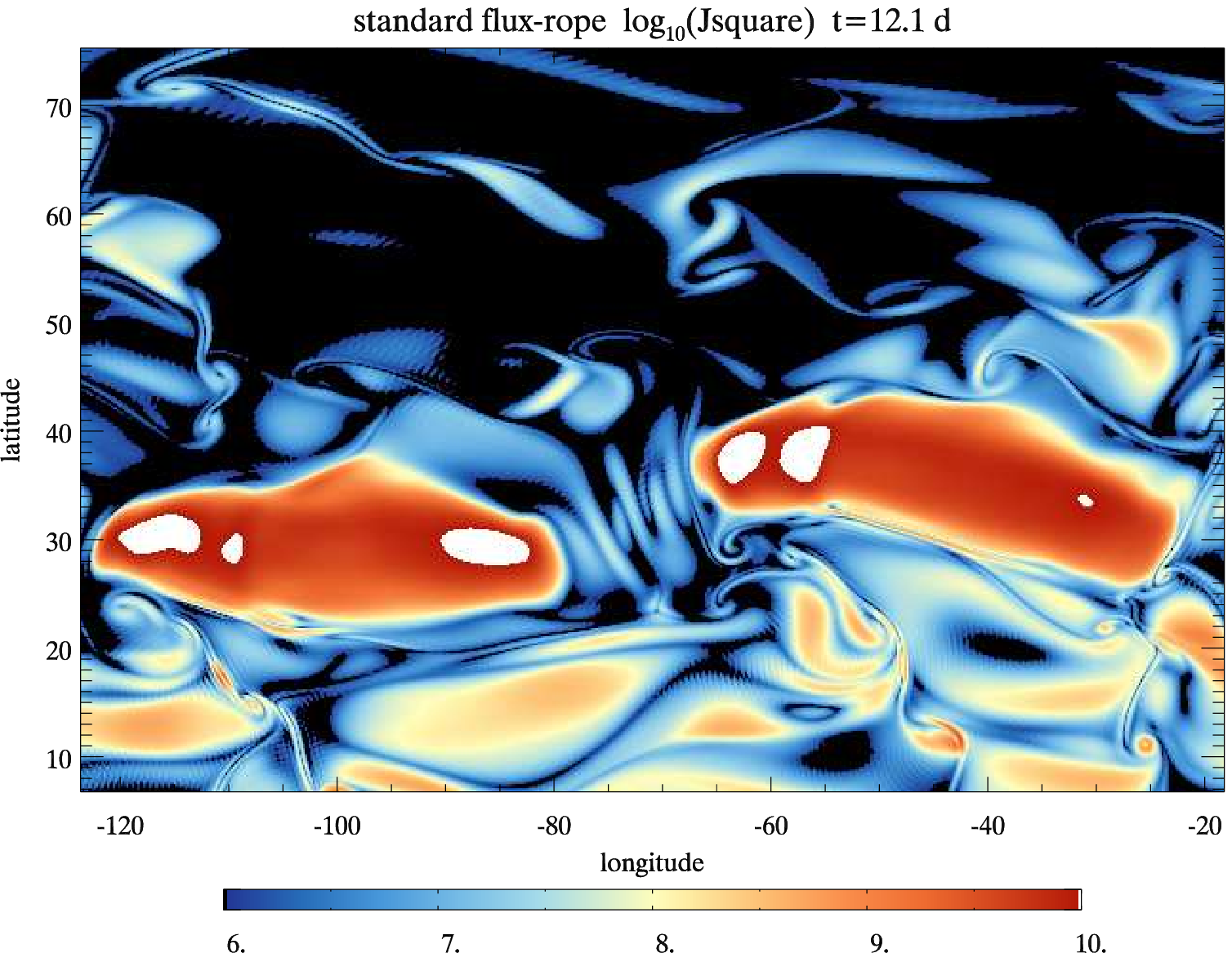}
  \includegraphics[width=0.5\picwd]{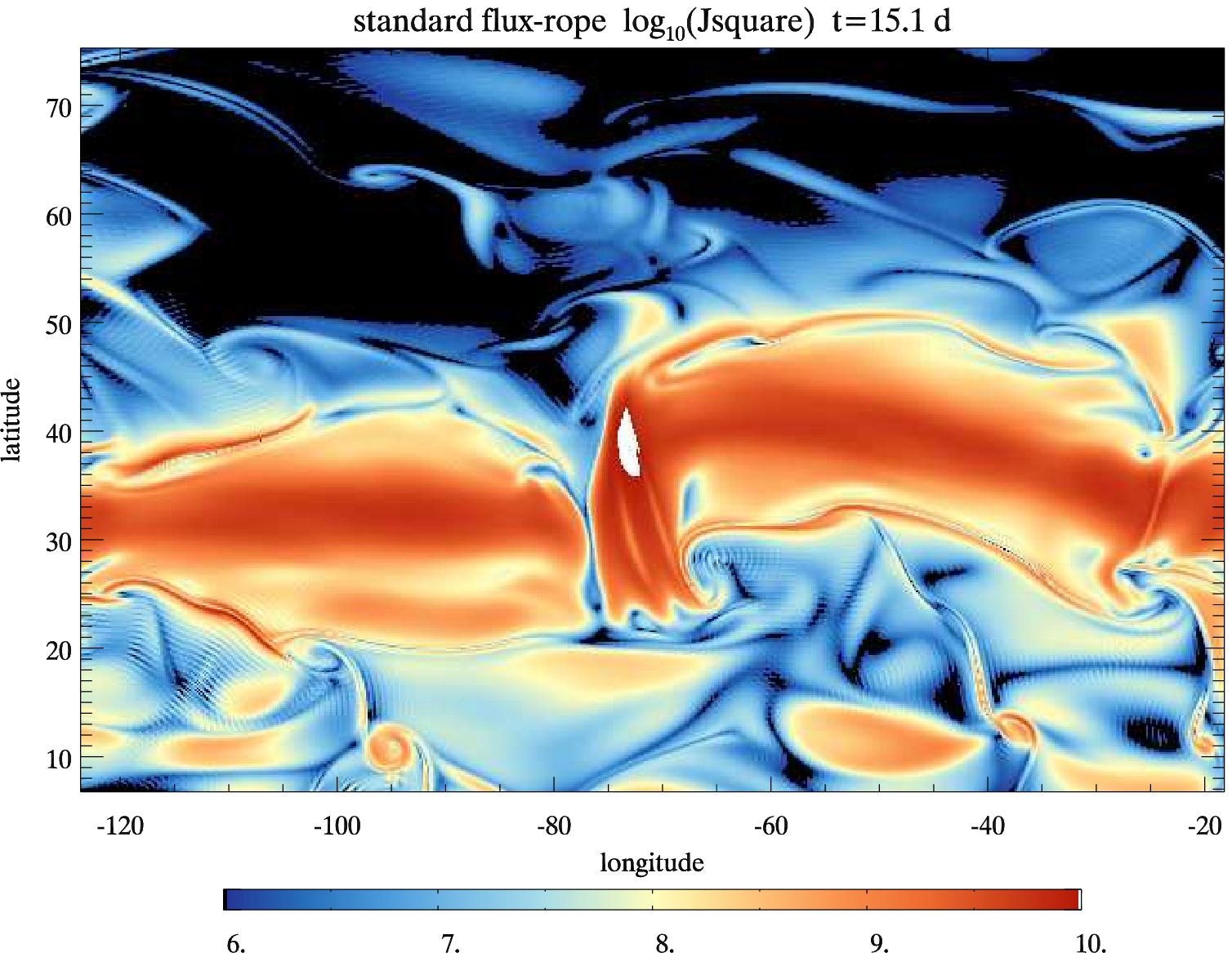}
  \includegraphics[width=0.5\picwd]{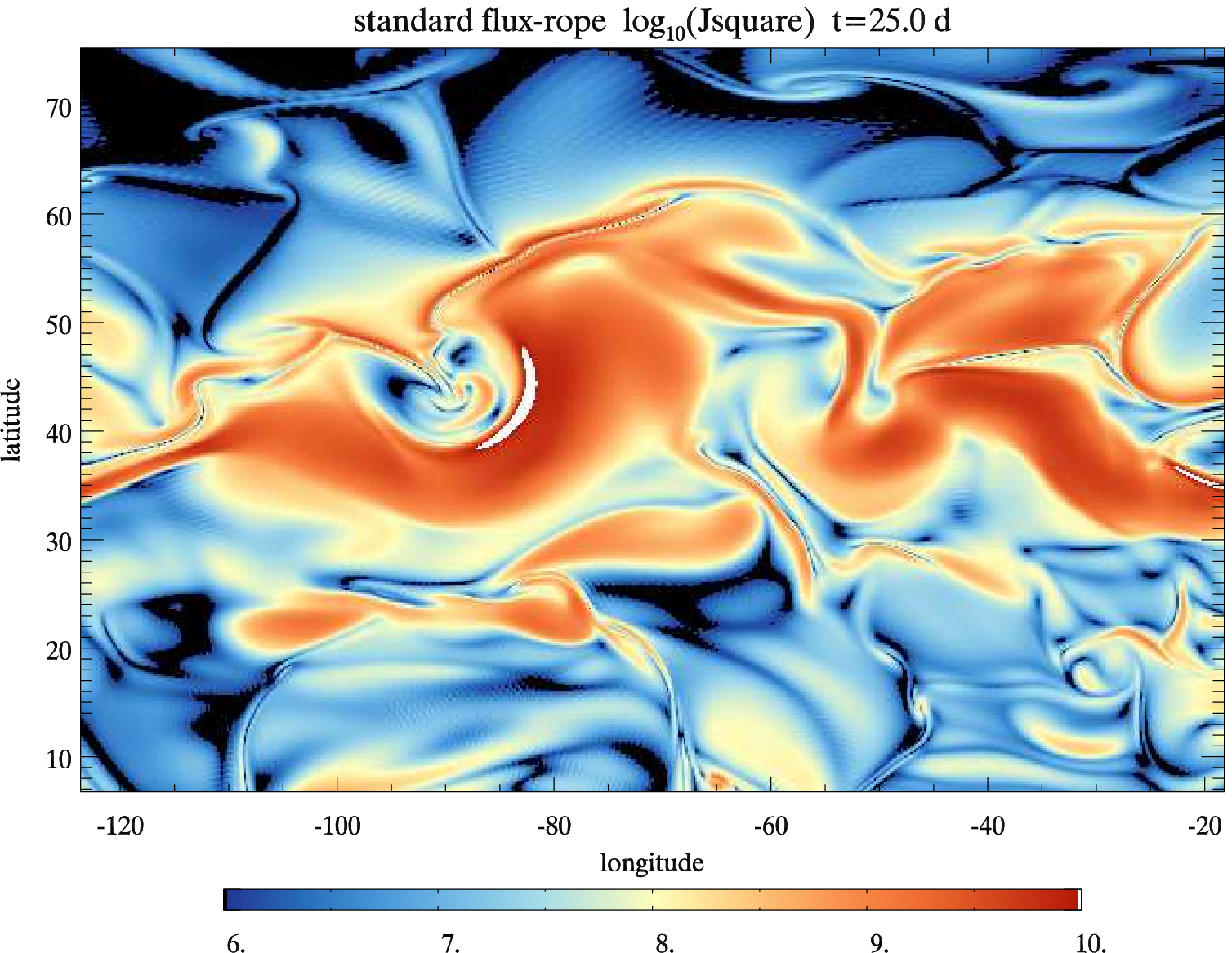}
  \includegraphics[width=0.5\picwd]{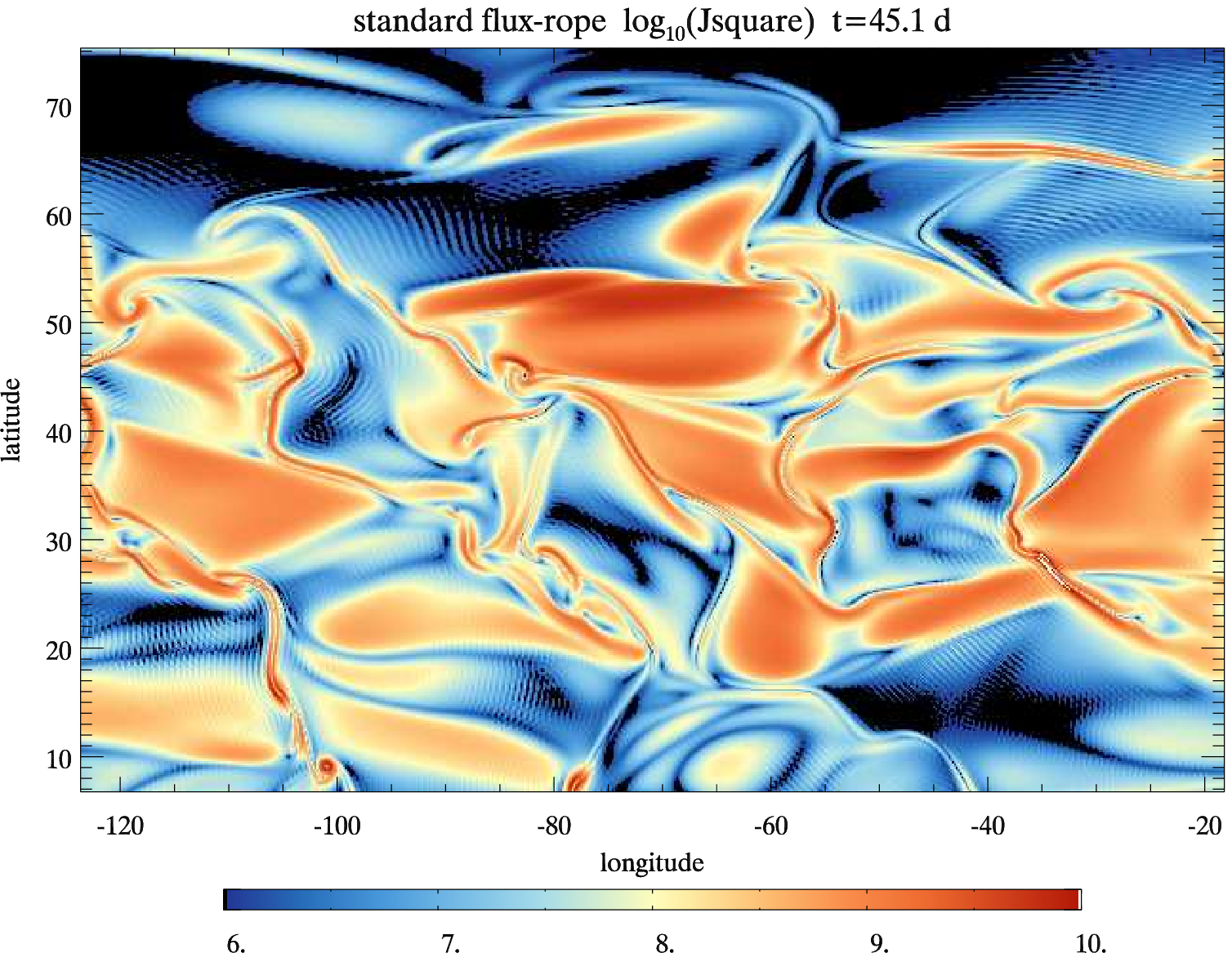} \\

  \vspace{0.02\picwd}

  \includegraphics[width=0.5\picwd]{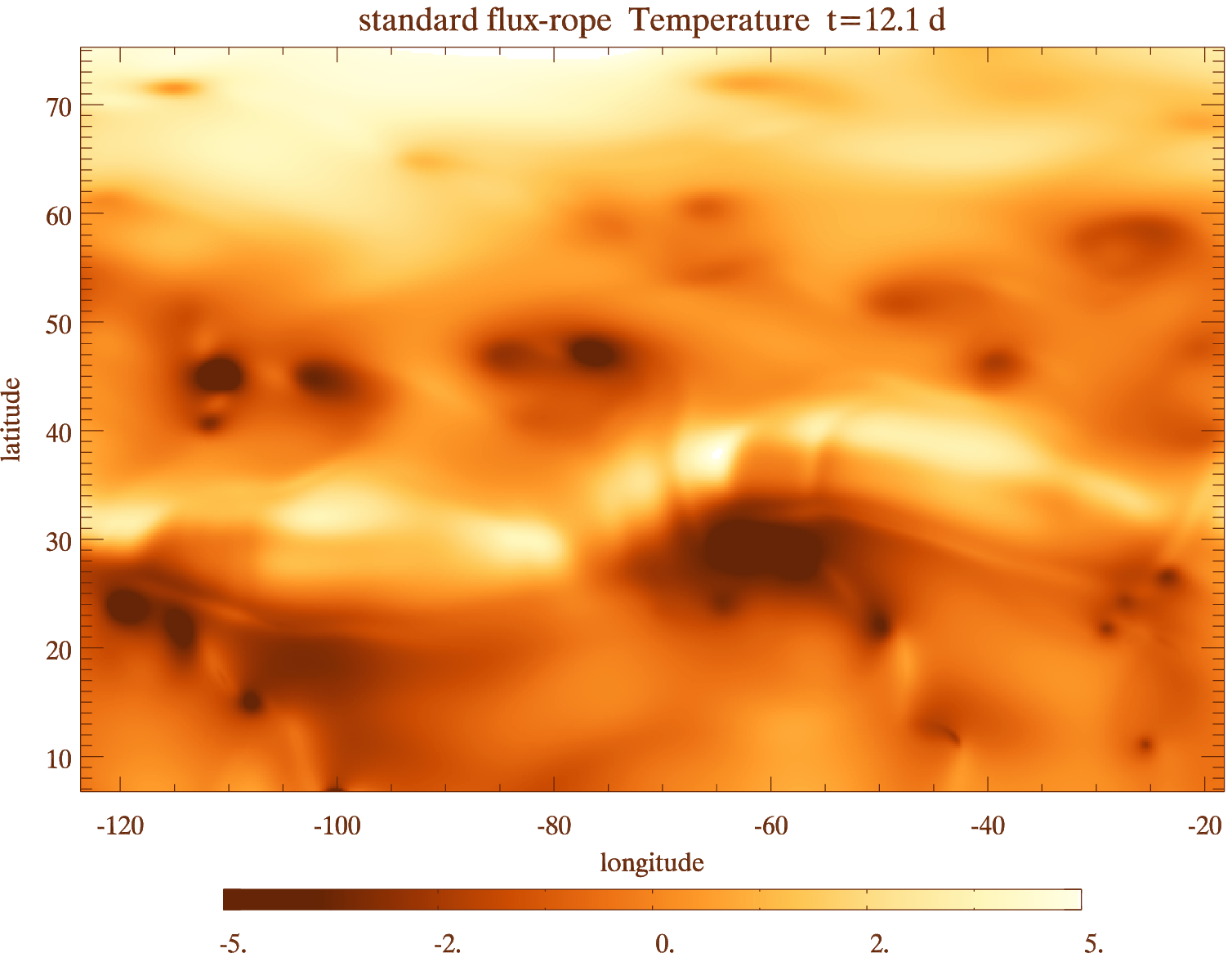}
  \includegraphics[width=0.5\picwd]{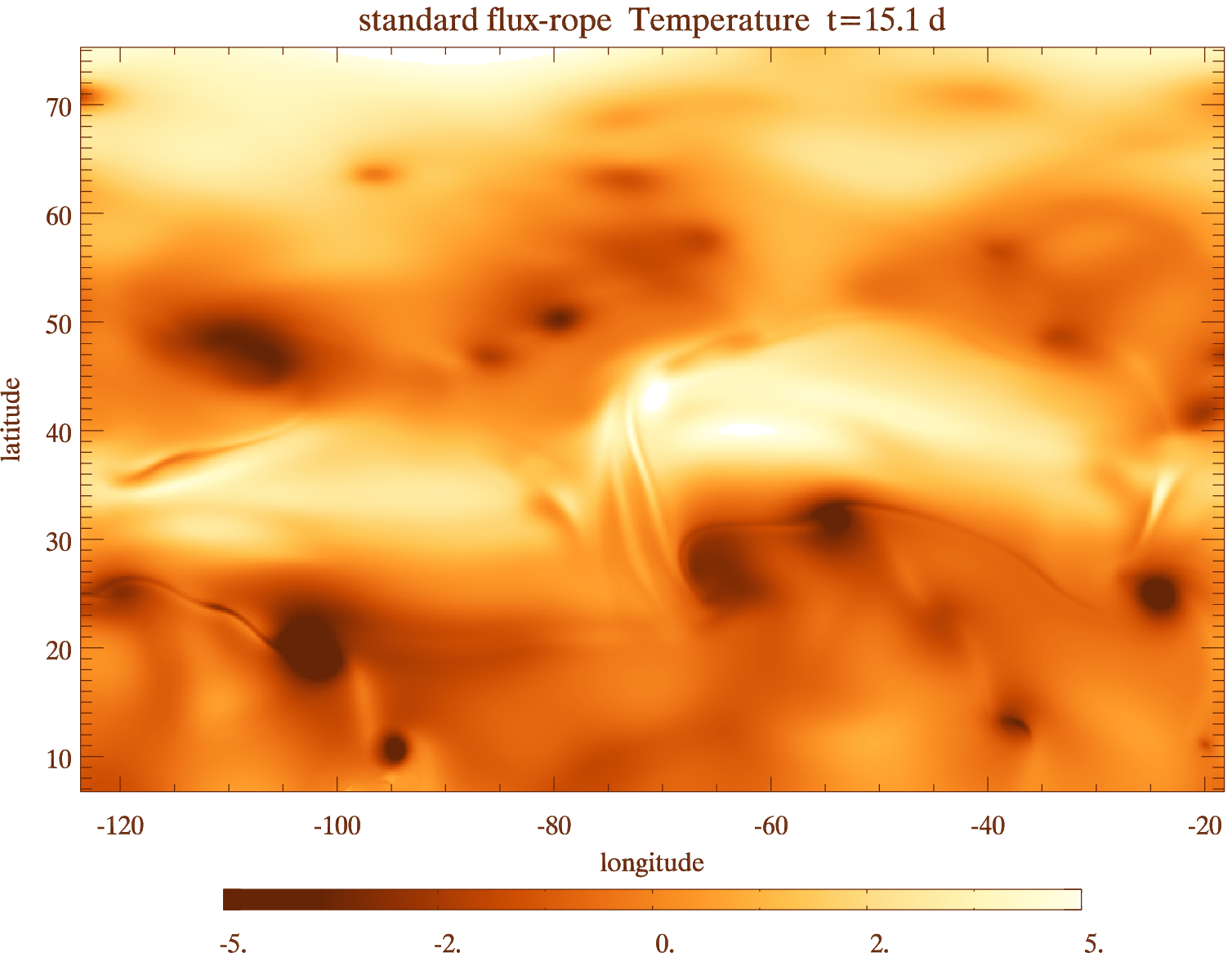}
  \includegraphics[width=0.5\picwd]{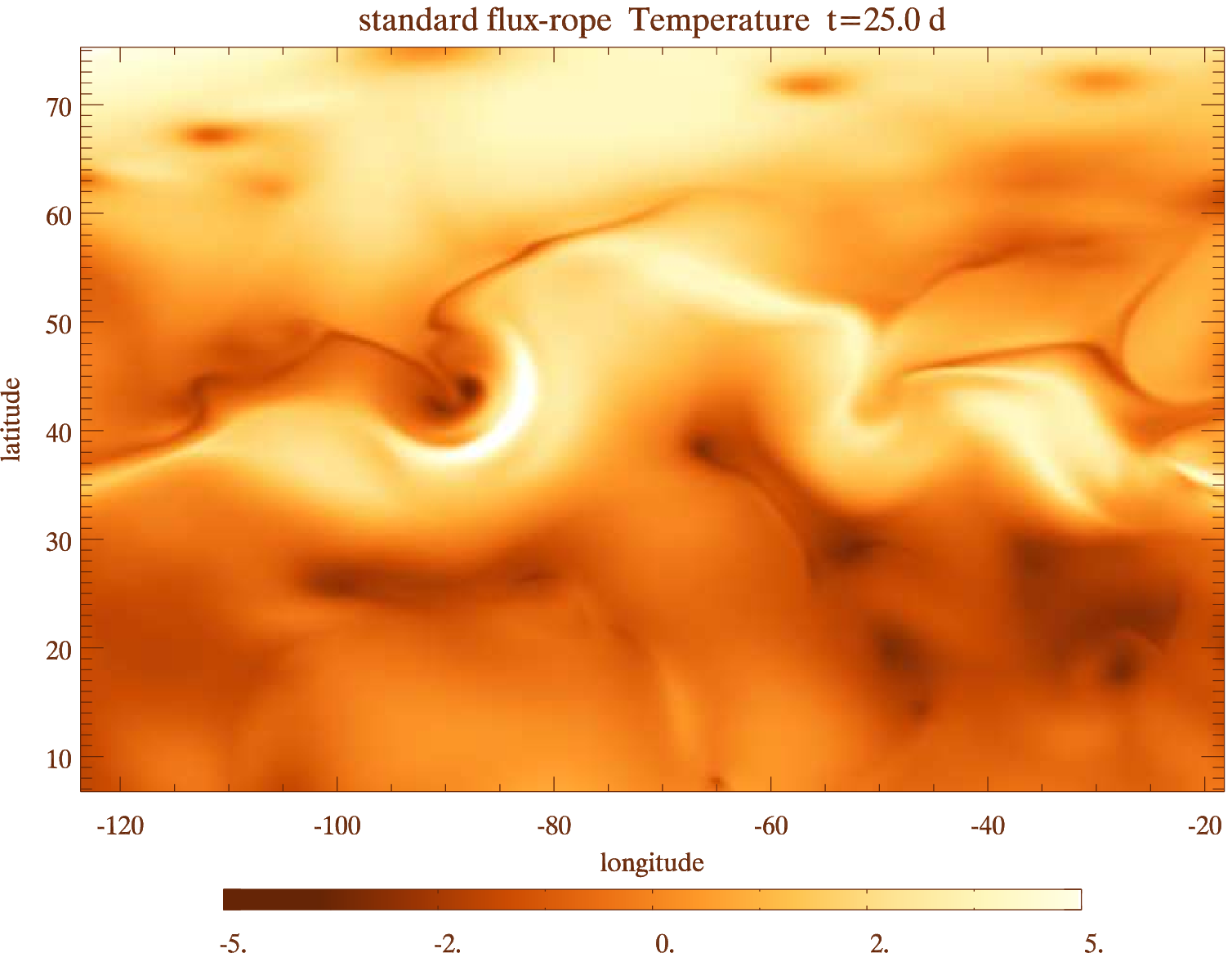}
  \includegraphics[width=0.5\picwd]{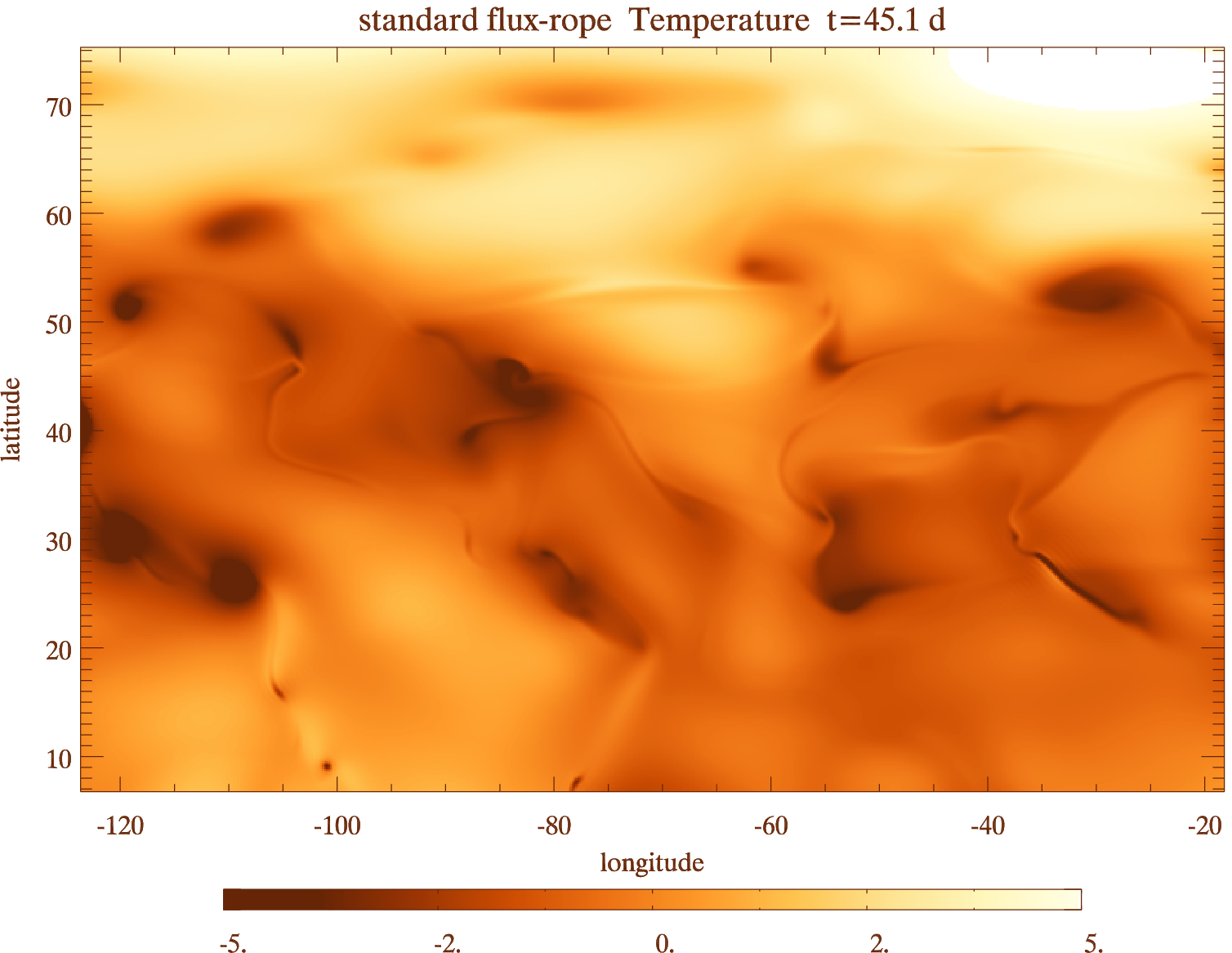} \\

  \caption{
    Close-up view of the radial velocity $v_r$ (1\textsuperscript{st} row), the square of the current density $\log\left(J^2\right)$ (2\textsuperscript{nd} row, in a log-scale plot to make its spatial distribution more visible), and the temperature fluctuations $T$ (3\textsuperscript{rd} row) for the standard case.
    The sub-domain and instants represented are the same as in Fig. \ref{fig:magnetogram}.
  }
  \label{fig:velogram}

\end{figure*}

We now turn our attention to understanding how the surface magnetic field and flows react to the emergence of a twisted magnetic flux-rope.
Figure \ref{fig:magnetogram} shows three time-series of synthetic magnetograms (that is, surface maps of $B_r$) for the standard case (top row), the case with a flux-rope with negative $B_0$ (second row) and the case with a left-handed flux-rope (third row).
The instants represented are, from left to right, $t=12,15,25$ and $45\un{d}$.
The panels only show a sub-domain of the magnetic active lane generated by the emergence of the flux-ropes.

The flux-ropes were initially placed near the bottom of the CZ at $30^{\circ}\un{N}$ and then rose almost radially, deflecting northwards by a small amount.
The activity lane then first appears at the surface between latitudes $30^{\circ}\un{N}$ and $40^{\circ}\un{N}$.
The slow poleward drift continues after the flux-rope reaches the surface and stops rising buoyantly.
The flux-rope is then advected by the sub-surface poleward meridional circulation (see Sect. \ref{sec:ropeevol}, and Figures \ref{fig:fluxrope_snapshots} and \ref{fig:fluxrope_massflux}, in particular the CCW circulation crossing the flux-rope in the last panel).
About $10-15\un{days}$ after the beginning of the emergence episode, the active lane occupies the latitudinal interval $40^{\circ}\un{N}$ to $50^{\circ}\un{N}$.
Besides the poleward drift, the active lane also broadens in latitude as the rope's emerged magnetic flux is sheared and advected by the convective flows, and undergoes resistive diffusion.

The flux-emergence episode starts with a series of discontinuous patches of magnetic field appearing at the surface ($t=12\un{d}$, first column in Fig. \ref{fig:magnetogram}), even though the flux-ropes are initially perfectly axi-symmetric.
This is due to the influence of the convective motions encountered by the flux-ropes during their buoyant rise.
The fractions of the flux-rope which cross strong downflow plumes are delayed with respect to the rest, while those encountering rising convective blobs are pushed forward.
The maximum delay between the first and the last emerging parts is of about $5$ days.
The distortion will be more accentuated for flux-ropes with weaker magnetic field strength.
The least buoyant ropes in our set of simulations do not fully emerge, having some sections which do not manage to overcome the local convective downflows they encounter.
%
A few days after the beginning of the flux-emergence episode the bulk of the flux-rope has stopped its buoyant rise and a larger fraction of its magnetic flux has reached or crossed the surface, making the azimuthal non-uniformities stated above fade away (second column in Fig. \ref{fig:magnetogram}, $t=15\un{d}$).
The orientation of the magnetic polarity is mainly North-South at this moment.
In other words, the polarity inversion line (PIL) is essentially aligned with the $\phi$ direction, and there is a well defined and uniform tilt angle of $\pm 90^{\circ}$ (the sign depending on the flux-rope's polarity and/or handedness).
%
The surface magnetic flux continues growing until $t=23\un{d}$ (\emph{cf.} bottom left panel in Fig. \ref{fig:precursors}).
During this last phase of the flux-rope emergence, the magnetograms seem to indicate that the emergence regions are rotating.
This effect is particularly noticeable in the northern and southern edges of the emerged structure (see online movie).
Further, the sense of rotation depends on the handedness of the flux-rope; right-handed flux-ropes exhibit a CCW rotation while left-handed flux-ropes turn CW.
The rotation patterns do not match the surface flows velocities; the observed rotation patterns are only apparent and are a signature of the flux-rope's own magnetic field as it crosses the surface.
For this reason, the flux-rope's handedness determines the sense of the apparent rotation of the emerging magnetic field.

As the simulation proceeds, the surface convection gradually overcomes the perturbation introduced by the flux-rope emergence.
The emerged upper fraction of the flux-rope is gradually twisted by CCW-rotating vortexes.
The PIL gets sheared accordingly, but it keeps a global continuous structure for a long period of time (third column in Fig. \ref{fig:magnetogram}, $t=25\un{d}$).
At about $t=30-40\un{d}$ all traces of the emerged flux-rope start to disappear.
At $t=45\un{d}$, all the flux-rope's emerged magnetic flux has been pushed into the network, between the convection cells (fourth column in Fig. \ref{fig:magnetogram}).
As discussed in Sect. \ref{sec:energy_balance}, the flux-rope's remaining magnetic field (\emph{i.e}, the part which has not emerged) as now been assimilated by the solar dynamo.


The emergence of a magnetic flux-rope is preceded by a local increase in radial velocity and current density, as described in Sect. \ref{sec:precursors}.
Figure \ref{fig:velogram} shows, in the first two rows, the temporal evolution of the radial velocity $v_r$ and the current density squared $J^2$ for the standard case (compare with first row 
of Fig. \ref{fig:magnetogram}).
The current density squared is displayed in a log-scale plot to make its spatial distribution more visible, as the contrast between the background current density and that of the emerging region can be very large.
The instants represented are the same as in Fig. \ref{fig:magnetogram}: from left to right, $t=12,15,25$ and $25\un{d}$. The surface sub-domain represented is also the same.


The delays between the maximum of $v_r$, $J^2$ and $\|B_r\|$ are well visible in the Figure.
The two former peak near $t=12\un{d}$ (first column) while the later only peaks at $t=23\un{d}$ (that is, close to the instant represented on the third column).
It is interesting to note how the amplitude of the $v_r$ signal fades away quickly but some traces of it remain till later on.
The current density at the surface evolves with a spatial and temporal pattern close to that of 
$\|B_r\|$, but its amplitude is at its maximum earlier in the flux-emergence episode.
The temperature fluctuations $T$ (relative to $\bar{T}$) for the standard case are also represented in the third row.
Hot spots (or bands) appear all along the emerging region.
Its spatial distribution correlates well with $v_r$; the hotter zones correspond to the strongest upflows, while the cold spots correspond to the top of the convective downflow plumes which form at the intersection of the convective cell boundaries.
The contribution of the current density (or rather the ohmic diffusion $\eta J^2$), does not suffice to explain the temperature variations found.


The case with a flux-rope placed at a higher latitude shows  behaviour very similar to that of the standard case.
In the low-latitude case, on the other hand, the emerged field organises into a much more discontinuous pattern, and the emergence episode lasts longer (almost $20$ days).
This is a consequence of the flux-rope's slower buoyant rise, and of it being more distorted by the convective flows during that period.
The total emerged magnetic flux is, nevertheless, equally high (mostly due to the larger circumference of a low-latitude flux-rope).

\subsection{Induced zonal flows and surface shear}
\label{sec:zonalflows}

\begin{figure}[]
  \centering
  \includegraphics[width=0.95\picwd]{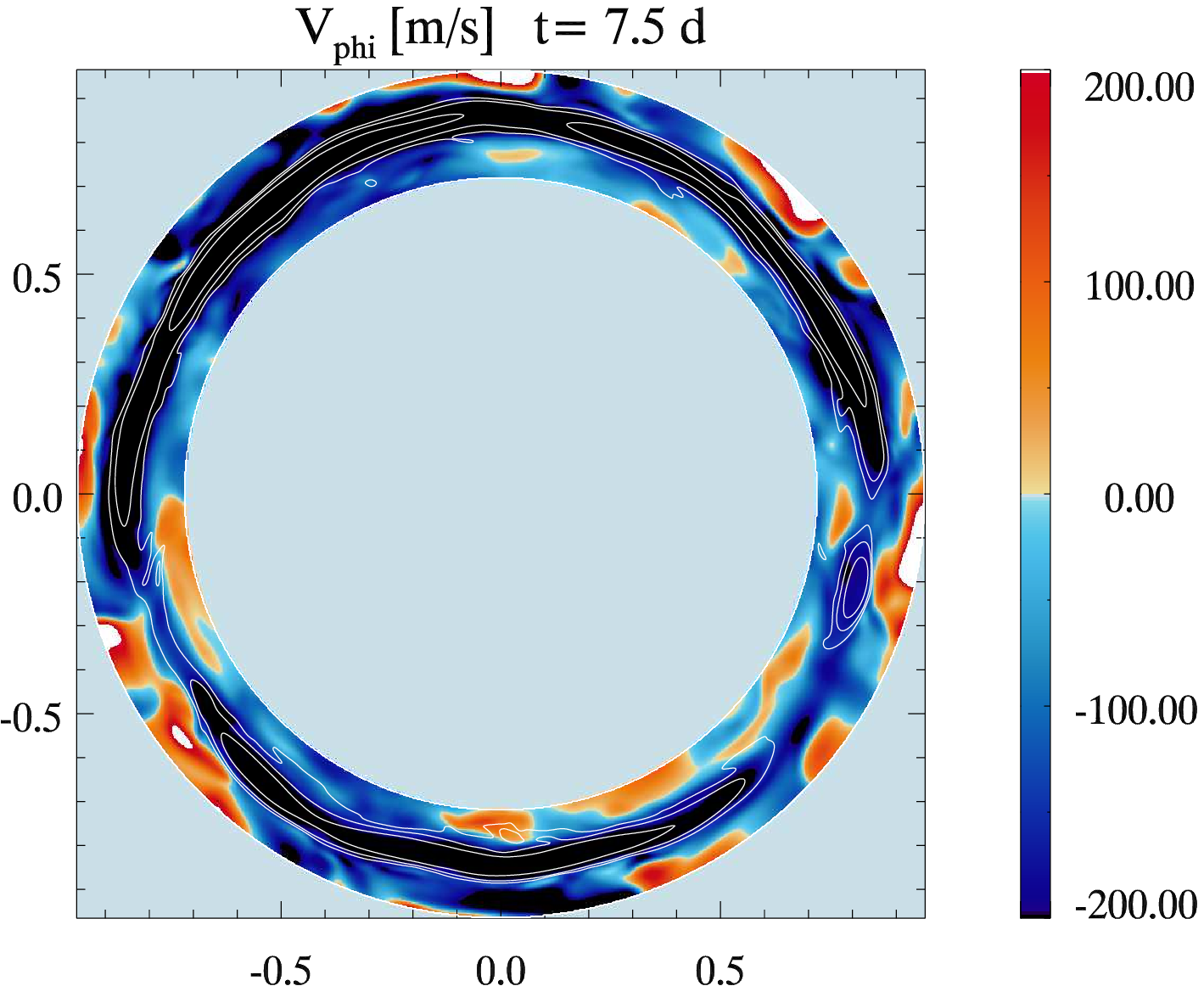}
  \caption{Zonal (azimuthal) flow generated inside the flux-rope at $t=7.5\un{d}$ for the standard case.
    The figure shows a conical slice taken at the latitude of introduction of the flux-rope.
    The colour-scale represent the azimuthal velocity $V_\phi$ (between $\pm 200\un{m/s}$) in the rotation reference frame and the white lines are contours of $B_\phi$.
    (movie in the electronic edition)
  }
  \label{fig:vphi_eqsl}

  \medskip

  \includegraphics[width=0.95\picwd]{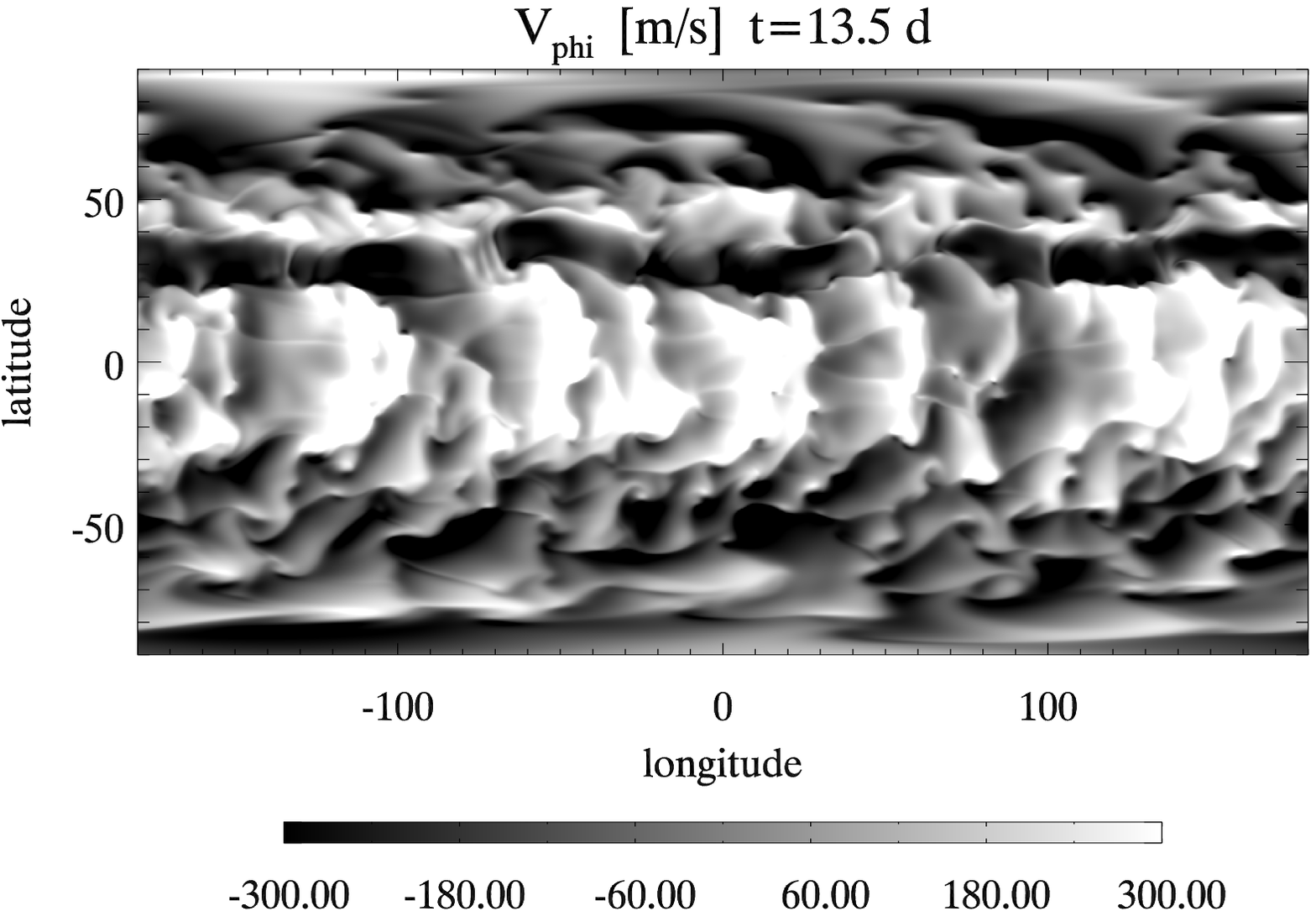}
  \caption{Azimuthal surface flows at $t=13.5\un{d}$ due to the zonal flow generated inside the flux-rope during the buoyant rise in the standard case.
    The colour-scale is saturated at $\pm 300\un{m/s}$.
  }
  \label{fig:surfvphi}

  \medskip

  \centering
  \includegraphics[width=\picwd]{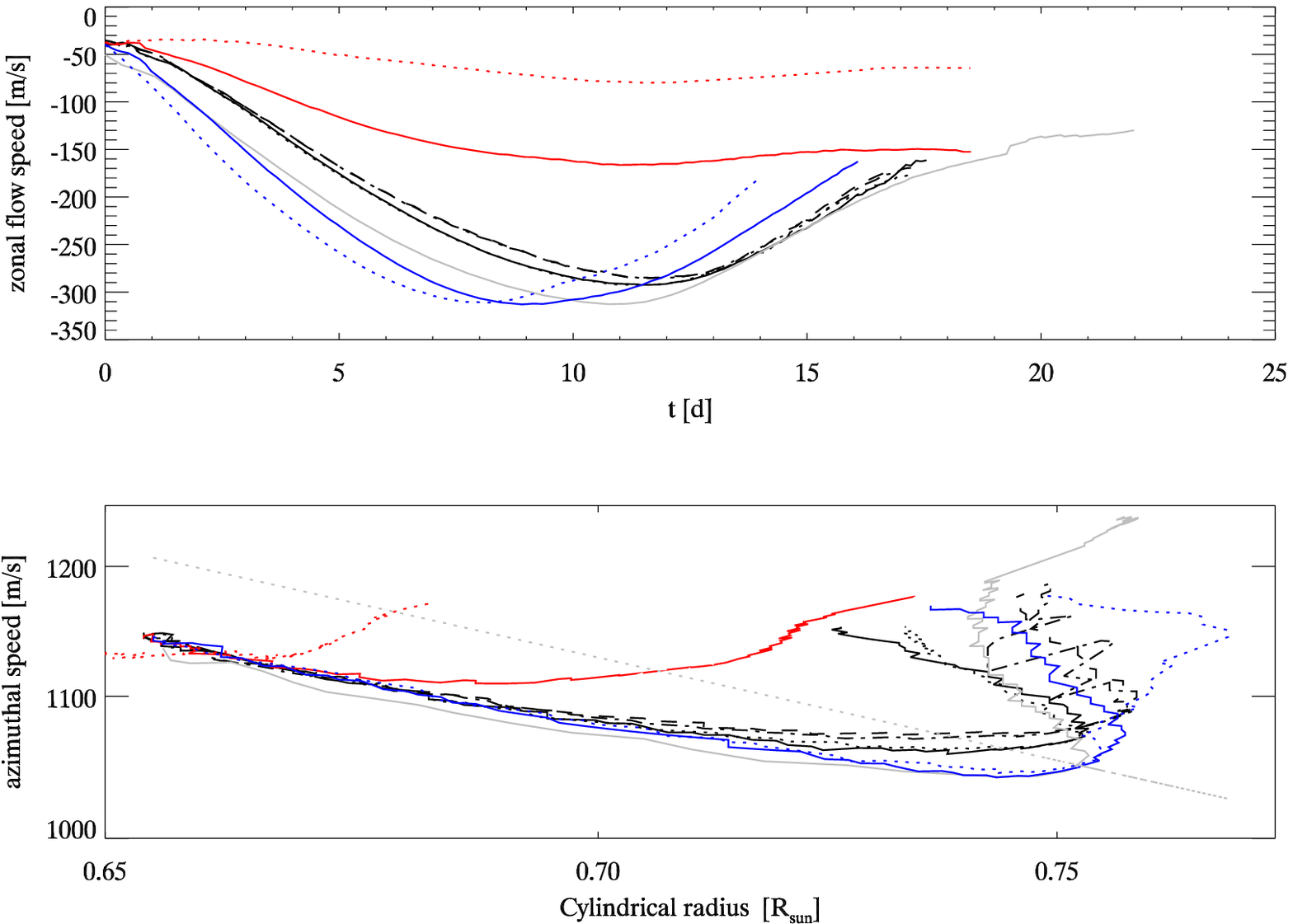}
  \caption{Evolution of the zonal (azimuthal) flow generated inside the flux-rope as a function of time (top panel) and cylindrical radius (bottom panel) for different runs.
    The velocity is measured at the axis of the flux-rope, which is defined here as the position of the maximum of $B_\phi$ on the meridional plane.
    The dotted grey line indicates the slope of a path corresponding to a zonal flow verifying exact angular momentum conservation (inside the flux-ropes).
}
  \label{fig:vphi_time}
\end{figure}

The rise of the magnetic flux-rope also perturbs the global background state by inducing a zonal (azimuthal) flow inside the flux-rope itself.
Initially, the rotation rate of the flux-rope perfectly matches the background rotation state.
The flux-rope then rises as a coherent and self-connected magnetic structure.
The plasma inside the flux-rope thus needs to decrease its azimuthal velocity with respect to the background in order to conserve angular momentum.
Figure \ref{fig:vphi_eqsl} shows a conical cut taken at $30\un{N}$ latitude (a $\{r,\phi\}$ map for a given constant $\theta$).
The colour-scale represents the azimuthal velocity $v_\phi$ (relative to the rotating frame, with $\Omega = \Omega_0$; see Sect. \ref{sec:equations}) at an intermediate phase of the flux-rope trajectory.
Contours of $B_\phi$ are superimposed to trace the position of the flux-rope.
Note that the relative amplitude of these ``backward'' flows is high ($200\un{m/s}$ in the figure, but becoming as high as $300\un{m/s}$ in some cases) and that they remain strong when the flux-ropes reach the surface.
At the surface, the $v_\phi$ signal shows an initial sharp increase, reaching its peak surface value in about $\sim 1$ day after the emergence episode starts, and then fades away more smoothly (the signal disappears in $\sim 5$ days).
Figure \ref{fig:vphi_time} shows the evolution of these retrograde zonal flows for different runs.
For the strong $B_0$ cases, the one evolving in a hydrodynamical background develops the strongest zonal flow. The positive polarity cases in a dynamo background then follow, and the negative polarity case produces the weakest zonal flow.
Comparison with Fig. \ref{fig:vphi_time} shows that the flux-ropes attaining a higher buoyant rise velocity also produce stronger retrograde flows.
Despite its variations during the buoyant rise, the amplitude of the induced zonal flow always seems to converge to the same surface value right after the flux-emergence episode, at least for the same flux-rope's buoyancy (same $\Delta\rho/\rho$).
We also note that the later (post-emergence) evolution seems to depend strongly on the flux-rope's polarity (\emph{cf.} Sect. \ref{sec:laterevol}).

We test now whether angular momentum conservation alone can be the sole (or the dominant) cause for the zonal flow generation.
If this is the case, the amplitude of the retrograde flow must only depend on position, and not on the actual buoyant rise velocity, for any given run.
The middle panel in Fig. \ref{fig:vphi_time} shows the zonal velocity amplitude as a function of radius.
It is clear how all runs follow a very close path in this plot, at least in the first half of the convection zone.
The flux-ropes brake down in a sub-surface of small radial extent, in the upper part of the convection zone.
Some of the runs show a clear departure from the main track in this braking region.
This is especially true for the run with a negative magnetic polarity flux-rope, meaning the other physical processes involved have to be related to the magnetic topology and the way the flux-rope connects with the dynamo field.
The bottom panel in Fig. \ref{fig:vphi_time} shows more precisely that the retrograde zonal flow amplitudes inside the flux-ropes are mostly a consequence of angular momentum conservation.
We define the cylindrical radius
\begin{equation}
  \label{eq:cylradius}
  R = r\sin\theta
\end{equation}
as the distance of a point to the rotation axis of the sun.
The azimuthal velocity in the inertial reference frame is
\begin{equation}
  v_\phi^{inertial} = v_\phi + \Omega_0 R\ .
\end{equation}
Conservation of angular momentum then implies simply
\begin{equation}
  v_\phi^{inertial} \propto R^{-1}\ .
\end{equation}
The bottom panel in Fig. \ref{fig:vphi_time} shows $v_\phi^{inertial}$ as a function of the cylindrical radius $R$ in a log-log plot for all runs.
The dotted grey line is a guideline indicating the slope of a $R^{-1}$ curve.
The sections of the plotted lines parallel to this guideline therefore correspond to parts of the flux-rope's trajectories where angular momentum inside them is strictly conserved.
We note that this holds for most of the buoyant rise, until the flux-rope starts braking near the top of the convection zone and angular momentum is necessarily exchanged with the external medium.
This moment corresponds to the flux-emergence episodes discussed previously and to the inflexion point present for all the curves plotted in the bottom panel Fig. \ref{fig:vphi_time}.
Past this episode, the flux-rope still maintains its spatial coherence for some time, and starts being advected poleward by the sub-surface meridional flow (hence the decrease in cylindrical radius, but not in height).
This phase corresponds to the upper branch (past the inflexion point) in the same plot.
It is interesting to note that the zonal flow inside the flux-ropes remains close to angular momentum conservation in these later phases.
The departure to a conservative path is carried out mostly by intermittent exchanges of momentum, plasma and magnetic flux with the environment.

Figure \ref{fig:surfvphi} shows surface maps similar to those in Figs. \ref{fig:magnetogram} but this time of $v_\phi$.
It is clear from this figure how a strong azimuthal shear is imposed at the surface during the flux-emergence episode.
This strong shearing is nevertheless transient.
Its amplitude decays and fades away quickly; the signal is almost indiscernible $5$ days after the beginning of the emergence episode.
Also, the surface $v_\phi$ perturbation is centered around the PIL, as it is carried upwards within the emerging flux-rope.
The latitudinal width of the perturbation is close to that of the $B_r$ and $v_r$ in Figs. \ref{fig:magnetogram} and \ref{fig:velogram}.
The maximum surface shearing amplitude ($\propto\partial_\theta v_\phi$) caused by the zonal flow therefore takes place near the boundaries of the emerging flux-rope.

We note that the retrograde zonal flows inside the flux-ropes are always predominantly azimuthal, showing little toroidal vorticity. 
The magnetic field inside the flux-rope remains twisted and rotates azimuthaly as a whole, rather than enforcing a noticeable inner helical zonal flow.

\section{Global magnetic field configuration}
\label{sec:global}

Let us now see in more detail how the global magnetic field is affected by the emergence of the flux-ropes, and how it evolves past the flux-emergence episode (as defined in Sect. \ref{sec:emergence}).
We will focus on the surface distribution of the magnetic field as a function of time and estimate its consequences for the external field evolution during the emergence episode by means of potential field extrapolations.
We will assume in what follows that the extrapolated magnetic field provides a reasonable qualitative indication of the potential component of the coronal field (hence not of the total magnetic field).

\subsection{Surface magnetic field distribution}
\label{sec:laterevol}

\begin{figure}[]
  \centering
  \includegraphics[width=\picwd,clip=true,trim=0 0 0 0]{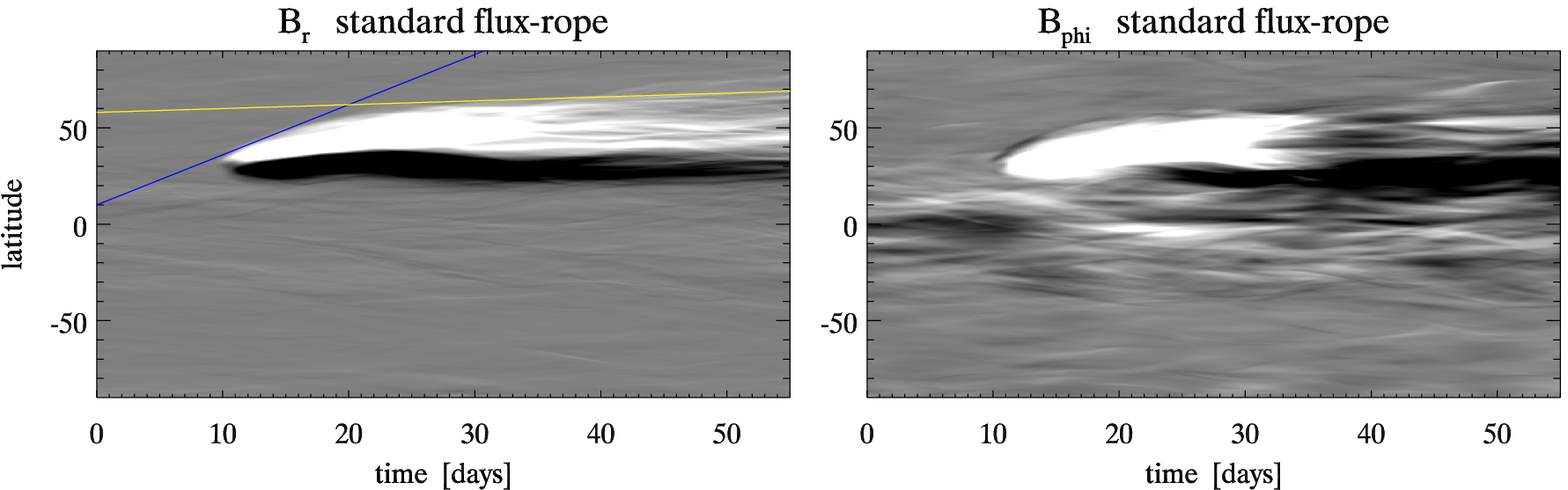}\\ \vspace{0.01\picwd}
  \includegraphics[width=\picwd,clip=true,trim=0 0 0 0]{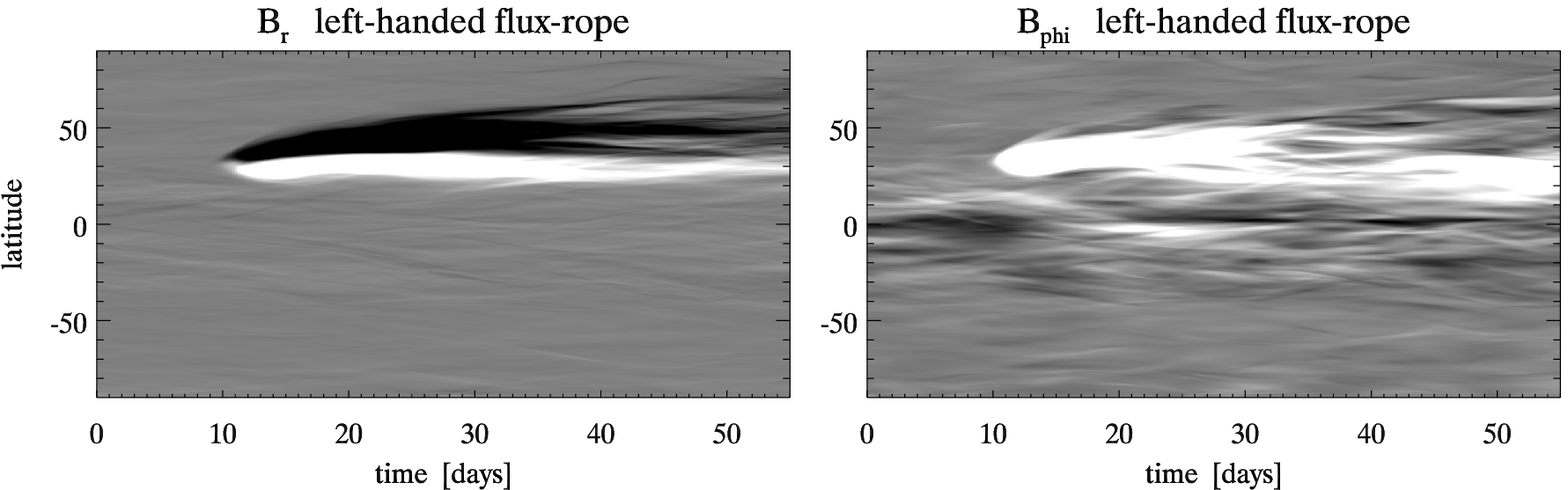}\\ \vspace{0.01\picwd}
  \includegraphics[width=\picwd,clip=true,trim=0 0 0 0]{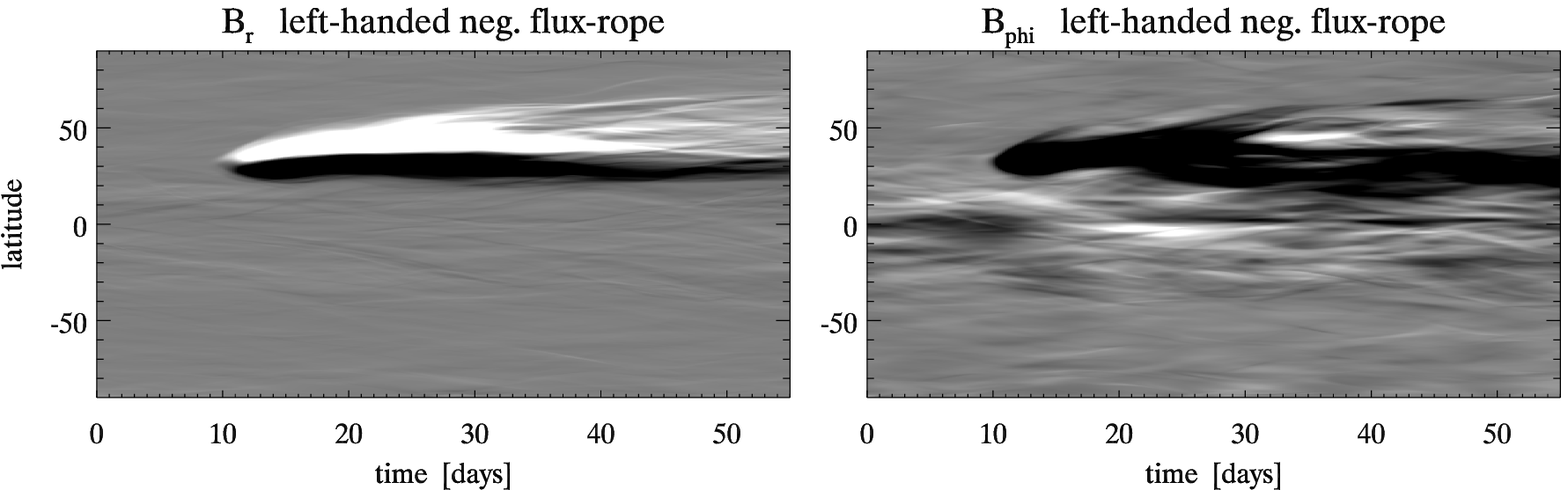}\\ \vspace{0.01\picwd}
  \includegraphics[width=\picwd,clip=true,trim=0 0 0 0]{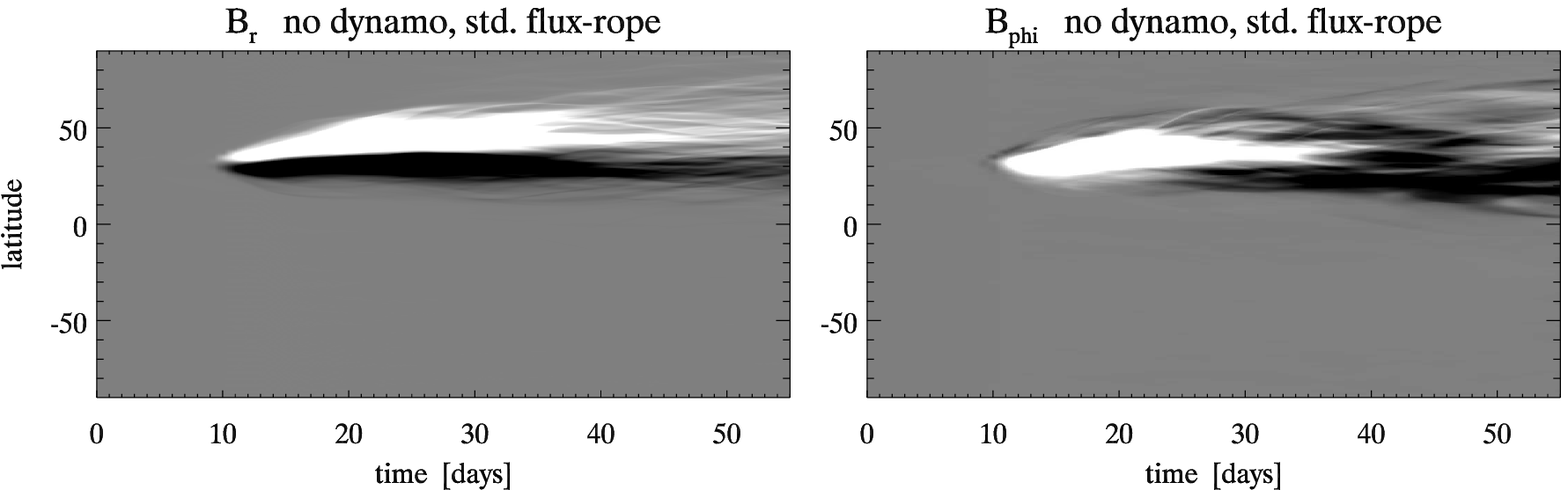}\\ \vspace{0.01\picwd}
  \includegraphics[width=\picwd,clip=true,trim=0 0 0 0]{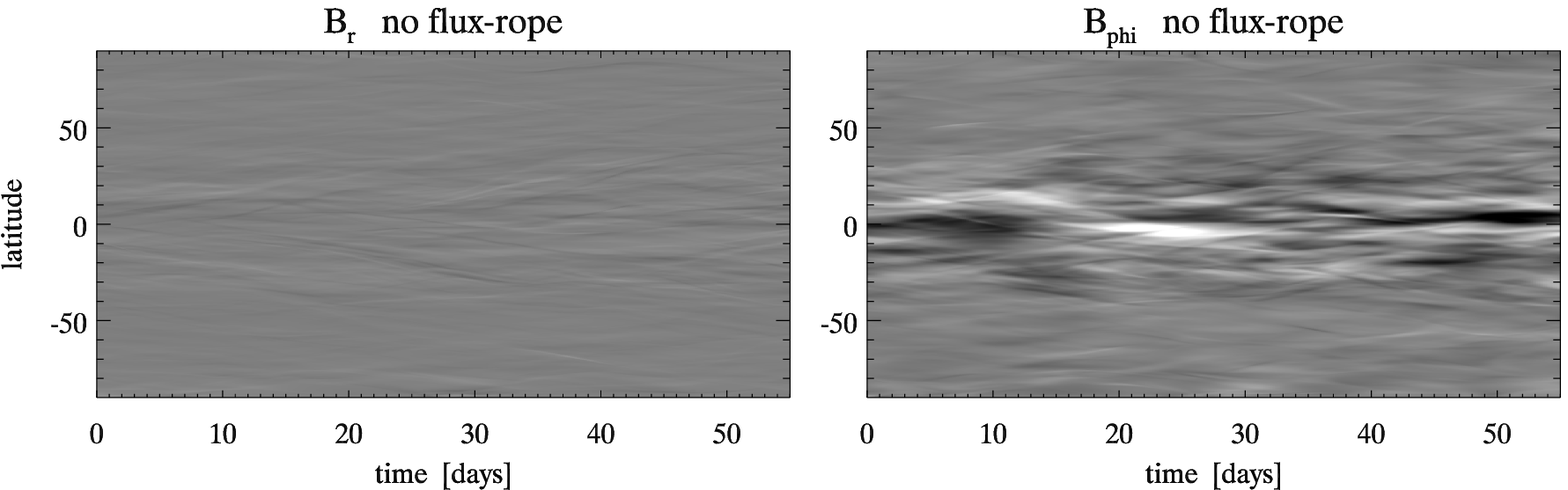}
  \caption{``Butterfly'' diagrams showing the latitudinal distribution of radial and azimuthal magnetic field for, respectively from top to bottom, the standard case, case with inverted handedness, case with inverted handedness and polarity, standard flux-rope in an hydrodynamical background and the background dynamo run (without flux-ropes).
    The blue and yellow lines in the upper left panel indicate, respectively, the poleward drift rates $2.6$ and $0.22\un{^{\circ}/day}$.
  }
  \label{fig:butterflies}
\end{figure}

We will consider primarily the latitudinal distributions of magnetic field at the surface.
Figure \ref{fig:butterflies} shows the temporal evolution of the azimuthally-averaged radial and azimuthal magnetic field at the surface of the Sun for different runs as a function of latitude and time.
The panel in the figure represent, respectively and from top to bottom rows, 
the standard case, 
the case with inverted handedness, 
the case with inverted handedness and polarity, 
the standard flux-rope in an hydrodynamical background and 
the background dynamo run (without flux-ropes).
The emergence episode initially produces a positive/negative magnetic polarity pair ($B_r$, on the left column) and a one-signed signature in $B_\phi$ (on the right column).
Its initial latitudinal extension grows quickly, up to about $20\degrees$ between $t=12\un{d}$ (beginning of the emergence episode) and $t=15\un{d}$.
At this stage, the northernmost limit of the emerged region expands at a rate of about $2.6\un{^{\circ}/day} \approx 365\un{m/s}$ (which is the slope of the blue line in Fig. \ref{fig:butterflies}).
In the standard case, the positive polarity occupies the northern half of the emerging region and $B_\phi$ is positive, as expected for the emergence of a right-handed twisted flux-rope \citep[see e.g][]{pevtsov_helicity_2003,zhang_helicity_2006,zhang_current_2012}.
The remaining cases display symmetric $B_r$ pairs and/or $B_\phi$, according their polarity and handedness.
The time-latitude diagrams show that the emerged magnetic flux is transported poleward globally (and not only in the longitude interval shown in Fig. \ref{fig:magnetogram}), and also that its latitudinal extent keeps growing in time.
Furthermore, the northern half spreads more strongly than the southern half (the asymmetry being slightly more pronounced for the right-handed flux-ropes).
In the case with an hydrodynamical background this magnetic polarity pattern evolves more smoothly than in the cases with a dynamo background.
In all cases, the bipolar pattern is progressively distorted by the surface flows and ends up being disrupted between $t=30$ and $t=40$ days (compare with the last two columns in Fig. \ref{fig:magnetogram}, showing the surface distribution of $B_r$).
The emerged azimuthal magnetic field also displays the same latitudinal spread and disruption patterns.
More surprisingly, a strong negative $B_\phi$ signal appears at the surface at $t\sim 20\un{d}$ in the standard dynamo case.
This signal outlives the original $B_\phi$ signal (with positive sign).
This feature is not introduced by the background dynamo, as it does not appear in the background dynamo run (without flux-rope; see bottom row in Fig. \ref{fig:butterflies}).
Moreover, the hydrodynamic (no dynamo) counterpart of the standard case shows the same behaviour.
The key parameter is the flux-rope's handedness: all right-handed cases display this behaviour, while left-handed flux-ropes do not.

The northernmost edge of the emerged region drifts at this moment northward (poleward) at a nearly uniform rate of $\sim 0.22\un{^{\circ}/day} \approx 30\un{m/s}$ for all cases (yellow line in Fig. \ref{fig:butterflies}).

\subsection{Consequences for the external field}
\label{sec:coronal}

\begin{figure}[] 
  \centering
  \includegraphics[width=\picwd,clip=true,trim=55 0 55 0]{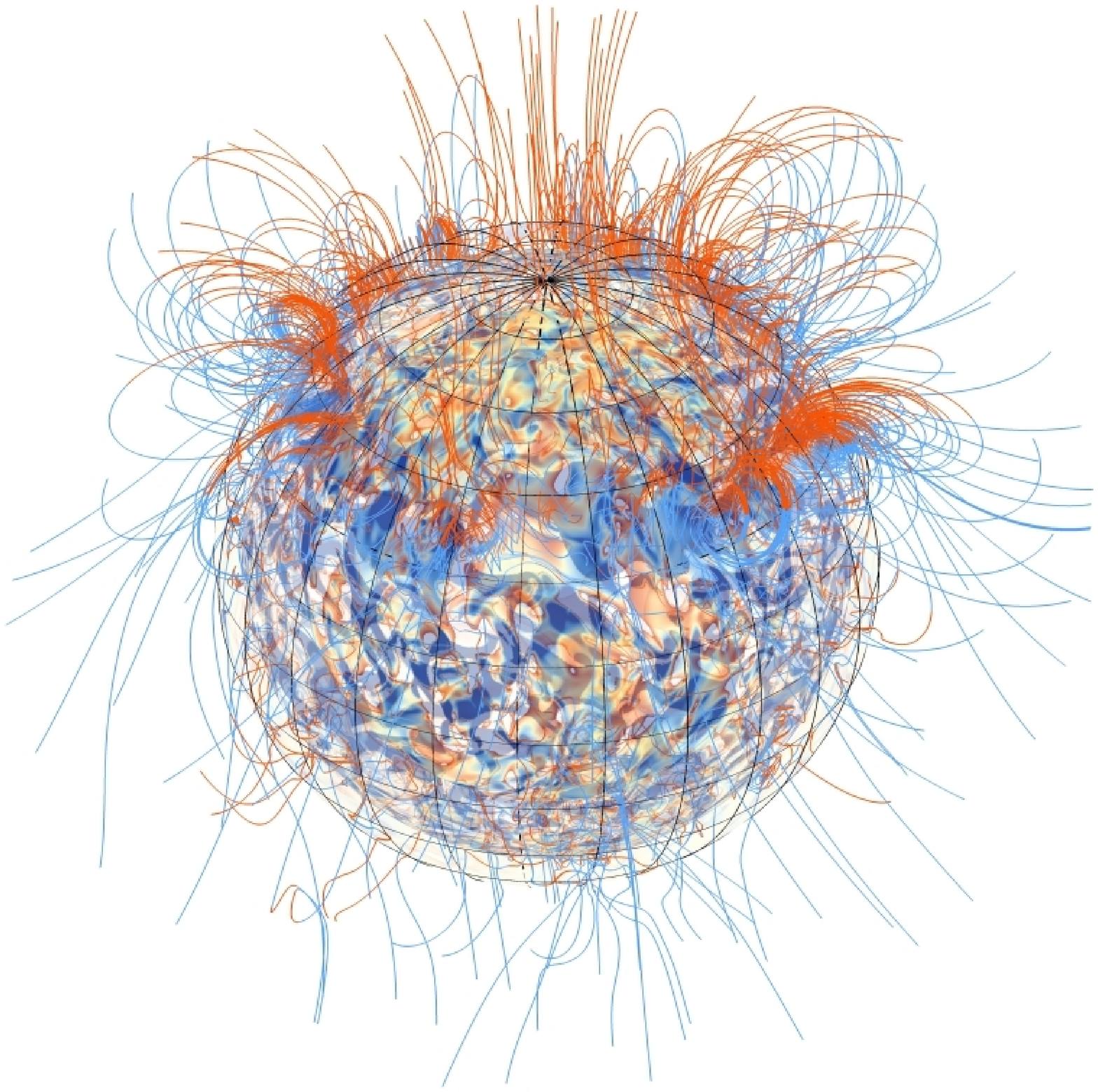}
  \includegraphics[width=\picwd,clip=true,trim=55 0 55 0]{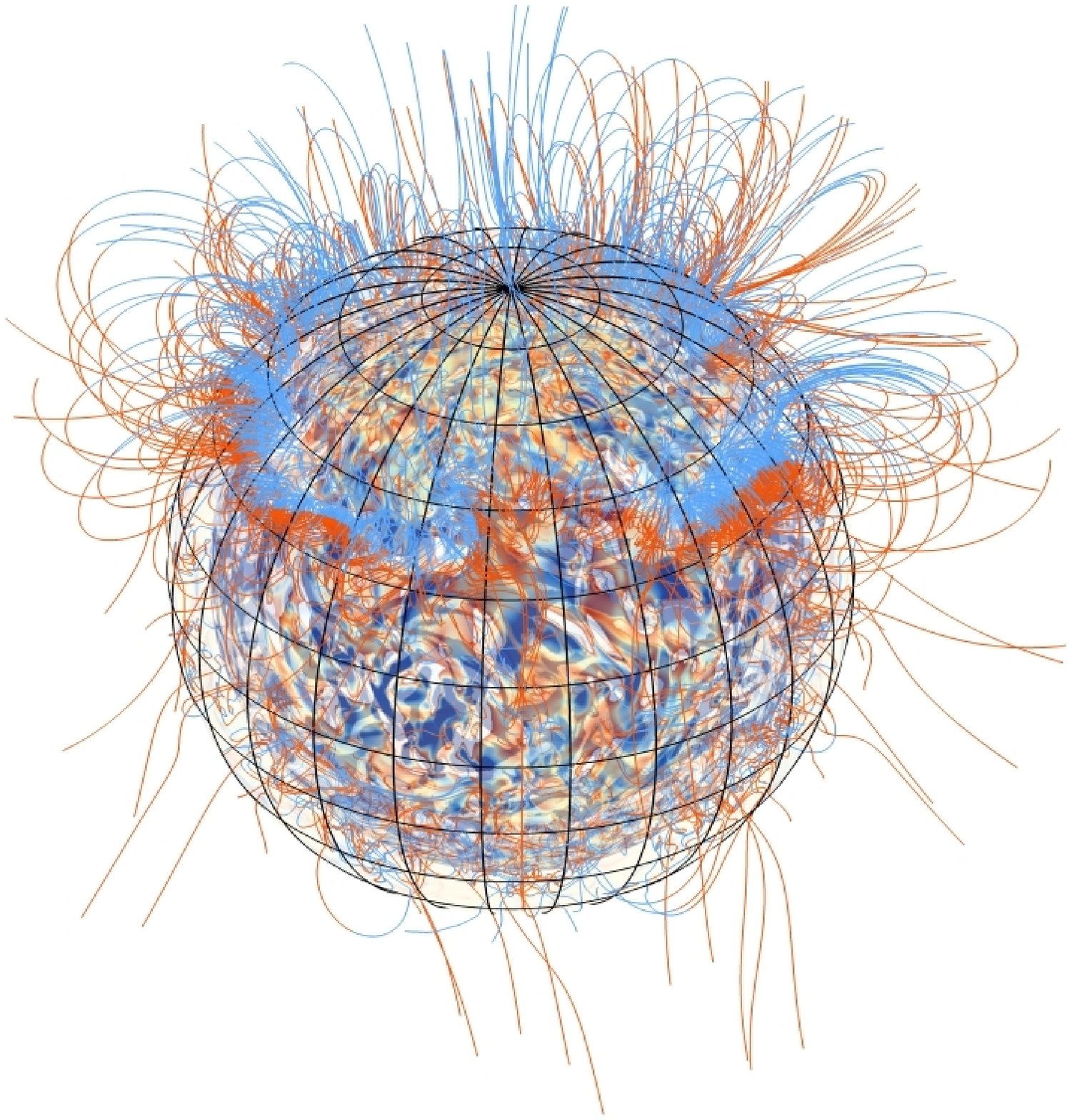}
  \caption{Three-dimensional renderings of the magnetic field in the convection zone and of the extrapolated magnetic field for the standard case (top) and the case with a right-handed flux-rope (bottom), both at $t=25\un{d}$.
  The lines represent magnetic field lines and the surfaces are spherical cuts in the CZ.
  The colours blue and orange represent, respectively, negative and positive values of $B_r$.}
  \label{fig:extrap_3D}
\end{figure}

\begin{figure*}[] 
  \centering

  \includegraphics[width=.19\linewidth]{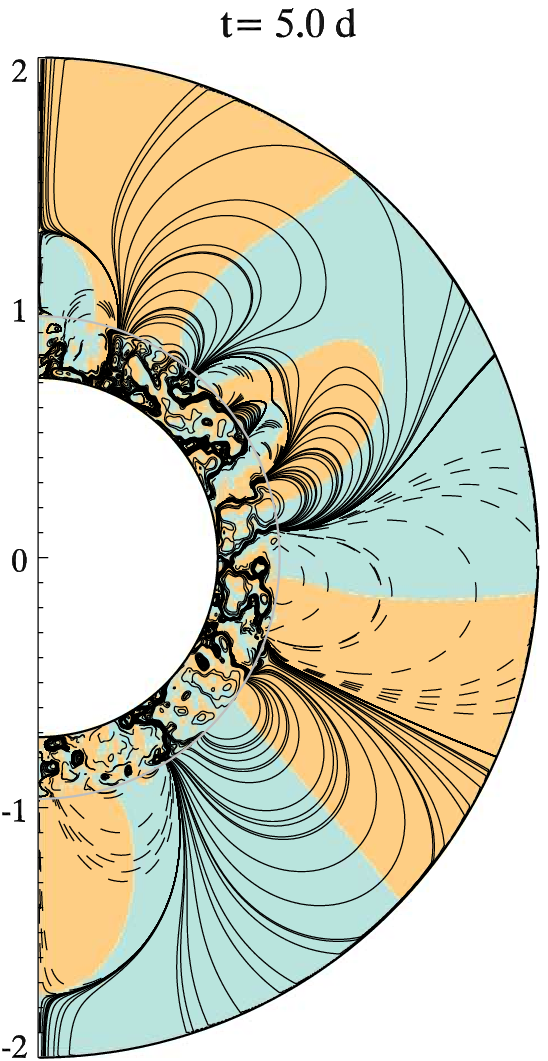}
  \includegraphics[width=.19\linewidth]{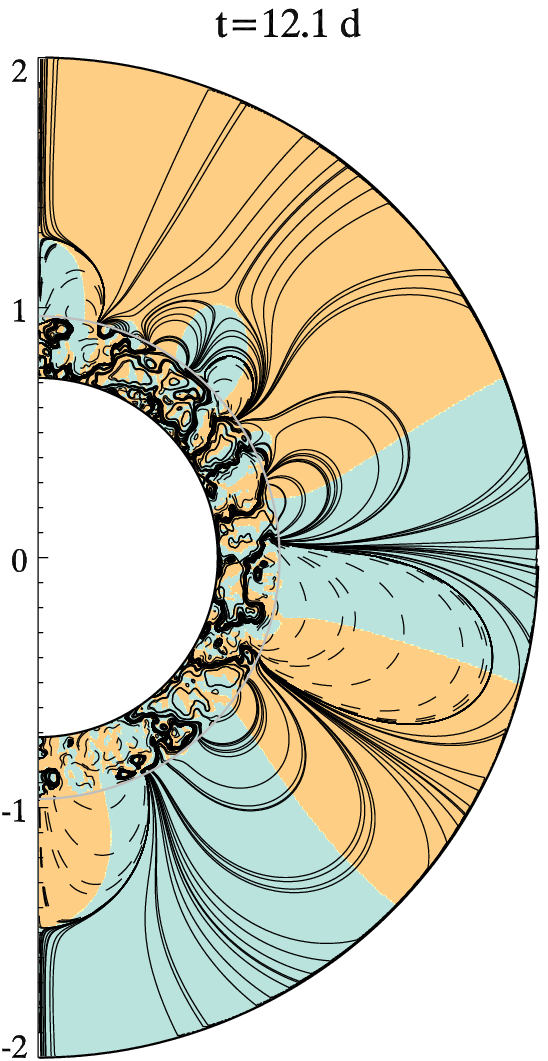}
  \includegraphics[width=.19\linewidth]{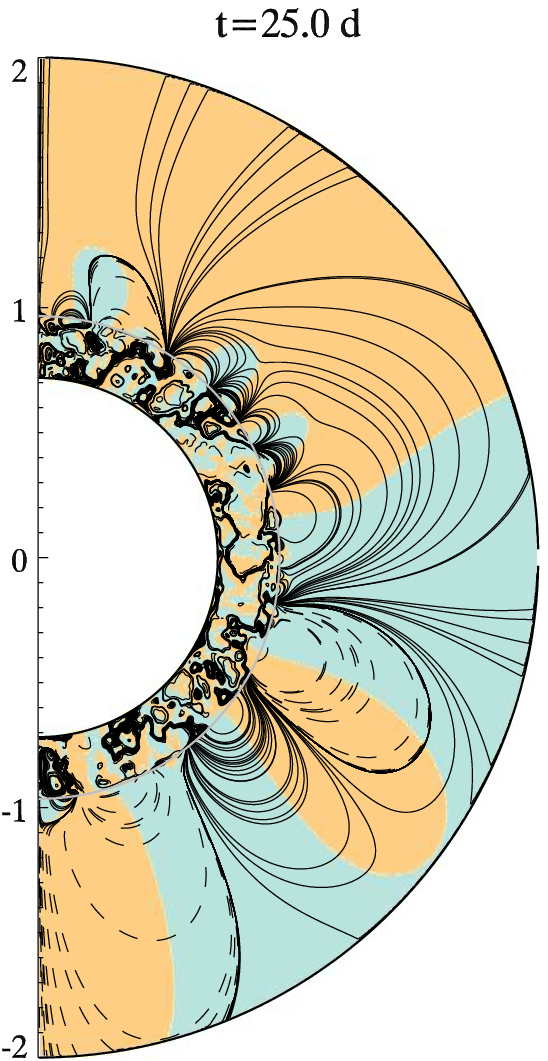}
  \includegraphics[width=.19\linewidth]{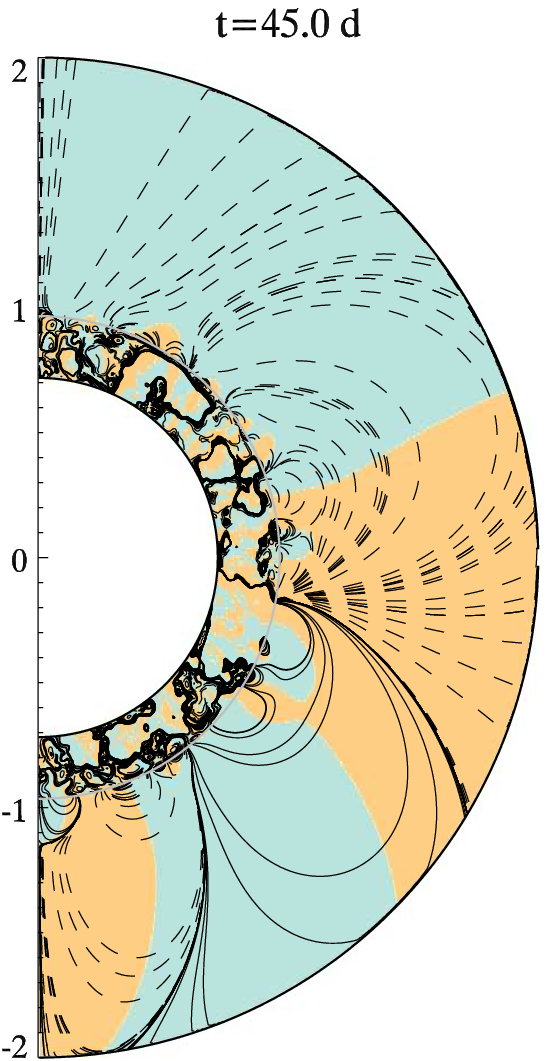}
  \includegraphics[width=.19\linewidth]{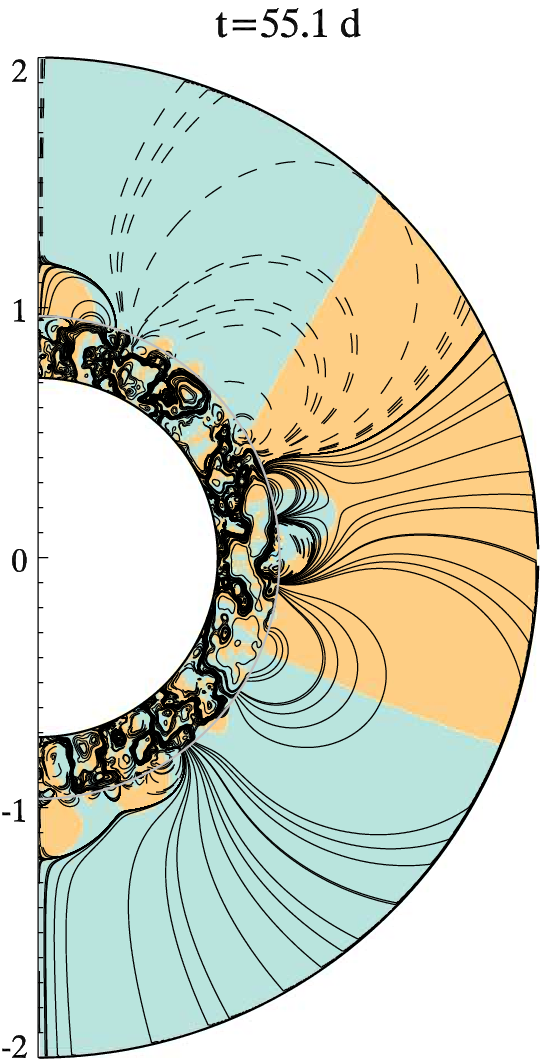} \\

  \includegraphics[width=.19\linewidth]{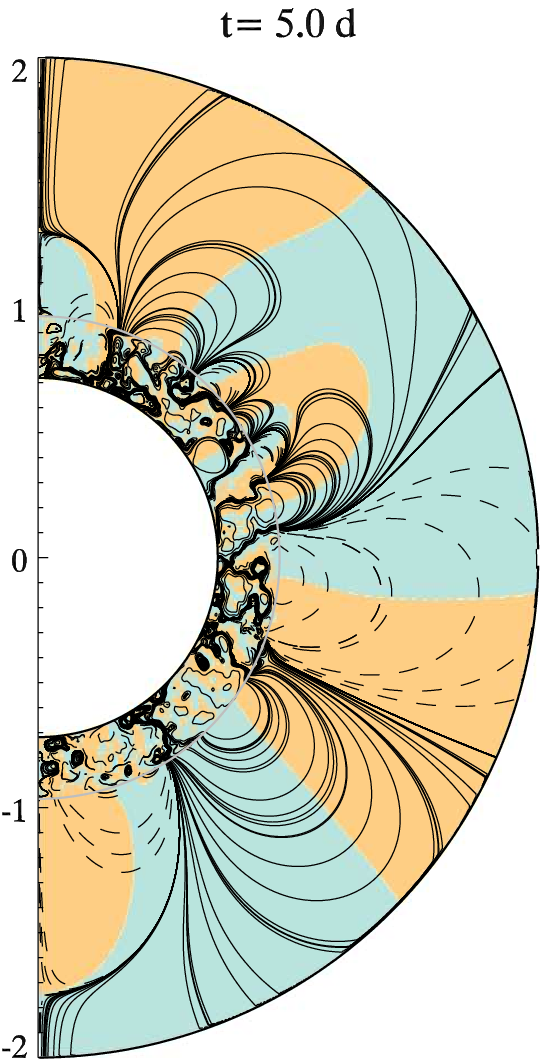}
  \includegraphics[width=.19\linewidth]{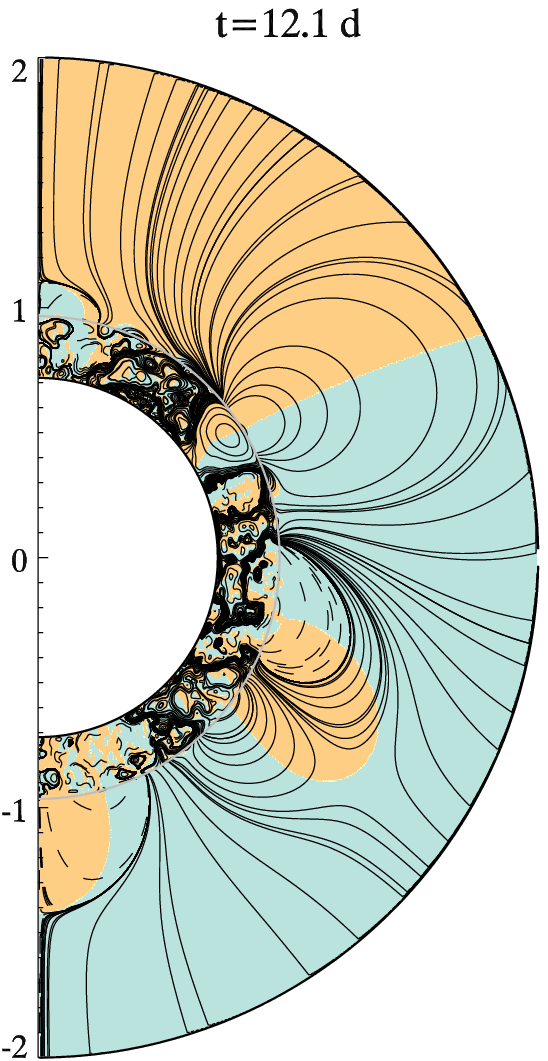}
  \includegraphics[width=.19\linewidth]{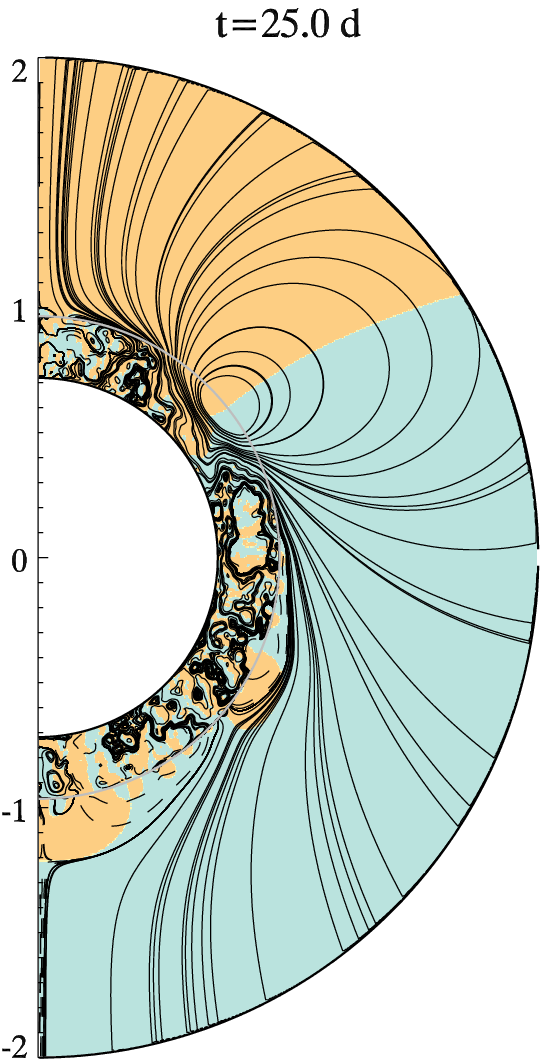}
  \includegraphics[width=.19\linewidth]{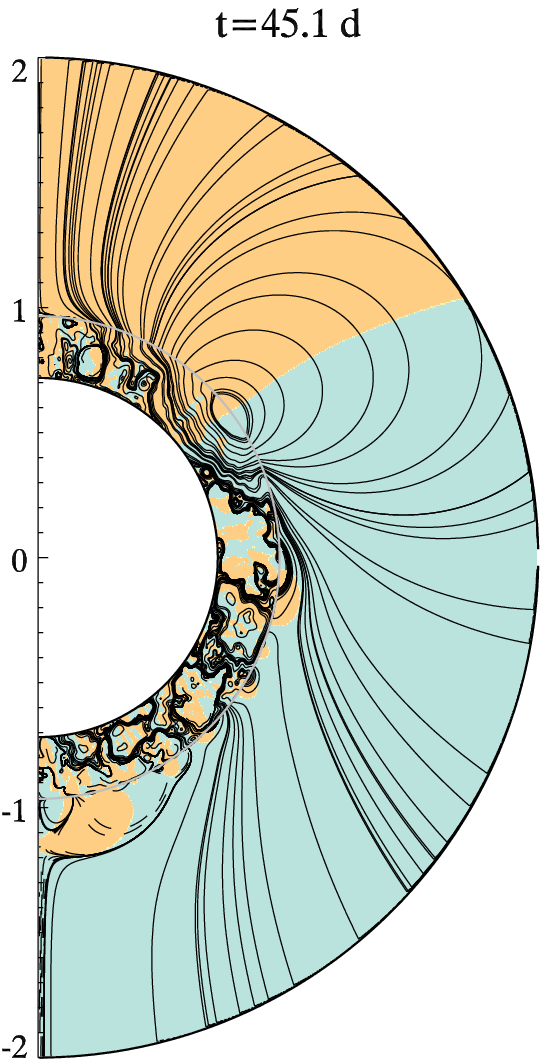}
  \includegraphics[width=.19\linewidth]{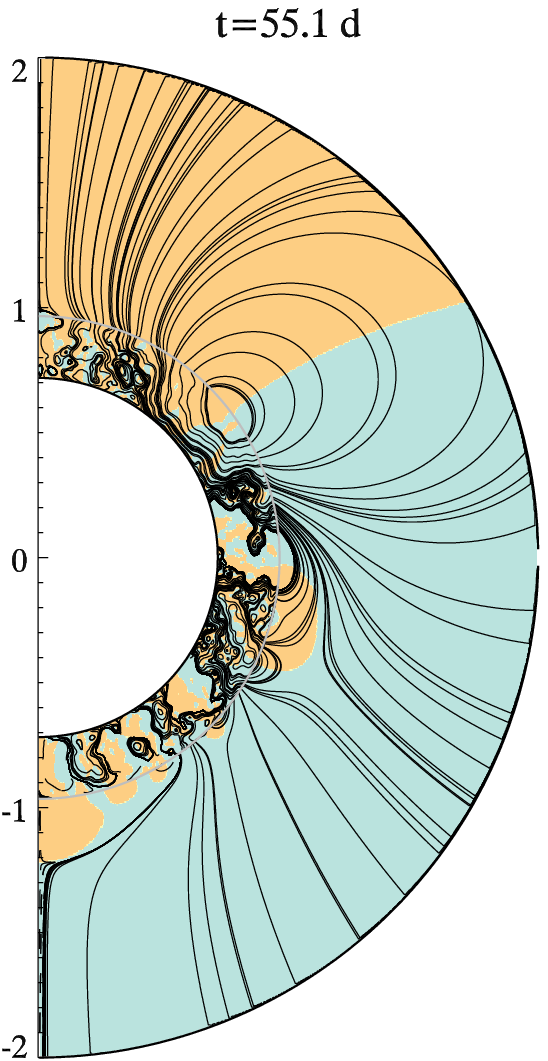} \\

  \includegraphics[width=.19\linewidth]{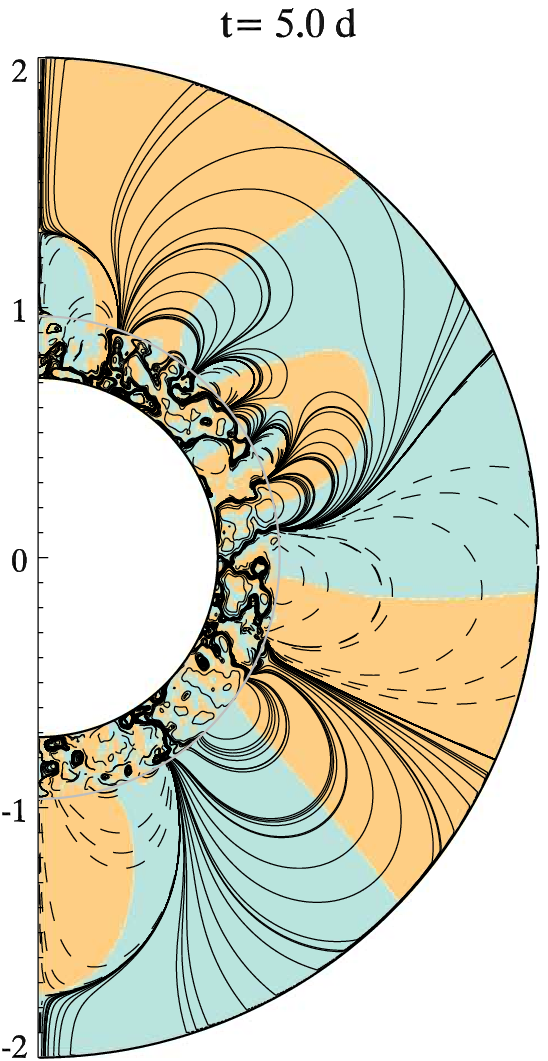}
  \includegraphics[width=.19\linewidth]{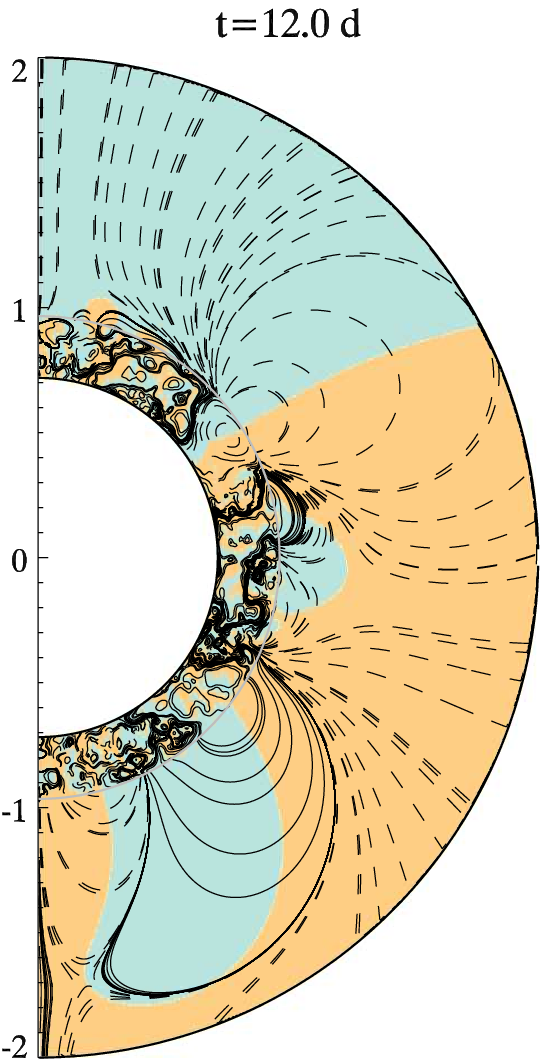}
  \includegraphics[width=.19\linewidth]{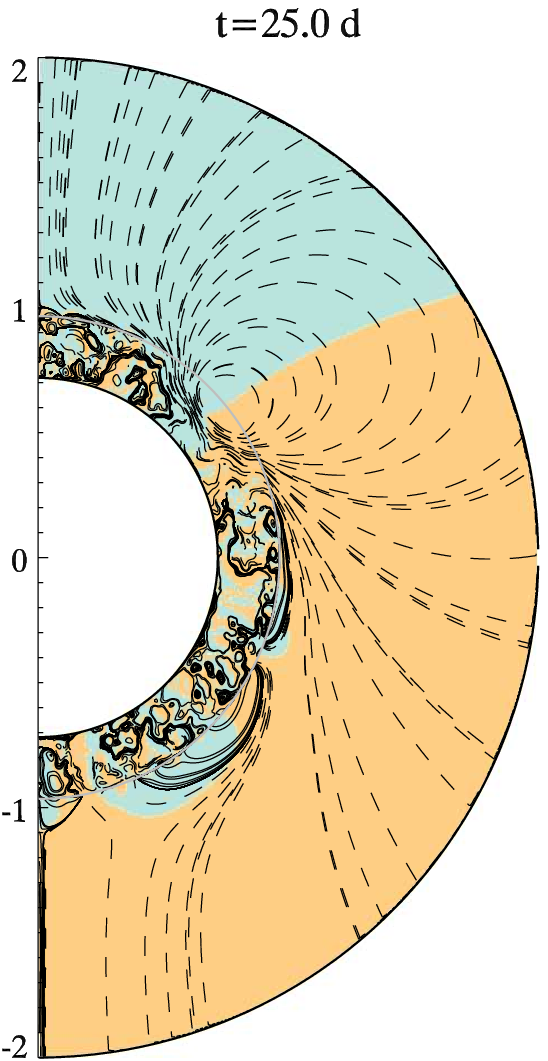}
  \includegraphics[width=.19\linewidth]{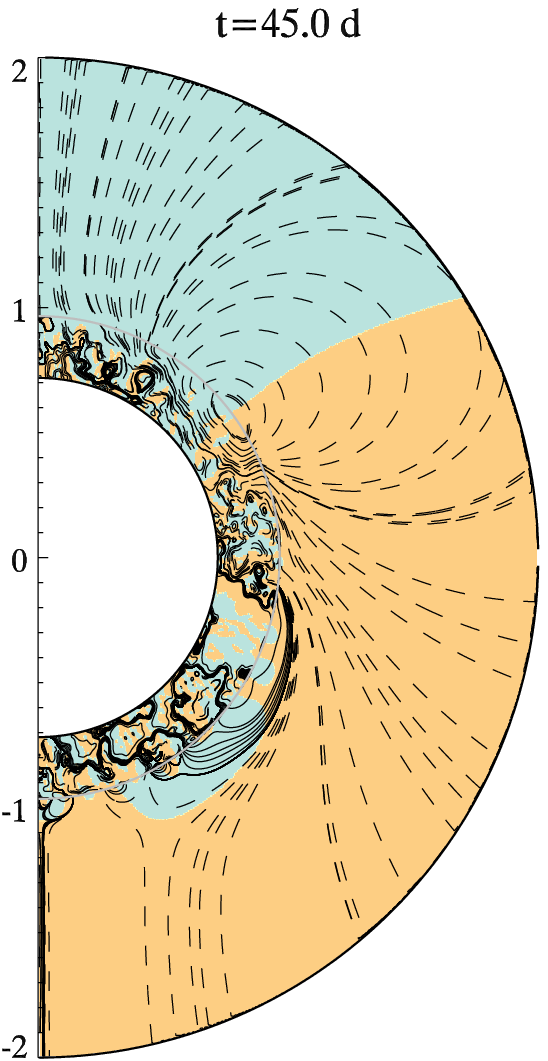}
  \includegraphics[width=.19\linewidth]{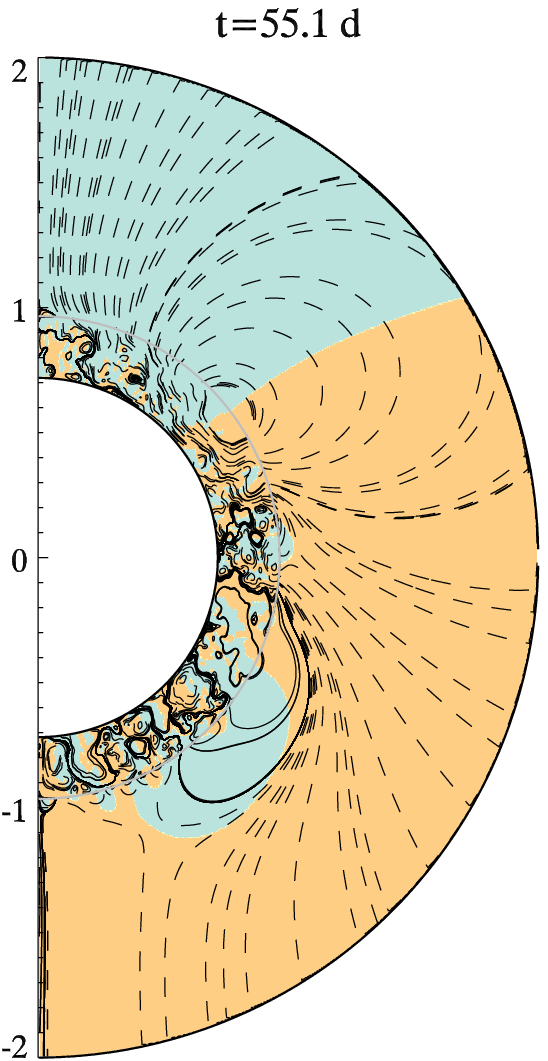}

  \caption{Time-series of the azymuthally-averaged poloidal magnetic field in the CZ and the extrapolated potential field in the corona for the dynamo run without flux-rope (top row), for the dynamo row with a standard flux-rope (middle row) and for the dynamo run with a flux-rope with the inverse polarity (bottom row).
    The figures show the convective zone (with the light grey curve representing the upper boundary of the numerical domain) and the extrapolated field up to $2\rsun$. 
    Black lines represent magnetic field lines, with continuous and dashed lines representing magnetic loops with opposite chirality (continuous lines correspond to CW-oriented magnetic loops and dashed lines to CCW-oriented loops).
    Orange and blue tones represent the regions where the radial magnetic field is, respectively, positive and negative.
}
  \label{fig:extrap_multi}
\end{figure*}

\begin{figure}[] 
  \centering
  \includegraphics[width=.38\linewidth]{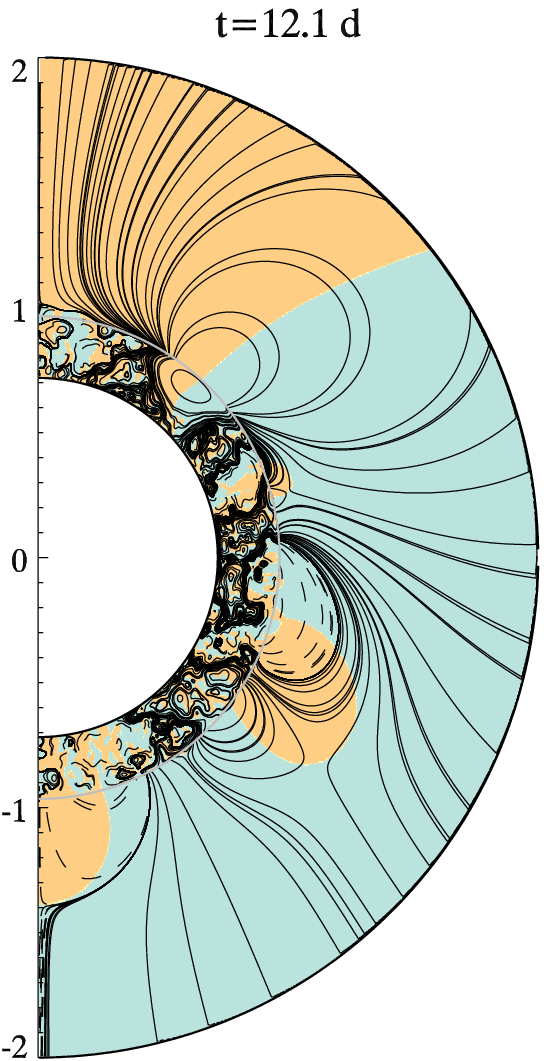}
  \includegraphics[width=.38\linewidth]{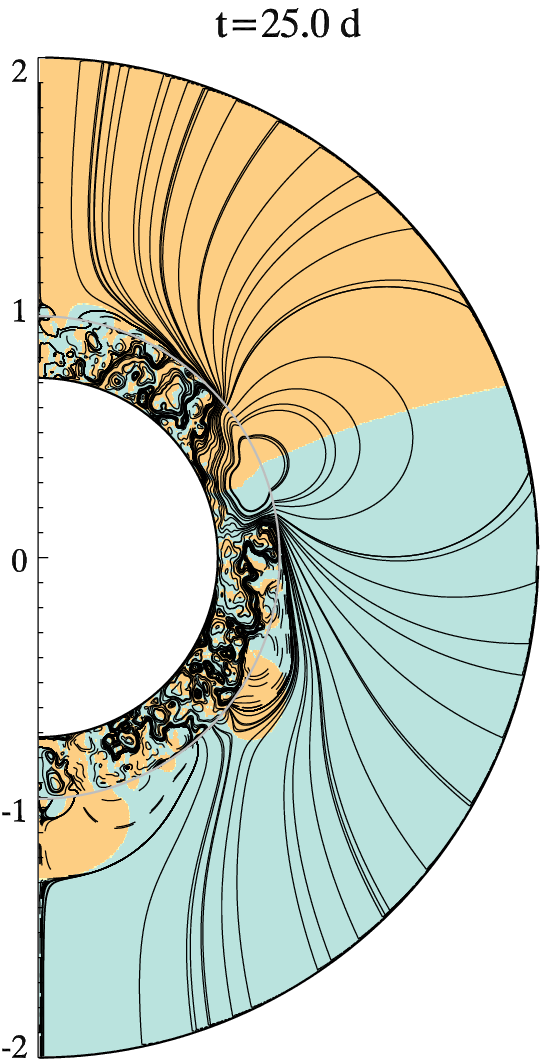}
   \caption{Snapshots of the high-latitude and low-latitude cases at the instants $t=12\un{d}$ and $t=25\un{d}$ (respectively).
     These instants correspond roughly to the stage in the evolution of the emergence episode represented in the second column in Fig. \ref{fig:extrap_multi} for the cases at standard latitude.}
   \label{fig:extrap_varlatitude}
 \end{figure}

\begin{figure}[] 
  \centering
  \includegraphics[width=\picwd]{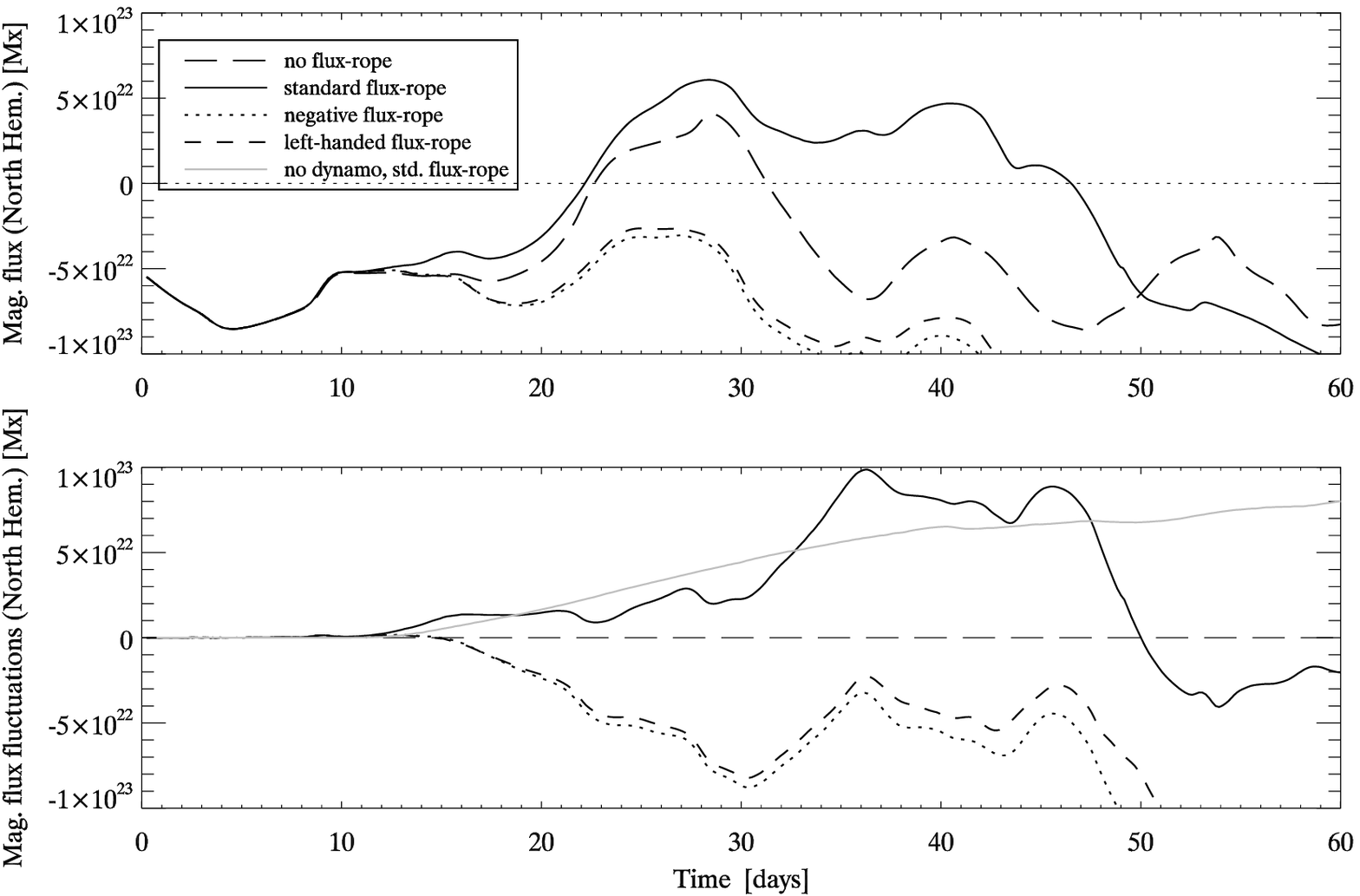}
  \caption{Signed magnetic flux $\int_0^{\pi/2} B_r\cdot dS$ at the surface as a function of time (over the northern hemisphere only).
    The top panel shows the total magnetic flux while the bottom panel shows the deviations relative to the run with background dynamo field but no flux-rope.
    The runs represented are the standard case,
    the cases with opposite polarity and handedness, 
    the hydro case (standard flux-rope without dynamo field) and 
    the dynamo run without flux-rope.
    The hydro case is only shown in the second panel.
    The figure shows that the global polarity of the magnetic field is strongly dominated by the emerged flux in the strong field cases.
    This is not so much the case for the weaker field cases.
    The difference between the hydro and standard cases (respectively, grey and black continuous lines) shows that the flux-rope's field interacts with the background dynamo field.
  }
  \label{fig:magnetic_flux}
\end{figure}

The flux emergence episodes produce, in our simulations, a strong impact in the global magnetic field topology.
In order to make this more evident, we show on Figure \ref{fig:extrap_3D} two three-dimensional renderings of the magnetic field in the CZ and of the extrapolated field for the standard case and for the case with a right-handed flux-rope.
The extrapolations were computed from the surface magnetic field (\emph{i.e}, from the data shown in the first column of Fig. \ref{fig:butterflies})
The magnetic field lines are represented as continuous lines coloured in blue where $B_r<0$ and in orange where $B_r>0$.
The instant represented is $t=25\un{d}$, close to the peak in emerged magnetic flux.
We clearly see how the overall polarity near the north pole is different in both cases, being strongly influenced by the emerged flux polarity.
In the low latitude case, trans-equatorial loops develop and the impact of the emergence extends over both hemispheres (not shown in Fig. \ref{fig:extrap_3D}; see Fig. \ref{fig:extrap_varlatitude}). 
In all cases, the polarity inversion lines does not remain everywhere aligned with the azimuthal direction, and covers a latitudinal interval of more than 15 degrees.

Figure \ref{fig:extrap_multi} shows the geometry of the poloidal axi-symmetric component of the coronal field as obtained by potential field extrapolation of the $\phi$-averaged surface magnetic field at different instants.
The figure shows three time-series (one per row) spanning the interval $t=5-55\un{d}$.
The top row shows the dynamo run, without magnetic flux-rope.
The middle row shows the standard case, and the bottom row the case with a flux-rope with negative polarity.
The colour-scale represents the radial magnetic field polarity, with $B_r > 0$ in orange and $B_r < 0$ in blue.
The continuous and dashed lines are magnetic field lines, the former corresponding to CW-oriented magnetic loops and the latter to CCW-oriented loops.
Although the coronal magnetic field is not expected to be potential, potential field extrapolations as those in Figs. \ref{fig:extrap_3D} to \ref{fig:extrap_varlatitude} are known to give a reasonable indication of its global topology, showing the location of the strong/weak flux concentrations and the coronal magnetic connectivity \citep{wang_topological_2003,schrijver_photospheric_2003}.
The study of the dynamical evolution of the corona requires a different type of approach, based on MHD descriptions or at least on non-potential approximations \citep[\emph{e.g},][]{yeates_solar_2010}.
Nevertheless, it as been shown that the slowly-varying global-scale coronal magnetic field structure is still well reproduced by potential field extrapolations, and that these produce estimates close to those from full MHD models \citep{riley_comparison_2006,hu_three-dimensional_2008}.
Potential field extrapolations are, therefore, useful tools to elaborate hypothesis on such dynamics, and can ultimately be used to initialise time-dependent MHD simulations \citep[e.g,][]{lionello_effects_2005}.

The background dynamo undergoes on its own a short global polarity reversal, taking place between $t=20\un{d}$ and $t=30\un{d}$ (see Fig. \ref{fig:magnetic_flux} to see this more precisely).
The global external magnetic field is strongly perturbed by the emergence of the flux-rope, which can reinforce or completely inhibit this polarity reversal (depending on the flux-rope's polarity).
This is visible in the fourth column of Fig. \ref{fig:extrap_multi}, for $t=25\un{d}$.
The final states for both cases with flux-ropes clearly display inverse polarities in the northern hemisphere, meaning that the emerged flux is indeed substantial compared to the background external field (generated by the dynamo).
More precisely, in the standard case the emerged flux contributes to an increase in the global magnetic flux and reinforces the underlying dynamo polarity reversal.
The magnetic topology in the northern hemisphere then becomes much simpler than the one due solely to the dynamo field.
A strong dipolar feature clearly dominates over the smaller scale higher-order magnetic structures which are visible in the low corona at the initial state.
In the southern hemisphere, the influence of the emerged magnetic flux is less dominant.
Some topological features resulting from the distribution of surface smaller-scale mixed polarity pairs prevail, such as the pseudo-streamer structure visible at a latitude of $\sim 45^\circ\un{S}$.
In the case with a negative polarity flux-rope, the emerged flux completely inhibits the system to undergo a global magnetic polarity reversal.
%
The latitude of emergence plays an important role on the perturbations to the external magnetic field, and does so in different ways.
  Magnetic flux-ropes placed at lower latitudes will rise more slowly than high latitude ones (see Sect. \ref{sec:buoyantrisespeed} and Fig. \ref{fig:fluxrope_height}), and hence emerge at different instants.
  The emerged magnetic field then interacts with different states of the time-dependent background dynamo global field.
  More specifically, the low-latitude case studied here ($15^\circ\un{N}$) emerges before the background dynamo reverses its global polarity, while the high-latitude case ($45^\circ\un{N}$) emerges afterwards.
  Figure \ref{fig:extrap_varlatitude} shows a snapshot of the poloidal magnetic field (CZ and extrapolated field) both for the high and the low-latitude cases.
  The instants represented were chosen to correspond roughly to the same stage in the evolution of the emergence episode (taking into account the different buoyant rise speeds).
  The low-latitude flux-rope introduces a strong topological perturbation to the external field, spreading well across the equator and into the opposite hemisphere.
  The influence of the high-latitude case is more restrained to the hemisphere of emergence.
  This is not only due to the different distances between the emergence site and the equator.
  In fact, the total amount of flux-rope related magnetic flux which crosses the upper boundary depends on the latitude of emergence.
  Similar flux-ropes, with the same cross-section and toroidal magnetic flux, will reach the top of the domain with different azimuthal extents according to the latitude of emergence.
  Hence, the amount of magnetic flux available to cross the surface scales as $\sin{\left(\theta\right)}$ (ignoring other effects).

It is worth noting that these effects are long-lasting; the external magnetic field geometry roughly maintains its newly acquired topology during the post-emergence phases, even though the signs of the emerged flux at the surface are already disappearing (\emph{cf.} the last column in Fig. \ref{fig:magnetogram}).
%
This supports the idea that much of the magnetic flux involved in the actual process of polarity reversal in the sun may be carried outwards by the emerging active region flux-ropes.

Figure \ref{fig:magnetic_flux} shows the signed magnetic flux integrated at the surface over the northern hemisphere as a function of time, which approximates the amplitude of the global solar dipole.
The top panel then compares the global dipole polarity of the dynamo run without introduction of a flux-rope (long-dashed line) and for the strong-field cases with different handedness and polarities (continuous, dotted and dashed black lines).
The bottom panel shows the same curves, but with the dynamo background run subtracted off.
The case with a hydrodynamical background (continuous grey line) is also represented here for comparison.
As discussed in the previous paragraph, the background dynamo undergoes, by itself, a short polarity reversal between $t=20 - 30$ days.
This figure lets us see more quantitatively how in the runs with magnetic flux-ropes the emerged flux is sufficient to either reinforce or completely inhibit the polarity inversion.
This effect depends only on the flux-rope's own poloidal magnetic polarity; the flux-rope's handedness is irrelevant to it.
The dotted and dashed curves -- which follow each other closely -- represent runs setup with the same polarities, but opposite handedness.
The case featuring a standard flux-rope evolving in an hydrodynamical background shows a strikingly simpler behaviour, associated with a very uniform growth in surface magnetic flux.
The background dynamo magnetic flux amplitude reaches a maximum of $10^{23}\un{Mx}$ during the time interval we are considering, with a time-averaged amplitude (r.m.s.) of about $4\e{22}\un{Mx}$.
The maximum contribution of the emerging flux-ropes themselves amounts up to $10^{23}\un{Mx}$.
The total magnetic flux amplitude depends on how both components (dynamo and emerging flux-rope) combine in time.
In our simulations, we found a maximum flux amplitude of about $1.6\e{23}\un{Mx}$.
If the flux-rope were to emerge at a different time, we could expect a slightly higher value, closer to $2\e{23}\un{Mx}$.
These values are on the upper side (but within range) of the observed distributions given by \citet{schrijver_photospheric_1994} and \citet{rempel_flux-transport_2006}.
Note that our simulations produce an azimuthal activity band rather than individual active regions (AR).
The net magnetic flux is then necessarily high, with contributions from a few simultaneous ARs.
Also keep in mind that Fig. \ref{fig:magnetic_flux} represents only the cases with the strongest $B_0$ and that the magnetic fluxes reported above correspond to those strong-field cases.
The simulation fluxes were computed at $r=0.97\rsun$; the actual photospheric fluxes are likely to be smaller.

As discussed in Sect. \ref{sec:energy_balance}, Figure \ref{fig:fluxrope_me} (bottom right panel) shows the total Poynting flux crossing the upper spherical boundary of the domain as a function of time.
Note that the sign of Poynting flux is defined such that an outward energy flux is negative in the figure.
That is, a negative Poynting flux means electromagnetic energy is being transferred from the CZ to the outside, while a positive flux means the opposite.
The figure shows that the whole flux-emergence episode ($t=12-20\un{d}$ for the standard case) is indeed associated with a strong increase in Poynting flux (into the corona).
After the emergence episode, the Poynting flux actually reverses for some time.
This time interval corresponds to the third and fourth columns in Figs. \ref{fig:magnetogram} and \ref{fig:extrap_multi}.
The emerged magnetic field is losing its spatial coherence and slowly decaying at this moment.

The maximum amplitude depends strongly on the flux-rope's strength (scaling roughly as $B^2$).
The emerging flux-rope's handedness also plays a role here.
In our runs, left and right-handed cases show differences of a factor $0.1$ in amplitude.
The polarity of the flux-rope (the sign of $B_\phi$) is unimportant, though.
The presence of a background dynamo field also conditions the results.
The grey curve in Fig. \ref{fig:fluxrope_me} represents the evolution of the hydrodynamical background case, for which there are some quantitative differences (specially in the later phases of the emergence episode).

\section{Discussion and conclusions}
\label{sec:discussion}

The results discussed in the previous sections are based on a series of global-scale numerical simulations using the ASH code to model flux emergence in a spherical convective shell possessing simultaneously differential rotation, meridional flows
and a dynamo generated magnetic field. 
We have investigated how buoyant magnetic flux-ropes are influenced by a three-dimensional background nonlinear dynamo during their rise through the convection zone and how they contribute to the global magnetic flux budget.
Initial position, flux-rope twist and radius were chosen following the conclusions in \citet{jouve_three-dimensional_2009}, who studied the buoyant rise of such magnetic structures in a fully developed but hydrodynamical CZ.
We focused on parameters which relate more directly to the interactions between the flux-rope's magnetic field and the background dynamo, namely the flux-rope strength, polarity and handedness, and the flux-rope's initial latitude.
This study is, to our knowledge, the first one addressing this problem systematically.
\citet{dorch_buoyant_2007} performed a first step on this direction, but
they limited their simulations to cartesian setups where the background field was fixed and uniform.

Our main results can be summarised as follows
\begin{enumerate}
\item \label{item:fluxropedynamo}
  The effects of the interaction between the flux-rope and the background magnetic field are negligible in the initial phases of the buoyant rise, but become progressively more important as the flux-ropes evolve.
  The overall buoyant rise speeds are marginally lower than for flux-ropes evolving in a purely hydrodynamical convective background.
  Specific orientations of the flux-rope's magnetic field lines with respect to the background field can produce large rise velocity disparities.
  For a given flux-rope configuration, the largest source of buoyant rise speed disparities is the flux-rope initial latitude.
\item \label{item:emergedflux}
  The fraction of the flux-rope's magnetic flux which emerges is large enough to strongly perturb the global topology of the external field.
  The emerged field is predominantly dipolar and may contribute to enhancing the global dynamo polarity reversal or prevent it from happening.
  The global magnetic topology remains affected by the emerged flux for a long time, well beyond the period of time during which the surface tracers of flux-emergence are visible.
\item \label{item:brhoscaling}
  The amplitude of the flux-rope's magnetic field $B_c$ and density $\rho_c$ were observed to scale as $B_c \propto \rho_c^\alpha$ with $\alpha\lesssim 1$ during the buoyant rise phase.
  This differs from the results of local-scale simulations of flux-emergence in the upper layers of the CZ only \citep{cheung_simulation_2010}.
\item \label{item:zonalflow}
  The flux-rope transports a retrograde zonal (azimuthal) flow, which shows a strong signature at the surface levels.
  This zonal flow manifests itself as a localised surface shearing whose actual amplitude depends on the latitude at which the flux-rope emerges (due to the difference between the mean differential rotation and the flux-rope's azimuthal velocity).
  The shearing amplitude is maximum for low latitude flux-ropes (up to a few hundreds of $\un{m/s}$) albeit with a limited duration.
\item \label{item:patches1}
  A set of discontinuous North-South aligned magnetic bipolar patches appear as the buoyant flux-ropes reach the top of the domain.
    These are then sheared and twisted by the surface flows, and the emerged magnetic flux is pushed into the convective cell boundaries.
    The bipolar patches have an intrinsic magnetic helicity which depends directly on the flux-rope's polarity and handedness.
    The helicity's amplitude and signal changes afterwards as a result of the underlying horizontal surface motions, namely the latitudinal shearing of the mean azimuthal velocity (differential rotation) and vertical vorticity at the latitude of emergence.
  \item \label{item:patches2}
    The bipolar patches first expand quickly as the buoyant flux-ropes slow down and come to a complete stop.
    From then on, this magnetic flux is slowly advected polewards by the underlying meridional flows.
    The nature of the distribution of the surface magnetic patches varies with the latitude of emergence, the low-latitude cases being particularly patchy (as they rise more slowly and hence are more affected by the interaction with the convective flows).
\item \label{item:precursors}
  Magnetic flux emergence is preceded by a strong and localised enhancement in $v_r$ (accompanied by a weaker density increase) at the place where the flux rope will emerge.
  A sharp current density increase is also observed immediately before the actual flux emergence episode.
  We suggest thus the following temporal sequence: 1) increase in $v_r$ and $\rho$; 2) sharp increase in $J^2$; 3) increase in magnetic flux amplitude.
  The exact temporal delays between these diagnostics depends on the specific properties of the emerging flux-ropes model, but the ordering seems to be general.
\item \label{item:doubletail}
    The typical double-tailed cross-section for the flux-rope is not so clearly found (as in simulations using an hydrodynamical background), as the peripheral magnetic flux is interchanged continuously with the magneto-convective environment.
\end{enumerate}


All of the points listed above (except points \ref{item:precursors} and \ref{item:doubletail}) describe global-scale consequences of the buoyant rise and emergence of magnetic flux-ropes, or effects which are affected by the underlying global-scale flow properties.
Please note that the top of the numerical domain (the ``surface'') is placed at $0.97\rsun$, 
and flux-emergence episodes are hence defined here as enhancements of the ``surface'' magnetic flux related to the arrival of a buoyant flux-rope.
For these reasons, our estimations concerning surface emergence diagnostics are meant to give an insight on the sub-surface dynamics and flux-emergence processes rather than to produce directly observable features.
\del{
It is possible that some of these diagnostics vanish at the surface, making them a target for indirect detection methods such as with local helioseismology techniques \citep{ilonidis_detection_2011,kosovichev_local_2012,hanasoge_seismic_2012} rather than for more direct methods.
We nevertheless believe that the predictions concerning the temporal ordering (item \ref{item:precursors} in the list above) and general qualitative properties of the flows and magnetic flux variations induced at the surface by the rising flux-ropes are robust in regards to the dynamics of the topmost layers of the Sun.}
Also note that we introduced buoyant twisted magnetic flux-ropes in pressure equilibrium directly at the base of the CZ with various tunable parameters (twist, location, amplitude), 
as the physical mechanisms at the origin of such buoyant magnetic structures are out of the scope of our study \citep[see ][]{nelson_magnetic_2013}.
The dynamical evolution of these flux-ropes can be described in terms of their buoyant rise speed, trajectory and interaction with the surroundings alone.
The interaction with the multi-scale dynamo magnetic field is at first weak but becomes more and more important as the flux-rope rises from the deep layer of the convective envelope to the top of the domain.
The added magnetic energy is reprocessed by the dynamo, yielding a higher magnetic energy but similar growth rate than the reference flux-rope free dynamo run.

%

We performed potential field extrapolations of the surface magnetic fields during the flux-rope buoyant rise, emergence and post-emergence phases in order to have an idea on how the emerged flux perturbs the external magnetic field.
\del{Extrapolations of this kind do not reproduce the complex dynamics of the real solar corona, but give a reasonable indication of the global-scale magnetic topology.
We also note that, as the topmost surface layers of the Sun are missing in our model, the amplitude of the emerged external field is probably overestimated.}
\add{Due to the caveats expressed in Section \ref{sec:scope_limits}, such extrapolations should not be compared with diagnostics of the solar atmosphere.}

The emergence episode (that is, from $t=12$ to $t=20$ days for our standard case) is characterised by the growth of a strongly dipolar magnetic arcade system which disrupts the background multipolar dynamo field.
Interestingly, a strong (yet transient) azimuthal flow appears at the surface as the flux-emergence episode proceeds, being a source of latitudinal surface shearing.
The azimuthal flow is centered right in the middle of the emerging arcades, as it corresponds to a zonal flow carried within the flux-ropes.
As a result, the shearing is strongest near the boundaries of the emerging region;  and it's not clear if there is net shearing between opposite footpoints of the arcades.

In the simulation presented in this work, the diffusive coefficients were kept fixed for all runs.
An effort should be made in future work to allow these coefficients to be lowered (increasing the turbulence levels in the domain)  and to explore different magnetic Prandtl numbers.
It would be extremely interesting to be able to reproduce the flux-rope formation mechanism of \citet{nelson_buoyant_2011} in simulations like ours.
This is to our knowledge still unattainable by the current global convection models rotating at the solar rotation rate, but work is being done to reach that goal, nevertheless.
Achieving much lower diffusivities is one of the key aspects of the problem.

Future work will also consider a better description of the photospheric layers \citep{pinto_3d_2011} and the presence of multiple buoyant flux-ropes in both hemispheres.

\acknowledgements
  This work was supported by the ERC Grant \#207430 (STARS2 project, PI: S. Brun, http://www.stars2.eu) and the CNRS PNST Interfaces group. %
  Computations were carried out using CNRS IDRIS and CEA's CCRT facilities (GENCI 1623 project).
  We thank L. Jouve for useful discussions.

\bibliographystyle{aa} 
\bibliography{refs}    

\end{document}